\documentclass[aps,amsfonts,pra,twocolumn,superscriptaddress,showpacs,eqsecnum]{revtex4}
\usepackage{epsfig,amsmath,amssymb,bm,epsf,graphics,psfrag,verbatim,subfigure,color}

\def\VOPGS{\Omega_{\textrm{GF}}}
\def\VOPSP{\Omega_{\textrm{SP}}}
\def\StraTensT{\boldsymbol{\epsilon}}

\def\MyZzero{\Xi_{0}}
\def\PartNumb{N}
\def\PartPosi{c}
\def\Volu{V}
\def\Pres{p}
\def\VExc{v_{E}}
\def\ExclVolu{\nu^2}
\def\ExclVoluRN{\tilde{\nu}^2}
\def\PartPosi{c}
\def\LinkNumb{M}
\def\LinkScal{a}
\def\DEDist{\mathcal{P}}
\def\RealDiso{\chi}
\def\LinkDens{\eta^2}
\def\LinkPote{\Delta^{(0)}}
\def\LinkPoteRepl{\Delta^{(1+n)}}
\def\BoltCons{k_B}
\def\HelmFreeEner{F}
\def\GibbFreeEner{\mathcal{G}}
\def\lda{\lbrack}
\def\rda{\rbrack}
\def\ltha{\langle}
\def\rtha{\rangle}
\def\lthal{\Bigg\langle}
\def\rthal{\Bigg\rangle}
\def\VLink{v_L}
\def\HamiExcl{H_0}
\def\ReplProd{\prod_{\alpha=0}^{n}}
\def\ReplProdOne{\prod_{\alpha=1}^{n}}
\def\ReplSum{\sum_{\alpha=0}^{n}}
\def\ReplSumOne{\sum_{\alpha=1}^{n}}
\def\REP0{^{0}}
\def\PartitionLiquid{Z_{\mathcal{L}}}
\def\FLiquid{F_{\mathcal{L}}}
\def\DensFunc{Q}
\def\BASE{\boldsymbol{\epsilon}}
\def\RealPart{\textrm{Re}}
\def\ImagPart{\textrm{Im}}
\def\ISLL{\tau}
\def\tISLL{\tilde{\tau}}
\def\DistISLL{\mathcal P}

\def\Contraction{\zeta}
\def\VOP{\Omega}
\def\REPX{\hat{x}}
\def\REPP{\hat{p}}
\def\REPa{^{\alpha}}
\def\REPb{^{\beta}}
\def\LocaPart{Q}
\def\LocaLeng{\xi}
\def\UnivPara{\theta}
\def\DefoPosi{R}
\def\deformation{u}
\def\DefoScalPsi{\Psi_R}
\def\KernOne{K_1}
\def\KernTwo{K_2}
\def\ContEner{h}
\def\FreeEnerPhen{\Gamma}
\def\NonLocaKern{G}
\def\BulkModuZero{\lambda_0}
\def\BulkModuZeroP{\lambda ' _0}
\def\BulkModu{\lambda}
\def\DefoGrad{\Lambda}
\def\tz{\tilde{z}}
\def\tp{\tilde{p}}
\def\RelaRand{v}
\def\MeanSheaModu{\rho}
\def\Dime{d}
\def\NonLocaKernZero{G^{(0)}}
\def\NonLocaKernOne{G^{(1)}}
\def\Vect{\underline}
\def\PLong{p^{\textrm{L}}}
\def\PPerp{p^{\textrm{T}}}
\def\PPerpT{{p}^{T}}
\def\RandForc{f}
\def\Jaco{\mathcal{J}}
\def\tNonLocaKern{\widetilde{G}}
\def\tu{\tilde{\deformation}}
\def\StraTens{\epsilon}
\def\tStraTens{\tilde{\epsilon}}
\def\tStraTensT{\tilde{\boldsymbol{\epsilon}}}

\def\tDefoGradT{\tilde{\boldsymbol{\Lambda}}}
\def\StreBulk{\sigma '}
\def\StreBulkT{\boldsymbol{\sigma} '}
\def\SheaModu{\mu}
\def\BulkModu{\lambda}
\def\tDefoPosi{\tilde{\DefoPosi}}
\def\Stre{\sigma}
\def\StreT{\boldsymbol{\sigma}}
\def\tq{\tilde{q}}
\def\ty{\tilde{y}}
\def\NLM{M}
\def\NLMZero{\NLM^{(0)}}
\def\NLMOne{\NLM^{(1)}}
\def\tNLM{\tilde{\NLM}}

\def\ProbG{\mathcal{P}_{\NonLocaKern}}
\def\Corr{\gamma}

\def\RelaRandL{\RelaRand_{\DefoGrad}}
\def\PLongL{\Tens{p}^{\textrm{L}}_{\DefoGrad}}
\def\PPerpL{\Tens{p}^{\textrm{T}}_{\DefoGrad}}
\def\Tens{\mathbf}
\def\MetrTens{\Tens{g}}
\def\RandForcL{\RandForc_{\DefoGrad}}
\def\DefoGradT{\Tens{\DefoGrad}}
\def\IdenT{\Tens{I}}
\def\PLongT{\Tens{p}^{\textrm{L}}}
\def\PPerpT{\Tens{p}^{\textrm{T}}}
\def\NADF{w_{\DefoGrad}}
\def\tzL{\tz_{\DefoGrad}}
\def\Tran{^{\textrm{T}}}
\def\tg{g}
\def\trOne{t_1}
\def\trTwo{t_2}

\def\PartDens{n_0}
\def\DiffG{D}
\def\BASEl{\hat{\BASE}_{\lambda}}
\def\GSDF{U}
\def\HVOPGF{H_{\VOP}}
\def\HVOPSP{H_{\VOP}^{(\textrm{SP})}}
\newcommand\T{\rule{0pt}{3.0ex}}

\newcommand\colorred{}
\newcommand\colorgreen{}

\begin{document}

\title{Soft random solids and their heterogeneous elasticity}

\author{Xiaoming Mao}
\affiliation{Department of Physics and Institute for Condensed Matter Theory,\\
University of Illinois at Urbana-Champaign,
1110 West Green Street, Urbana, Illinois 61801}
\affiliation{Department of Physics and Astronomy, University of Pennsylvania, Philadelphia, Pennsylvania 19104}

\author{Paul M.~Goldbart}
\affiliation{Department of Physics and Institute for Condensed Matter Theory,\\
University of Illinois at Urbana-Champaign,
1110 West Green Street, Urbana, Illinois 61801}

\author{Xiangjun Xing}
\affiliation{Department of Physics,
Syracuse University, Syracuse, New York 13244}

\author{Annette Zippelius}
\affiliation{Institut f\"ur Theoretische Physik,
Georg-August-Universit\"at G\"ottingen,
37077 G\"ottingen, Germany}
\affiliation{Max Planck Institute for Dynamics and Self Organization, Bunsenstrasse 10, 37073 G\"ottingen, Germany}
\date{\today}


\begin{abstract}
Spatial heterogeneity in the elastic properties of soft random solids is examined via vulcanization theory.
The spatial heterogeneity in the \emph{structure} of soft random solids is a result of the fluctuations locked-in at their synthesis, which also brings heterogeneity in their \emph{elastic properties}.
Vulcanization theory studies semi-microscopic models of random-solid-forming systems, and applies replica field theory to deal with their quenched disorder and thermal fluctuations.
The elastic deformations of soft random solids are argued to be described by the Goldstone sector of fluctuations contained in vulcanization theory, associated with a subtle form of spontaneous symmetry breaking that is associated with the liquid-to-random-solid transition.
The resulting free energy of this Goldstone sector can be reinterpreted as arising from a phenomenological description of an elastic medium with quenched disorder. Through this comparison, we arrive at the statistics of the quenched disorder of the elasticity of soft random solids, in terms of residual stress and Lam\'e-coefficient fields. In particular, there are large residual stresses in the equilibrium reference state, and the disorder correlators involving the residual stress are found to be long-ranged and governed by a universal parameter that also gives the mean shear modulus.
\end{abstract}

\maketitle

\section{Introduction}\label{SEC:Intr}

Random solids, such as chemical gels, rubber, glasses and amorphous silica, are characterized by their {\it structural\/} heterogeneity, which results from the randomness \lq\lq locked-in\rq\rq\ at the time they are synthesized.  The mean positions of the constituent particles exhibit no long-range order, and every particle inhabits a unique spatial environment.  The {\it elasticity\/} of random solids also inherits heterogeneity from this locked-in randomness.  For example, the Lam\'e coefficients and the residual stress vary from point to point throughout the elastic medium.  The central goal of this paper is to develop a statistical characterization of random elastic media, via the mean values of the Lam\'e coefficients and the residual stress as well as the two-point spatial correlations amongst the fluctuations of these quantities, which we shall name as the \lq\lq disorder correlator\rq\rq.  These mean values and correlations are to be thought of as averages taken over realizations of the sample fabrication for a given set of fabrication parameters.  We expect these characteristic quantities to coincide with the volume-averages of their single-sample counterparts.

Our focus will be on {\it soft\/} random solids.  These are network media that include chemical gels~\cite{Addad1996}, which are formed by the permanent random chemical
bonding of small molecules, as well as rubber~\cite{Treloar1975}, which is formed via the introduction of permanent random chemical cross-links between nearby monomers in melts or solutions of flexible long-chain polymers.  Soft random solids are characterized by their \emph{entropic elasticity}.  These are media in which the shear modulus originates in the strong thermal fluctuations of the configurations of the constituent particles and is much smaller than the bulk modulus, which is energetic in nature and originates in the excluded-volume interactions between the particles.  The concept of entropic elasticity forms the basis of the classical theory of rubber elasticity, developed long ago by Kuhn, Flory, Wall, Treloar and others (see Ref.~\cite{Treloar1975}).

As we discuss soft random solids we shall take chemical gels as our prototype media.  When the density of the introduced links exceeds the percolation threshold, an infinite cluster of linked molecules forms, spanning the system, and the network acquires a thermodynamic rigidity with respect to shear deformations~\footnote{This realization of rigidity should not be confused with \emph{rigidity percolation}, which captures the rigidity of athermal (i.e., mechanical, rather than thermodynamic) networks~\cite{Thorpe1999}.  In the case of soft random solids, shear rigidity results from the entropy of thermal fluctuations of the positions of the constituent particles, and is proportional to the temperature}. This event is often called the \lq\lq gelation transition\rq\rq\ or the \lq\lq vulcanization transition\rq\rq~\cite{Goldbart1996}.

The {\it geometrical\/} or {\it architectural\/} aspects of the gelation/vulcanization transition can be well captured by the theory of percolation~\cite{Stauffer1994}.  However, to study the {\it elasticity\/} that emerges at the gelation/vulcanization transition, and especially its heterogeneity, one needs a theory that incorporates not only the geometrical aspects, but also the equilibrium thermal fluctuations of the particle positions and the strong, qualitative changes that they undergo at the gelation/vulcanization transition.  In the setting of rubber elasticity, although the classical theory is successful in explaining the entropic nature of the shear rigidity of rubber, it is essentially based on a single-chain picture and, as such, is incapable of describing the consequences of the long scale random structure of rubbery media, e.g., the random spatial variations in their local elastic parameters and the resulting nonaffinity of their local strain-response to macroscopic applied stresses.

The general problem of heterogeneous elasticity and nonaffine deformations has been studied in the setting of flexible polymer networks~\cite{Rubinstein1997,Glatting1995,Holzl1997,Svaneborg2005}, and also semi-flexible polymer networks~\cite{Head2003,Heussinger2007}, glasses~\cite{Wittmer2002}, and granular materials~\cite{Utter2008}. Particularly noteworthy is the recent investigation by DiDonna and Lubensky of the general relationship between the spatial correlations of the nonaffine deformations and those of the underlying quenched random elastic parameters~\cite{DiDonna2005}.

The mission of the present work is to develop a statistical characterization of the heterogeneous elasticity of soft random solids by starting from a semi-microscopic model and applying a body of techniques that we shall call vulcanization theory to it.  In particular, we aim to obtain the mean values and disorder correlators of elastic parameters, such as the Lam\'e coefficients and the residual stress, in terms of the parameters of the semi-microscopic model, such as the density of cross-links, the excluded-volume interactions, etc.  One of our key findings is that the disorder correlator of the residual stress is long ranged, as are all cross-disorder correlators between the residual stress and the Lam\'e coefficients.  We also find that these disorder correlators are controlled by a universal scale parameter---independent of the microscopic details---that, moreover, controls the scale of the mean shear modulus.  In addition, we characterize the nonaffininity of the deformations in terms of these parameters.

The strategy we adopt for accomplishing our goals involves a \lq\lq handshaking\rq\rq\ between two different analytical schemes.~\footnote{
{\colorred{%
This handshaking is the analog of that between
the Born-Huang expansion for crystals and the continuum theory of elasticity,
or that between the Newtonian equations of motion for particles and the Navier-Stokes equations of hydrodynamics.
}}}
The first scheme follows a well-trodden path.
We begin with a semi-microscopic model, the Randomly Linked Particle Model (RLPM)~\cite{Mao2007,Ulrich2006,Broderix2002,Mao2005}, involving particle coordinates and quenched random interactions between them that represent the randomness that is locked-in at the instant of cross-linking.  In order to account for this quenched randomness, as well as the thermal fluctuations in particle positions which are the origin of the entropic elasticity, we adopt the framework of vulcanization theory~\cite{Goldbart1996}.  This framework includes the use of the replica method to eliminate the quenched randomness, followed by a Hubbard-Stratonovich transformation to construct a field-theoretic representation in terms of an order parameter field---in this case, the random solidification order parameter.  We analyze this representation at the stationary-point level of approximation, and then focus on the gapless excitations around the stationary point---the Goldstone fluctuations---observing that these excitations can be parameterized in terms of a set of replicated shear deformation fields~\cite{Mao2007,Ulrich2006,Goldbart2004}.

The second prong of our approach is less conventional.  It begins with our introduction of a phenomenological model free energy of an elastic continuum, characterized by a nonlocal kernel of quenched random attractions between mass-points.
To obtain statistical information about this quenched random kernel, we use the replica method to eliminate the randomness, and obtain a pure model of replicas of the deformation field with couplings controlled by the disorder-moments of the kernel; these disorder-moments are then treated as unknown quantities to be determined.
This pure model has precisely the same structure as the Goldstone theory mentioned in the previous paragraph has.  Thus, by comparing the two models we can learn the moments of the quenched random kernel from the (already-computed) coupling functions of the Goldstone model.  Then we analyze a particular realization of the phenomenological model having a fixed value of the quenched randomness.  We observe that the natural reference state (i.e., the state of vanishing displacement field) of this model is not in fact an equilibrium state for any given realization of disorder, due to the random attractive interactions embodied by the kernel. We analyze how these attractions compete with the near-incompressibility of the medium to determine the displacement to the new, equilibrium configuration, which we shall term the \lq\lq relaxed state\rlap.\rq\rq\thinspace\
(This process can be understood in the setting of a hypothetical, instant process of preparing a sample of rubber: the cross-links introduce attractions and random stresses, and the system then undergoes relaxation, including global contraction and local deformation.)\thinspace\  We then explore shear deformations around this relaxed state, pass to the local limit, and arrive at the standard form of continuum elasticity theory, expressed in terms of the strain around the relaxed state, but with coefficients that are explicit functions of the quenched random kernel.  Thus, using the information about the statistics of the quenched random kernel obtained via the comparison with the RLPM, we are able to infer statistical information about the elastic properties of the random elastic medium in the relaxed state, which is of experimental relevance.

Why is it legitimate to identify the Goldstone theory arising from the microscopic model with the replica theory of the phenomenological model?  The reason is that, within the schemes that we have chosen to analyze them, both models describe shear deformations not of the equilibrium state of the system but, rather, of the system immediately after cross-linking has been done but before any cross-linking-induced relaxation has been allowed to occur.
{\colorred{
The equivalence between these two schemes is not based solely on the equality of free energies of the two models; it is also based on the identity of the physical meaning of the (replicated) deformation fields in the two theories, and thus the way that these fields couple to externally applied forces.  Both schemes are descriptions of the elasticity of soft random solids displaying heterogeneous elastic properties, one from a semi-microscopic viewpoint, the other invoking phenomenological parameters.  Thus, the equivalence of the two schemes provides the values of the phenomenological parameters as functions of the semi-microscopic parameters.
}}

{\colorred Before concluding this introduction, let us emphasize that this work is a first attempt to {\em derive} the elastic heterogeneities of vulcanized matters.  To date, the prevalent strategy in studies of disordered systems is to assume a particular structure for the quenched disorder on phenomenological grounds (such as Gaussian, short ranged, etc.)
and explore the consequences.  Assumptions about disorder structure are usually based on symmetry arguments and also the preference for simplicity, but otherwise lack theoretical substantiation.  It is one of the main advantage of vulcanization theory that it can {\em predict\/} some generic properties of the disordered structure in vulcanized matter, as is shown in the present work, which can be used to support and sharpen the assumptions underlying more phenomenological theories.

The classical theory of rubber elasticity, which was shown to be derivable from the saddle-point approximation of vulcanization theory, is known to fail to describe rubber elasticity in the intermediate and large deformation regimes \cite{Treloar1975}.  While a recent study \cite{Xing2007} shows that long wave-length thermal elastic fluctuations account qualitatively for this failure, on general grounds, we expect that elastic heterogeneities should play an equally important role.  It would therefore be interesting to explore how the elastic heterogeneities discovered in the present work modify the macroscopic elasticity of rubbery materials.  Such a program is left for a future work.  }

The outline of this paper is as follows.
In Section~\ref{SEC:RLPM}, we analyze the semi-microscopic RLPM using the tools of vulcanization theory. Specifically, we use the replica method~\cite{Mezard1987} to study a model network consisting of randomly linked particles, which exhibits a continuous phase transition from the liquid state to the random solid state, paying particular attention to the Goldstone fluctuations of the random solid state, which, as we have mentioned above, are related to the elastic shear deformations of the random solid state.
In Section~\ref{SEC:Phen}, we propose a nonlocal phenomenological model of a random elastic medium, and subsequently derive it from the RLPM, by identify that this phenomenological model is the low energy theory (i.e., it captures the Goldstone fluctuations) of the RLPM in the random solid state. Through this correspondence we learn information of the statistics of the quenched random nonlocal kernel.
In Section~\ref{SEC:relaxation}, we study the relaxation of the phenomenological model to a stable state for any fixed randomness (i.e., any realization of disorder), due to random stresses and attractive interactions. We re-expand the free energy about this relaxed state to obtain the true elastic theory.  This relaxed reference state is, however, still randomly stressed~\cite{Alexander1998}; nevertheless, the stress in this state---the so-called \emph{residual stress}---does satisfy the condition of mechanical equilibrium, viz., $\partial_i \Stre_{ij}(x)=0$.
In its local limit, the proposed phenomenological model reproduces a version of Lagrangian continuum elasticity theory that features random Lam\'e coefficients and residual stresses.   In this section, we also use the phenomenological model to explore the related issue of elastic heterogeneity, viz., the nonaffine way in which the medium responds to external stress.
In Section~\ref{SEC:Heterogeneity}, we arrive at predictions for the statistics of the quenched random elastic parameters that feature in the phenomenological model in the relaxed state, along with the statistics of nonaffine deformations. Thus we provide a \lq\lq first principles\rq\rq\ account of the heterogeneous elasticity of soft random solids.  We conclude, in Section~\ref{SEC:ConcDisc}, with a brief summary and discussion of our results.

\section{Semi-microscopic approach: The randomly linked particle model}
\label{SEC:RLPM}

\subsection{Randomly linked particle model}
\label{SEC:RLPMBasics}
The Randomly Linked Particle Model (RLPM) consists of $\PartNumb$ particles in a volume $\Volu$ in $\Dime$ dimensions. In order to study elasticity, including bulk deformations, $\Volu$ is allowed to fluctuate under a given pressure $\Pres$.  The positions of the particles in this fluctuating volume are denoted by $\{\PartPosi_j\}_{j=1}^{\PartNumb}$.
The particles in the RLPM interact via two types of interactions: a repulsive interaction $\VExc$ between all pairs of particles (either direct or mediated via a solvent);
and an attractive interaction $\VLink$ between the pairs of particles that are chosen at random to be linked.  We take the latter to be a soft link (as opposed to the usual hard constraint of vulcanization theory).  Thus, the Hamiltonian can be written as
\begin{eqnarray}\label{EQ:HRLPM}
	H_{\RealDiso} =
	\sum_{1\le i<j\le N}^{\PartNumb}
	\VExc (\PartPosi_i-\PartPosi_j)+
	\sum_{e=1}^{\LinkNumb}
	\VLink \big(\vert \PartPosi_{i_e}-\PartPosi_{j_e}\vert\big) .
\end{eqnarray}
The label $e$, which runs from $1$ to the total number of links $M$, indexes the links in a given realization of the quenched disorder, and specifies them via the quenched random variables $M$ and $\{i_{e},j_{e}\}_{e=1}^{M}$.

We take $\VExc$ to be a strong, short-ranged repulsion, which serves to penalize density fluctuations and thus render the system nearly incompressible, as is appropriate for regular or polymeric liquids.  As we shall describe in Section~\ref{SEC:FTMF}, we address the interactions in Eq.~(\ref{EQ:HRLPM}) by eliminating the particle coordinates in favor of collective fields, which have the form of joint densities of the replicas of the particles.  This continuum approach enables us to focus on the physics of random particle localization, particularly at lengthscales that are relevant for such localization, which (except when the density of links is extremely high) are long compared with the ranges of $\VExc$ and $\VLink$.  With a focus on these longer lengthscales in mind, we see that it is adequate to replace the repulsive interaction $\VExc(\PartPosi)$ by the model Dirac delta-function excluded-volume interaction $\ExclVolu\,\delta(\PartPosi)$, characterized by the strength $\ExclVolu$~\cite{Gennes1979,Deam1976,Doi1986}.  This procedure amounts to making a gradient expansion in real space (or, equivalently, a wave vector expansion in Fourier space) of $\VExc$ and retaining only the zeroth-order term; it gives for the strength $\ExclVolu$ the value ${\int}d\PartPosi\,\VExc(\PartPosi)$.  Terms of higher order in the gradient expansion would have a non-negligible impact on the suppression of density fluctuations only at lengthscales comparable to or shorter than the range of $\VExc$, and fluctuation modes at such lengthscales are not the ones driven via random linking to the instability associated with random localization (and thus are not modes in need of stabilization via $\VExc$).  We remark that the approximate interaction $\ExclVolu\,\delta(\PartPosi)$ is not, in practice, singular, and is instead regularized via a high wave vector cut-off.

At our coarse-grained level of description, the particles of the RLPM can be identified with polymers or small molecules, and the soft links can be identified with molecular chains that bind the molecules to one another.  The potential for the soft links can be modeled as Gaussian chains
\begin{eqnarray}\label{EQ:VGC}
	\VLink^{(\textrm{GC})}(\vert r \vert) = \frac{\BoltCons T \vert r \vert^2}{2\LinkScal^2} ,
\end{eqnarray}
i.e., a harmonic attraction, or a \lq\lq zero rest-length\rq\rq\ spring, of lengthscale $\LinkScal$ between the two particles.  In making this coarse-graining one is assuming that microscopic details (e.g., the precise locations of the cross-links on the polymers, the internal conformational degrees of freedom of the polymers, and the effects of entanglement) do not play significant roles for the long-wavelength physics. In part, these assumptions are justified by studying more detailed models, in which the conformational degrees of freedom of the polymers are retained~\cite{Goldbart1996}.  However, we should point out that the precise form of $\VLink$ is not important for long-wavelength physics and, hence, for the elastic properties that we are aiming to investigate [cf.~the discussion at the end of Sec.~\ref{SEC:FTMF}].

%

From the discussion above, the RLPM is a convenient minimal model of soft random solids, inasmuch as it adequately captures the necessary long-wavelength physics. It can be regarded as either a model of a chemical gel, or as a caricature of vulcanized rubber or other soft random solid. The RLPM can be viewed as a simplified version of vulcanization theory~\cite{Castillo1994,Goldbart2004}, with microscopic details, such as polymer chain conformations, being ignored. Nevertheless, it is able to reproduce the same universality class as vulcanization theory at the liquid-to-random-solid transition. For the study of elasticity, we shall consider length-scales on which the system is a well-defined solid (i.e., scales longer than the \lq\lq localization length\rlap,\rq\rq\ as we shall see later in this paper). Both of these scales are much larger than the characteristic linear dimension of an individual polymer.
The RLPM is a model very much in the spirit of lattice percolation, except that it naturally allows for particle motion as well as particle connectivity, and is therefore suitable for the study of continuum elasticity and other issues associated with the (thermal or deformational) motion of the constituent entities.

Equation~(\ref{EQ:HRLPM}) is a Hamiltonian for a given realization of quenched disorder $\RealDiso \equiv \{ i_e, j_e\}_{e=1}^{\LinkNumb}$, which describes the particular random instance of the linking of the particles. These links are the quenched disorder of the system, which are specified at synthesis and do not change with thermal fluctuations. This is because there is a wide separation between the timescale for the linked-particle system to reach thermal equilibrium and the much longer timescale required for the links themselves to break. Therefore, we treat the links as permanent. Later, we shall apply the replica technique~\cite{Mezard1987} to average over these permanent random links.

\subsection{Replica statistical mechanics of the RLPM}
\label{SEC:RLPMReplica}
For a given volume and a given realization of disorder $\RealDiso$ we can write the partition function $Z_{\RealDiso}$ for the RLPM as
\begin{eqnarray}\label{EQ:partition}
	Z_{\RealDiso} (\Volu)
	&\equiv& \int _{\Volu} \prod_{i=1}^{\PartNumb} d \PartPosi_i
	\exp\Big(-\frac{H_{\RealDiso}}{{\BoltCons}T}\Big) \nonumber\\
	&=& \PartitionLiquid(V) \Bigg\ltha \prod _{e=1}^{\LinkNumb}
	\LinkPote\big( \vert \PartPosi_{i_e} - \PartPosi_{j_e} \vert \big)
	\Bigg\rtha_{1}^{\HamiExcl},
\end{eqnarray}
where $\HamiExcl \equiv \frac{\ExclVolu }{2}\sum_{i,j=1}^{\PartNumb} \delta (\PartPosi_i-\PartPosi_j)$ is the excluded-volume interaction part of the Hamiltonian, and $\PartitionLiquid(V) \equiv \int _{\Volu} \prod_{i=1}^{\PartNumb} d \PartPosi_i \exp\big(-\HamiExcl/{\BoltCons}T\big)$ is the partition function of the liquid in the absence of any links. The issue of the Gibbs factorial factor, which is normally introduced to compensate for the overcounting of identical configuration, is a genuinely subtle one in the context of random solids (for a discussion, see Ref.~\cite{Goldbart1996}). However, our focus will be on \lq\lq observables\rq\rq~such as order parameter rather than on free energies, and thus the omission of the Gibbs factor is of no consequence.
The factor
\begin{eqnarray}\label{EQ:DeltReal}
	\LinkPote\big( \vert \PartPosi_{i_e} - \PartPosi_{j_e} \vert \big)
	\equiv e^{  -\frac{\vert \PartPosi_{i_e} - \PartPosi_{j_e} \vert^2}
	{2  \LinkScal^2}}
\end{eqnarray}
is associated with the link-induced attractive interaction term in the Hamiltonian. The average $\ltha \cdots \rtha_{1}^{\HamiExcl}$, taken with respect to a Boltzmann weight involving the excluded-volume interaction Hamiltonian $\HamiExcl$, is defined as
\begin{eqnarray}
	\ltha \cdots \rtha_{1}^{\HamiExcl} \equiv
	\frac{1}{\PartitionLiquid(V)}\int _{\Volu} \prod_{i=1}^{\PartNumb} d \PartPosi_i \,
		e^{-\frac{\HamiExcl}{{\BoltCons}T}}\ldots\, .
\end{eqnarray}
The corresponding Helmholtz free energy is then given by
\begin{eqnarray}
	\HelmFreeEner _{\RealDiso} (\Volu) \equiv -\BoltCons T \ln Z_{\RealDiso} (\Volu).
\end{eqnarray}

To perform the average of the free energy over the quenched disorder, we shall need to choose a probability distribution that assigns a sensible statistical weight $\DEDist (\{ i_e, j_e\}_{e=1}^{\LinkNumb})$ to each possible realization of the total number $\LinkNumb$ and location $\{i_e,j_e\}_{e=1}^{\LinkNumb}$ of the links. Following an elegant strategy due to Deam and Edwards~\cite{Deam1976}, we assume a version of the normalized link distribution as follows:
\begin{eqnarray}
	\DEDist(\RealDiso)
	= \frac{
			\Big(\frac{\LinkDens \Volu_0}{2\PartNumb \LinkPote_0}\Big)^{\LinkNumb}
		 	Z_{\RealDiso}(\Volu_0)}
		{\LinkNumb ! Z_1},
\end{eqnarray}
where $\LinkDens$ is a parameter that controls the mean total number of links. We assume that the preparation state (i.e., the state in which the links are going to be introduced) is in a given volume $\Volu_0$.
The $Z_{\RealDiso}(\Volu_0)$ factor is actually the partition function, as given in Eq.~(\ref{EQ:partition}), and can be regarded as probing the equilibrium correlations of the underlying unlinked liquid.
The factor $\LinkPote_0=\big(2\pi \LinkScal^2 \big)^{d/2}$ is actually the $p=0$ value of the Fourier transform of the $\LinkPote$ function defined in Eq.~(\ref{EQ:DeltReal}), and we shall see later that these factors ensure that the (mean-field) critical point occurs at $\LinkDens_C =1$. The normalization factor $Z_1$ is defined to be $\sum_{\RealDiso} \big(\frac{\LinkDens \Volu_0}{2\PartNumb \LinkPote_0}\big)^{\LinkNumb} Z_{\RealDiso}(\Volu_0)/ \LinkNumb !$. The calculation for $Z_1$ is straightforward, and is given in Appendix~\ref{APP:DisoAver}.

The Deam-Edwards distribution can be understood as arising from a realistic vulcanization process in which the links are introduced simultaneously and instantaneously into the liquid state in equilibrium.  Specifically, it incorporates the notion that all pairs of particles that happen (at some particular instant) to be nearby are, with a certain probability controlled by the link density parameter $\LinkDens$, linked. Thus, the correlations of the link distribution reflect the correlations of the unlinked liquid, and it follows that realizations of links only acquire an appreciable statistical weight if they are compatible with some reasonably probable configuration of the unlinked liquid.

The factor $\big(\frac{\LinkDens \Volu_0}{2\PartNumb \LinkPote_0}\big)^{\LinkNumb} /\LinkNumb !$ in the Deam-Edwards distribution introduces a Poissonian character to the total number $\LinkNumb$ of links. These links are envisioned to be the product of a Poisson chemical linking process. The factor $Z_{\RealDiso}(\Volu_0)$ assures that the probability of having a given random realization of links is proportional to the statistical weight for, in the unlinked liquid state, finding the to-be-linked pairs to be co-located in the liquid state to within the shape function $\exp \big(- \vert \PartPosi_{i_e}-\PartPosi_{j_e}\vert ^2  / 2\LinkScal^2\big)$.

As a result of the Deam-Edwards distribution, the mean number of links per particle is given by $\lda \LinkNumb \rda/\PartNumb = \LinkDens/2$.
Thus, $\LinkDens = 2 \lda \LinkNumb \rda/N$ is the \emph{mean coordination number}, i.e., the average number of particles to which a certain particle is connected, {\colorred{the factor of $2$ results from the fact that each link is shared by two particles}}. For a detailed discussion of the Deam-Edwards distribution, see Ref.~\cite{Deam1976,Broderix2002}.

By using this distribution of the quenched disorder, we can perform the disorder average of the Helmholtz free energy via the replica technique, thus obtaining
\begin{eqnarray}\label{EQ:HFReplica}
	\lda \HelmFreeEner \rda
	&\equiv& \sum_{\RealDiso} \DEDist(\RealDiso)
		\HelmFreeEner_{\RealDiso} (\Volu) \nonumber\\
	&=& -\BoltCons T \sum_{\RealDiso} \DEDist(\RealDiso) \ln Z_{\RealDiso} (\Volu)
		\nonumber\\
	&=& -\BoltCons T  \lim _{n \to 0} \sum_{\RealDiso} \DEDist(\RealDiso)
		\frac{Z_{\RealDiso}(\Volu)^{n}-1}{n}.
\end{eqnarray}
We now insert the Deam-Edwards distribution to get
\begin{eqnarray}\label{EQ:HFReplicaDE}
	\lda \HelmFreeEner \rda
	&=& -\BoltCons T  \lim _{n \to 0} \sum_{\RealDiso}
		\frac{
			\Big(\frac{\LinkDens \Volu_0}{2\PartNumb \LinkPote_0}\Big)^{\LinkNumb}
		 	Z_{\RealDiso}(\Volu_0)}{\LinkNumb ! Z_1} \nonumber\\
	&&\quad\times	\frac{Z_{\RealDiso}(\Volu)^{n}-1}{n}.
\end{eqnarray}
This disorder-averaged free energy differs from the form traditionally obtained via the replica technique, in that there is an extra replica $Z_{\RealDiso}(\Volu_0)$, which originates in the Deam-Edwards distribution. We shall call this extra replica the $0^{\textrm{th}}$ replica, and note that it represents the \emph{preparation state} of the system.~\footnote{{\colorred{%
In this preparation state, the temperature and the strength of the excluded-volume interaction can differ from those characterizing the measurement ensemble,
and this has been discussed in Ref.~\cite{Xing2004}}}}

The summation over the realizations of the quenched disorder $\RealDiso$ can be performed, following the calculation in Appendix~\ref{APP:DisoAver}; thus we arrive at the form
\begin{eqnarray}\label{EQ:HFEDisoAver}
	\lda \HelmFreeEner \rda
	&=& -\BoltCons T  \lim _{n \to 0} \frac{1}{n}\Big( \frac{Z_{1+n}}{Z_1}-1 \Big),
\end{eqnarray}
which can also be expressed as
\begin{eqnarray}\label{EQ:HFEDisoAverDeri}
	\lda \HelmFreeEner \rda = -\BoltCons T \lim _{n \to 0}
		\frac{\partial}{\partial n} \ln Z_{1+n} \, ,
\end{eqnarray}
where
\begin{widetext}
\begin{eqnarray}\label{EQ:ReplPart}
	Z_{1+n} &\equiv& \sum_{\RealDiso}
			\frac{\Big(\frac{\LinkDens \Volu_0}{2\PartNumb \LinkPote_0}\Big)^{\LinkNumb}}
			{ \LinkNumb !} Z_{\RealDiso}(\Volu_0) Z_{\RealDiso}(\Volu)^n \nonumber\\
		&=& \PartitionLiquid(\Volu_0)\PartitionLiquid(\Volu)^n 
		\lthal \exp \Big(\frac{\LinkDens \Volu_0}
			{2\PartNumb \LinkPote_0} 
			\sum_{i\ne j}^{\PartNumb} \ReplProd
			\LinkPote  \big(
			\vert \PartPosi_{i}\REPa - \PartPosi_{j}\REPa \vert
			\big)\Big)\rthal_{1+n}^{\HamiExcl} .
\end{eqnarray}
\end{widetext}
Notice that, here, the preparation state (i.e., $0^{\textrm{th}}$ replica) has a fixed volume $\Volu_0$ because, for convenience, we have assumed that the linking process was undertaken instantaneously in a liquid state of fixed volume and thus the pressure is fluctuating, whereas the measurement states (replicas $1$ through $n$) are put in a fixed-pressure $\Pres$ environment, the volume $\Volu$ of which is allowed to fluctuate. In the latter parts of the paper we shall set the pressure $\Pres$ to be the average pressure measured in the preparation state at volume $\Volu_0$. In particular, for a given volume of the liquid state in which the links are made, the average pressure is given by
\begin{eqnarray}
	\Pres = - \frac{\partial \FLiquid (\Volu_0)}{\partial \Volu_0}
					\Big\vert_{T} \, ,
\end{eqnarray}
where we have introduced the Helmholtz free energy of the unlinked liquid $\FLiquid (\Volu_0) \equiv - \BoltCons T \ln \PartitionLiquid (\Volu_0)$. We suppose that the excluded-volume interactions are so strong that the density fluctuations are suppressed, and the density of the unlinked liquid is just $\PartNumb/\Volu_0$.~\footnote{For (unswelled) rubbery materials this is an appropriate assumption.}
Thus, the mean-field value of Helmholtz free energy in the unlinked liquid state is
\begin{eqnarray}\label{EQ:FLiquid}
	\FLiquid (\Volu_0) &=& -\PartNumb \BoltCons	T \ln \Volu_0
		+\frac{\ExclVolu \PartNumb^2}{2\Volu_0}.
\end{eqnarray}
Therefore, the mean pressure in the unlinked liquid state is given by
\begin{eqnarray}\label{EQ:AverPres}
	\Pres = \frac{\PartNumb \BoltCons	T}{\Volu_0}
		+\frac{\ExclVolu \PartNumb^2}{2\Volu_0^2},
\end{eqnarray}
{\colorgreen{from which we can identify---by the standard way in which the second virial coefficient $B_2$ appears in the free energy of the equation of state for a fluid---that $B_2$ for the unlinked liquid is $\ExclVolu/2\BoltCons T$.  Thus, without having actually performed a cluster expansion, we see that the Dirac delta-function interaction with coefficient $\ExclVolu$ indeed leads to a virial expansion with a suitable excluded volume, viz., $\ExclVolu/(\BoltCons T)$.
}}
As mentioned above, we shall apply this pressure in the measurement states (described by $1^{\textrm{th}}$ through $n^{\textrm{th}}$ replicas), and let their volumes $\Volu$ fluctuate, in order to obtain an elastic free energy that can describe volume variations. In particular, by choosing the pressure $p$ to be exactly the mean pressure of the liquid state, we shall obtain an elastic free energy that takes the \emph{state right after linking}, which has the same volume $\Volu_0$ as the liquid state, as the elastic reference state. This issue of the state right after linking and the elastic reference state will be discussed in detail in Section~\ref{SEC:PhenFE}.

In light of this construction of the pressure ensemble, we have the capability of learning about the bulk modulus of the system, and to characterize volume changes caused by linking, a process that has the effect of eliminating translational degrees of freedoms.

To establish an appropriate statistical mechanics for the fixed-pressure ensemble, we shall make the following Legendre transformation of the Helmholtz free energy, which leads to the Gibbs free energy $\GibbFreeEner (\Pres,T)$:
\begin{subequations}\label{EQ:LegeTran}
\begin{eqnarray}
	\Pres &=& -\frac{\partial \HelmFreeEner(\Volu,T)}
		{\partial \Volu}\big\vert_{T} \label{EQ:LegeTran1}  \, , \\
	\GibbFreeEner (\Pres,T) &=& \HelmFreeEner(\Volu,T) + \Pres \Volu \, . \label{EQ:LegeTran2}
\end{eqnarray}
\end{subequations}
In Eq.~(\ref{EQ:LegeTran2}) the volume $\Volu$ takes the value (in terms of $\Pres$) that satisfies Eq.~(\ref{EQ:LegeTran1}), i.e., the volume that minimizes the Gibbs free energy at a given pressure $p$.

In the following sections, we shall first calculate the disorder-average of the Helmholtz free energy, and then make this Legendre transformation to obtain the disorder-averaged Gibbs free energy. This will allow us to explore the elasticity of the RLPM in detail.

\subsection{Field-theoretic description of the RLPM}
\label{SEC:FTMF}
We shall use field-theoretic methods to analyze the disorder-averaged free energy $\lda\HelmFreeEner\rda$ and, more specifically, the replicated partition function $Z_{1+n}$. To do this, we introduce a joint probability distribution for the particle density in the replicated space, i.e., the replicated density function
\begin{eqnarray}\label{EQ:DensReal}
	\DensFunc(\REPX)\equiv \frac{1}{\PartNumb} \sum_{i=0}^{\PartNumb}
		\ReplProd \delta^{(d)}(x\REPa-\PartPosi_i\REPa) ,
\end{eqnarray}
where $\REPX \equiv (x^{0},x^{1},\ldots,x^{n})$ is a short-hand for the $(1+n)$-replicated position $\Dime$-vector. For convenience, we introduce a complete orthonormal basis set in replica space $\{\BASE\REPa\}_{\alpha=0}^{n}$, in terms of which a vector $\REPX$ can be expressed as
\begin{eqnarray}
	\REPX=\ReplSum x\REPa \BASE\REPa .
\end{eqnarray}
Note that the components $x\REPa$ are themselves $\Dime$-vectors.  With this notation, the density function of a single replica $\alpha$ is given by $\DensFunc_{p\REPa\BASE\REPa}$, which is the Fourier transform of the $\DensFunc(\REPX)$ field, $\DensFunc_{\REPP}$, with momentum nonzero only in replica $\alpha$, corresponding to integrating over the normalized densities in other replicas in real space.

The replicated partition function~(\ref{EQ:ReplPart}) can be written as a functional of the replicated density function $\DensFunc$ in momentum space as
\begin{eqnarray}\label{EQ:ZQ}
	Z_{1+n} = \int_{\Volu_0} \prod_{i=1}^{\PartNumb} d\PartPosi_{i}^{0}
						\int_{\Volu} \ReplProd \prod_{j=1}^{\PartNumb} d\PartPosi_{j}\REPa
						\, e^{-\frac{H_{\DensFunc}\lbrack \DensFunc_{\REPP}\rbrack}{\BoltCons T}} ,
\end{eqnarray}
with
\begin{eqnarray}\label{EQ:HQ}
	H_{\DensFunc}\lbrack \DensFunc_{\REPP}\rbrack
	&\equiv& -\frac{\PartNumb \LinkDens \BoltCons T}{2 \Volu^n \LinkPote_0}
			 \sum_{\REPP}\DensFunc_{\REPP}\DensFunc_{-\REPP}\LinkPoteRepl_{\REPP}\nonumber\\
		&&+\frac{\ExclVolu \PartNumb^2}{2\Volu_0} \sum_{p}\DensFunc_{p\BASE^{0}}
				\DensFunc_{-p\BASE^{0}}	\nonumber\\
		&&+\frac{\ExclVolu \PartNumb^2}{2\Volu} \sum_{p} \ReplSumOne
				\DensFunc_{p\BASE\REPa} \DensFunc_{-p\BASE\REPa} ,
\end{eqnarray}
where the factor
\begin{eqnarray}
	\LinkPoteRepl_{\REPP} = \big(\LinkPote_0\big)^{1+n}
	e^{-\LinkScal^2 \vert \REPP \vert^2/2}
\end{eqnarray}
is the replicated version of the Fourier transform of the function $\LinkPote(x)$, defined in Eq.~(\ref{EQ:DeltReal}).
{{\colorred The summation $\sum_{\REPP}$ denotes a summation over all momentum $\Dime$-vectors $p\REPa$, one for each replica, with $\REPP$ taking the values $\REPP=\ReplSum p\REPa \BASE\REPa$.  The cartesian components of the $p\REPa$ take the values $2\pi m/L$, where $L$ is the linear size of the system and $m$ is any integer.  Similarly, summations $\sum_{p}$ over $\Dime$-vectors $p$ include components having the values $2\pi m/L$.}}

The first term on the right hand side of Eq.~(\ref{EQ:HQ}) arises from the attractive, link-originating, interaction part [see Eq.~(\ref{EQ:ReplPart})]; the next two terms represent the excluded-volume interaction in $\HamiExcl$, for the $0^{\textrm{th}}$ replica and for replicas $1$ through $n$, respectively.

The excluded-volume interaction is taken to be very strong, and thus the density fluctuations in any single replica are heavily suppressed. This means that $\DensFunc_{p\REPa\BASE\REPa}$ are very small for all $p\REPa\ne 0$, and $\DensFunc_{p\REPa\BASE\REPa}\vert_{p\REPa=0}=1$, corresponding to a nearly homogeneous particle density.
To manage this issue, we separate the replicated space into a Lower Replica Sector (LRS), in which the density fluctuations are suppressed, and a Higher Replica Sector (HRS), which captures the correlations between different replicas, and develops an instability at the liquid-to-random-solid transition. The definitions of the LRS and HRS are (in momentum space) as follows:
if two or more components of a replicated momentum vector $\REPP \equiv (p^{0},p^{1},\ldots,p^{n})$ are non zero then $\REPP$ is an element of the HRS;
on the other hand, if $\REPP$ has zero or only one component $p\REPa$ being nonzero, and all other $p\REPb=0$, then $\REPP$ is an element of the LRS. In addition, the LRS can be separated into a 1RS part, in which vectors $\REPP$ has exactly one component $p\REPa$ being nonzero, and a 0RS, which consists of only $\REPP=0$.
With this separation we can rewrite the effective Hamiltonian as
\begin{eqnarray}\label{EQ:HQHL}
	H_{\DensFunc}\lbrack \DensFunc_{\REPP}\rbrack
	&=& -\frac{\PartNumb \LinkDens \BoltCons T}{2 \Volu^n \LinkPote_0}
			\sum_{\REPP\in HRS}\DensFunc_{\REPP}
			\DensFunc_{-\REPP}\LinkPoteRepl_{\REPP}\nonumber\\
		&&+\frac{ \PartNumb^2}{2\Volu_0} \sum_{p}\ExclVoluRN_0(p)\DensFunc_{p\BASE^{0}}
				\DensFunc_{-p\BASE^{0}}	\nonumber\\
		&&+\frac{ \PartNumb^2}{2\Volu} \sum_{p}\ExclVoluRN(p) \ReplSumOne
				\DensFunc_{p\BASE\REPa} \DensFunc_{-p\BASE\REPa} ,
\end{eqnarray}
with the renormalized coefficients
\begin{eqnarray}\label{EQ:ExclVoluRN}
	\frac{\ExclVoluRN_0(p) \PartNumb^2}{2\Volu_0}
	&\equiv&\frac{\ExclVolu \PartNumb^2}{2\Volu_0}
	-\frac{\PartNumb \LinkDens\BoltCons T\LinkPoteRepl_{\REPP}}{2\Volu^n\LinkPote_0} ,
	\nonumber\\
	\frac{\ExclVoluRN(p) \PartNumb^2}{2\Volu}
	&\equiv&\frac{\ExclVolu \PartNumb^2}{2\Volu}
	-\frac{\PartNumb \LinkDens\BoltCons T\LinkPoteRepl_{\REPP}}{2\Volu^n\LinkPote_0} .
\end{eqnarray}
We suppose that $\frac{\ExclVolu N}{\BoltCons T \Volu}\gg \LinkDens$ (i.e., the excluded-volume repulsion is very strong, relative to the attractive effects of the links), so these coefficients $\frac{\ExclVoluRN(p) \PartNumb^2}{2\Volu}$ are always positive and large, relative to the energy-scale of the HRS that we are interested in.

The interactions in Eq.~(\ref{EQ:HQHL}) can be decoupled using a Hubbard-Stratonovich (HS) transformation (for details see Appendix~\ref{APP:HSTransformation}).  Thus, we arrive at a field-theoretic formulation of the replicated partition function, in terms of the order parameter field $\VOP$:
\begin{eqnarray}\label{EQ:HS}
	Z_{1+n} = \int \mathcal{D}\VOP_{\hat{p}}\ReplProd\mathcal{D}\VOP_{p\BASE\REPa}
	e^{-\frac{H_{\VOP}\lbrack \VOP_{\hat{p}},\VOP_{p\BASE\REPa} \rbrack}{\BoltCons T}} ,
\end{eqnarray}
where the effective Hamiltonian is given by
\begin{widetext}
\begin{eqnarray}\label{EQ:HEffcVOP}
	\! H_{\VOP}\lbrack \VOP_{\hat{p}},\VOP_{p\BASE\REPa} \rbrack
	&=& \frac{\PartNumb \LinkDens \BoltCons T}{2 \Volu^n \LinkPote_0}
			\sum_{\REPP\in HRS}\!\VOP_{\REPP}
			\VOP_{-\REPP}\LinkPoteRepl_{\REPP}
					+\frac{ \PartNumb^2}{2\Volu_0} \sum_{p}\ExclVoluRN_0(p)\VOP_{p\BASE^{0}}
				\VOP_{-p\BASE^{0}}	\nonumber\\
	&&		+\frac{\PartNumb^2}{2\Volu} \sum_{p} \ExclVoluRN(p) \ReplSumOne
				\VOP_{p\BASE\REPa} \VOP_{-p\BASE\REPa} \!-\! N\BoltCons T \ln \MyZzero \, .
\end{eqnarray}
and
\begin{eqnarray}
	\MyZzero &=& \int_{\Volu_0} \! d\PartPosi^{0} \!
						\int_{\Volu} \!\ReplProd  \! d\PartPosi\REPa
			\exp\!\Big\lbrack
			\frac{ \LinkDens}{ \Volu^n \LinkPote_0}
			\!\sum_{\REPP\in HRS}\!\!\VOP_{\REPP}\,\LinkPoteRepl_{\REPP}e^{i\REPP \cdot \hat{c}} \nonumber\\
	&&		\quad\quad +  \frac{i \PartNumb}{\Volu_0\BoltCons T}
			\sum_{p}\ExclVoluRN_0(p)\VOP_{p\BASE^{0}}\,e^{ip^{0}c^{0}}
		+\frac{i \PartNumb}{\Volu\BoltCons T} \!\sum_{p} \ExclVoluRN(p)\! \ReplSumOne
				\VOP_{p\BASE\REPa} \,e^{ip\REPa c\REPa} \! \Big\rbrack .
\end{eqnarray}
The form of this HS transformation [see Appendix~\ref{APP:HSTransformation}, especially Eq.~(\ref{EQ:HSAver})] ensures that the mean value of the order parameter field $\VOP$ is related to the mean value of the replicated density function field $\DensFunc$ as
\begin{subequations}
\begin{eqnarray}
	&&\textrm{HRS:} \quad\quad
	\langle \DensFunc_{\REPP} \rangle_{H_{\DensFunc}}  \label{EQ:HSrelationHRS}
	= \langle \VOP_{\REPP} \rangle_{H_{\VOP}} , \\
	&&\textrm{LRS:} \quad\quad
	i\langle \DensFunc_{p\BASE\REPa} \rangle_{H_{\DensFunc}}
	= \langle \VOP_{p\BASE\REPa} \rangle_{H_{\VOP}} ,
\end{eqnarray}
\end{subequations}
where the averages on either sides are defined via
\begin{subequations}
\begin{align}
	\langle \, \cdots \rangle_{H_{\DensFunc}}
	& \equiv \frac{1}{Z_{1+n}}
			\int_{\Volu_0} \prod_{i=1}^{\PartNumb} d\PartPosi_{i}^{0}
						\int_{\Volu} \ReplProdOne \prod_{j=1}^{\PartNumb} d\PartPosi_{j}\REPa 		\label{EQ:HQAver}
						\,\, e^{-\frac{H_{\DensFunc}\lbrack \DensFunc_{\REPP}\rbrack}{\BoltCons T}} \cdots , \\
	\langle \, \cdots \rangle_{H_{\VOP}}	
	& \equiv \frac{1}{Z_{1+n}}
			\int \mathcal{D}\VOP_{\hat{p}}\ReplProd\mathcal{D}\VOP_{p}\REPa 						 \label{EQ:HVOPAver}
					 \,\,e^{-\frac{H_{\VOP}\lbrack \VOP_{\hat{p}},\VOP_{p\BASE\REPa} \rbrack}{\BoltCons T}}	\cdots .				 
\end{align}
\end{subequations}

The leading-order terms in $H_{\VOP}\lbrack \VOP_{\hat{p}},\VOP_{p\BASE\REPa} \rbrack$ can be constructed by expanding the $\ln \MyZzero$ term in Eq.~(\ref{EQ:HEffcVOP}) in powers of the fields $\VOP_{\REPP}$ and $\VOP_{p\BASE\REPa}$, and thus we can obtain the leading-order terms in the Landau-Wilson effective Hamiltonian.  To leading order this expansion gives
\begin{eqnarray}\label{EQ:HQExpansion}
	H_{\VOP}\lbrack \VOP_{\hat{p}},\VOP_{p\BASE\REPa} \rbrack
	= \frac{\PartNumb \LinkDens \BoltCons T}{2 \Volu^n \LinkPote_0}
			&& \sum_{\REPP\in HRS}\VOP_{\REPP}\VOP_{-\REPP}\LinkPoteRepl_{\REPP}
			\Bigg(1-\LinkDens \frac{\LinkPoteRepl_{\REPP}}{V^{n}\LinkPote_{0}}\Bigg)
			+\frac{\ExclVoluRN_0 \PartNumb^2}{2\Volu_0} \sum_{p}\VOP_{p\BASE^{0}}
				\VOP_{-p\BASE^{0}}	
			\Bigg(1+\frac{\ExclVoluRN_0 \PartNumb}{\Volu_0\BoltCons T}\Bigg)	
				\nonumber\\ \noalign{\medskip}
		+\frac{\ExclVoluRN \PartNumb^2}{2\Volu} &&\sum_{p} \ReplSumOne
				\VOP_{p\BASE\REPa} \VOP_{-p\BASE\REPa}
			\Bigg(1+\frac{\ExclVoluRN \PartNumb}{\Volu_0\BoltCons T}\Bigg)
				+O\big((\VOP_{\hat{p}})^3,(\VOP_{p\BASE\REPa})^3\big).
\end{eqnarray}
\end{widetext}
For the LRS fields $\VOP_{p\BASE\REPa}$ we see the coefficients of the corresponding quadratic term are always positive (given that $\ExclVoluRN_0,\,\ExclVoluRN>0$, i.e., the excluded-volume repulsion is very strong), so this sector of the field theory does not undergo an instability. Furthermore, because these coefficients (the masses, in particle-physics language) are very large [see Eq.~(\ref{EQ:ExclVoluRN})], the fluctuations of these LRS fields $\VOP_{p\BASE\REPa}$ are heavily suppressed. For this reason, we ignore these fluctuations and, for all $\alpha =0,1,\ldots,n$, we take
\begin{eqnarray}\label{EQ:LRScons}
	\VOP_{p\BASE\REPa}\vert_{p=0}&=& i , \nonumber\\
	\VOP_{p\BASE\REPa}\vert_{p\ne 0}&=& 0 ,
\end{eqnarray}
as a hard constraint.

Having implemented this constraint, we arrive at the HRS Hamiltonian (the full form, not just the leading-order expansion):
\begin{eqnarray}\label{EQ:HVOPHRS}
	H_{\VOP}\lbrack\VOP_{\REPP}\rbrack
	&=& \frac{\PartNumb \LinkDens \BoltCons T}{2 \Volu^n \LinkPote_0}
			\sum_{\REPP\in HRS}\VOP_{\REPP}
			\VOP_{-\REPP}\LinkPoteRepl_{\REPP}\nonumber\\ \noalign{\medskip}
		&&\quad - N\BoltCons T \ln \MyZzero ,
\end{eqnarray}
where
\begin{eqnarray}\label{EQ:Z_zero}
	\MyZzero&\equiv& \int_{\Volu_0} \! d\PartPosi^{0} \!\!
						\int_{\Volu} \!\ReplProdOne  d\PartPosi\REPa
			\exp \Big\lbrack
			\frac{ \LinkDens}{ \Volu^n \LinkPote_0} \nonumber\\ \noalign{\medskip}
			&&\quad\quad\times \sum_{\REPP\in HRS}\VOP_{\REPP}\LinkPoteRepl_{\REPP}e^{i\REPP \cdot \hat{c}}
		 \Big\rbrack .
\end{eqnarray}
The Landau theory of the vulcanization transition~\cite{Peng1998} can be recovered by making the expansion of this HRS Hamiltonian that keeps only the leading-order terms in the order-parameter $\VOP$ and the momentum $\REPP$. Up to an additive constant and an appropriate rescaling of the order parameter, this expansion reads
\begin{eqnarray}\label{EQ:Landau}
	H_{\VOP}\lbrack\VOP_{\REPP}\rbrack
	&=& \frac{1}{2}\sum_{\REPP\in HRS} \big( r+ \vert \REPP \vert^2 \big)
			\VOP_{\REPP} \VOP_{-\REPP} \nonumber\\
	&&		-\frac{v}{3!} \sum_{\REPP_1,\REPP_2 \in HRS} \VOP_{\REPP_1} \VOP_{\REPP_2} \VOP_{-\REPP_1-\REPP_2} \, ,
\end{eqnarray}
where the potential of the links $\LinkPoteRepl_{\REPP}$ has been momentum-expanded. This is precisely the form of the Landau free energy that was constructed via \emph{symmetry arguments} in Ref.~\cite{Peng1998}.
In the limit $n\to 0$, the coefficients become
\begin{eqnarray}
	r &\propto& \LinkDens(1-\LinkDens) , \nonumber\\
	v &\propto& (\LinkDens)^3 .
\end{eqnarray}

It is straightforward to see that the $r$ term leads to an \emph{instability} for the link density parameter $\LinkDens$ larger than the critical value $\LinkDens_C=1$,
and the lowest unstable modes are long-wavelength modes (i.e., $\REPP \to 0$). [One should, however, keep in mind that the component $\REPP=0$ itself, which is the 0RS, is excluded from this HRS-only field theory; see Eq.~(\ref{EQ:LRScons}).]
This instability corresponds to the \emph{liquid-to-soft-random-solid transition}, because the liquid state corresponds to the $\VOP_{\REPP}=0$ (in the HRS) state, and becomes unstable when the link density parameter $\LinkDens$ exceeds $1$.

{\colorred{
There are two points that we would like to discuss about this Hamiltonian.  First, it can be seen from the expansion in Eq.~(\ref{EQ:Landau}) that near the critical point the exact form of interaction is irrelevant, because we have only kept terms to $\vert\REPP\vert^2$ in $\LinkPoteRepl_{\REPP}$, and this governs the long-distance physics.  We have used the Gaussian-chain potential~(\ref{EQ:VGC}) in this calculation. It is clear that if we were to change to a different potential, such as a finite-rest-length spring $\VLink(\vert r \vert) = \frac{k}{2} \big(\vert r \vert -l \big)^2$, the long-distance (i.e., small-momentum) physics would be unchanged.
Second, as we shall see in Sec.~\ref{SEC:Heterogeneity}, neither do the statistics of the elastic modulus depend on the details of the interaction potential; instead, they only depend on the density of links.  This results from the fact that the elasticity originates in the entropy of the \emph{network}.  The critical point for the vulcanization transition, and thus the entropic rigidity in the presence of thermal fluctuations, occurs at the same point as \emph{connectivity percolation} does, rather than at the \emph{rigidity percolation} critical point.
The fact that we have a shear modulus scaling as $T$ results from the entropy of the network, and not from the factor of $T$ in the Gaussian-chain potential, is clear because if we were to change to another attractive potential, the same results would hold for the long-distance physics.
}}

\subsection{Mean-field theory of the RLPM}
To understand the physics of the order parameter in vulcanization theory, and thus obtain the form of the stationary value of the order parameter, we need to recall the property of the HS transformation, Eq.~(\ref{EQ:HSrelationHRS}), which relates the average of the order parameter field $\VOP$ to the average of the replicated density function $\DensFunc$. According to this relation, we have
\begin{eqnarray}
	\langle \VOP_{\REPP} \rangle_{H_{\VOP}}
	= \Big\langle \frac{1}{N}\sum_{j=1}^{\PartNumb}e^{i\REPP \cdot \hat{c}_j}\Big\rangle_{H_{\DensFunc}}
	-\delta^{((1+n)d)}_{\REPP,0} .
\end{eqnarray}
Here, the $\delta^{((1+n)d)}_{\REPP}$ removes the 0RS part of $\VOP_{\REPP}$. Equivalently, in real space we have
\begin{eqnarray}
	\langle \VOP(\REPX)\rangle_{H_{\VOP}} \!\!
	=\!\Big\langle \frac{1}{N}\!\sum_{j=1}^{\PartNumb}\delta^{((1+n)d)}(\REPX-\hat{c}_j)\!\Big\rangle_{H_{\DensFunc}}
	\!\!\!-\frac{1}{\Volu_0 \Volu^n} ,
\end{eqnarray}
which can be interpreted as
\begin{eqnarray}\label{EQ:VOPinte}
	\langle \VOP(\REPX) \rangle_{H_{\VOP}}
	&=&\!\frac{1}{N}\!\sum_{j=1}^{\PartNumb} \big\lbrack
		\langle \delta^{(d)}(x^0-c^{0}_{j})\rangle
		\langle \delta^{(d)}(x^1-c^{1}_{j})\rangle \cdots \nonumber\\
	&&\quad\times
		\langle \delta^{(d)}(x^n-c^{n}_{j})\rangle
		\big\rbrack - \frac{1}{\Volu_0 \Volu^n} \, .
\end{eqnarray}
This average consists of the following two steps: One first constructs \emph{independent thermal averages in each replica} (denoted by $\langle\cdots\rangle$) with a common given realization of disorder $\RealDiso$; one then forms the product over all replicas, and finally one \emph{averages over all realizations of disorder} (an average denoted by $\big\lbrack\cdots\big\rbrack$).
This interpretation can be understood from the definition of $H_{\DensFunc}$ via $Z_{1+n}$, as in Eq.~(\ref{EQ:ZQ}). Recall that $Z_{1+n}$, as defined in Eq.~(\ref{EQ:ReplPart}), contains thermal averages of the $(1+n)$ replicas, represented by the factor $Z_{\RealDiso}(\Volu_0)\, Z_{\RealDiso}(\Volu)^n$, together with an overall disorder average. This validates the interpretation given in Eq.~(\ref{EQ:VOPinte}). For a strict proof, see Ref.~\cite{Goldbart1996}.

The structure of Eq.~(\ref{EQ:VOPinte}) allows us to relate the value of the order parameter $\VOP$ to measurements on the system. In the liquid state, the single-particle densities $\ltha \delta^{(d)}(x\REPa-c\REPa_{j}) \rtha$ in each replica are simply $1/\Volu$ (or $1/\Volu_0$ for the $0^{\textrm{th}}$ replica), and thus the order parameter $\VOP$ vanishes.
In the soft random solid state, it is hypothesized that \emph{a finite fraction $\LocaPart$ of the particles become localized around random positions}. This happens when the density of links exceeds the \emph{percolation threshold}, and the particles that constitute the infinite, percolating cluster become localized.

In the language of replica field theory, a localized particle remains near the same spatial position in each replica\footnote{
Strictly speaking, the localized infinite cluster can undergo global translations or rotations from replica to replica, and this corresponds to a distinct equilibrium states of the field theory, which are connected by relative translations and/or rotations of the replicas.}.
This is because, as we discussed earlier, \emph{each replica corresponds to a copy of the same disordered system but with independent thermal fluctuations}, and for a particle localized as a part of the percolating cluster, it fluctuates around its fixed mean position, which is common to all replicas.
According to these considerations, it is reasonable to hypothesize the following form for the stationary value of the order parameter in real space:
\begin{eqnarray}\label{EQ:VOPAnsatzR}
	\!\!\!\!\VOPSP(\REPX)\!\!\!\!&=&\!\!\!\LocaPart \int \frac{dz}{\Volu_0}\int d\ISLL \DistISLL(\ISLL)
	\Big(\frac{\ISLL}{2\pi}\Big)^{\frac{(1+n)\Dime}{2}}
	\nonumber\\
	&& \!\!\times e^{-\frac{\ISLL}{2}
	\{ \vert x^0-z\vert^2 +\ReplSumOne\vert x\REPa-\Contraction z \vert^2 \}} \!\!
	-\!\frac{\LocaPart}{\Volu_0 \Volu^n}  .\,
\end{eqnarray}
To arrive at this form we have assumed that the fraction of particles that become localized is $\LocaPart$, and the localized single-particle density function is proportional to $e^{-\frac{\ISLL}{2}\vert x\REPa-\Contraction z \vert^2}$ in replica $\alpha$($=1,\ldots,n$), where $\Contraction z$ is the random position near to which the particle is localized, and $\Contraction$ corresponds to the uniform contraction of the entire volume in the measurement state with respect to the preparation state, due to linking [see the discussion following Eq.~(\ref{EQ:ContRLPM})]. Correspondingly, this particle was near the position $z$ in the preparation state, so, for replica $0$ the corresponding single-particle density function is proportional to $e^{-\frac{\ISLL}{2}\vert x^0-z \vert^2}$. For convenience of notation we define $\hat{z}\equiv (z,\Contraction z, \Contraction z,\ldots)$ as the mean position vector in the replicated space. The contraction $\Contraction$ is related to the change of volume as
\begin{align}\label{EQ:VContraction}
	\frac{\Volu}{\Volu_0}=\Contraction^{d} .
\end{align}
The localization of the particle is characterized by the localization length $\LocaLeng$, although for notational convenience we exchange this variable for the inverse square localization length $\ISLL\equiv 1/\LocaLeng^2$. Because the network is heterogeneous, the particles can have widely different localization lengths. This heterogeneity is characterized by the distribution $\DistISLL(\ISLL)$.

We can also write this stationarity-point order parameter in momentum space:
\begin{eqnarray}\label{EQ:VOPAnsatzM}
	\!\!\!\!(\!\VOPSP\!)_{\REPP}\!\!\!&=&\!\!\!\LocaPart \int \frac{dz}{\Volu_0}\int d\ISLL \DistISLL(\ISLL)
	 \nonumber\\  \noalign{\medskip}
	&& \!\!\times e^{-\frac{\vert\REPP\vert^2}{2\ISLL}
			\!-\! ip^0 \cdot z-i\ReplSumOne p\REPa \cdot (\Contraction z) }
	\!\!-\!\LocaPart\delta^{((1+n)d)}_{\REPP,0}\! .\,\,\,
\end{eqnarray}
The parameters that characterize this order parameter, $\LocaPart$ and $\DistISLL(\ISLL)$, have been obtained by solving the stationarity condition for the Hamiltonian:
\begin{align}
	\frac{\delta H_{\VOP}}{\delta \VOP_{\REPP}}=0 .
\end{align}
The form of the order parameter $(\VOPSP)_{\REPP}$ given in Eq.~(\ref{EQ:VOPAnsatzM}) exactly solves the above stationarity condition, thus we arrive at self-consistency equations of $\LocaPart$ and $\DistISLL(\ISLL)$.
In particular, the equation for $\LocaPart$ is
\begin{eqnarray}\label{EQ:SPQ}
	1-\LocaPart=e^{-\LinkDens \LocaPart}.
\end{eqnarray}
For all values of $\LinkDens$, Eq.~(\ref{EQ:SPQ}) has a solution $\LocaPart=0$, corresponding to the liquid state. However, for $\LinkDens> 1$, an additional root appears, emerging continuously from $\LocaPart=0$ at $\LinkDens= 1$, and describing the equilibrium amorphous solid state. In Fig.~\ref{FIG:ThetaGamma} we show the dependence of the localized fraction $\LocaPart$ on the link density, which we characterize by $\LinkDens$. {\colorred{The critical point $\LinkDens_C=1$ corresponds to mean coordination number $z$ of $1$, and this agrees with the classic work on the statistical properties of random graphs by Erd{\H o}s and R{\'e}nyi~\cite{Erdos1960}.}} For a detailed discussion and for the stationary-point distribution of inverse square localization lengths $\DistISLL(\ISLL)$, see Refs.~\cite{Goldbart1996,Castillo1994}.

The contraction $\Contraction$, which is relevant to the elasticity of the random solid state, can be investigated by inserting the form (\ref{EQ:VOPAnsatzM}) of the order parameter into the Hamiltonian $H_{\VOP}$, Eq.~(\ref{EQ:HVOPHRS}), which yields the dependence of the Hamiltonian on the parameters $\LocaPart$, $\DistISLL(\ISLL)$ and $\Contraction$. Through a lengthy derivation, and by keeping terms to $O(n)$, we arrive at the following Hamiltonian for the stationary point (cf.~Appendix~\ref{APP:HSP}):
\begin{widetext}
\begin{eqnarray}\label{EQ:HSP}
	\HVOPSP&=&\frac{\ExclVoluRN_0(0)\PartNumb^2}{2\Volu_0}
		+\frac{n\ExclVoluRN(0)\PartNumb^2}{2\Volu}
	-\PartNumb \BoltCons T \ln \Volu_0 -n\PartNumb \BoltCons T \ln \Volu
	+ n\PartNumb \BoltCons T \Bigg\lbrace
		\UnivPara \Big\lbrack
			\frac{d}{2}\big(\ln(2\pi)+\Contraction^2\big)-\ln\Volu
		\Big\rbrack \nonumber\\
	&&-	\frac{\LinkDens \LocaPart^2}{2}\cdot\frac{d}{2}\int_{\ISLL_1,\ISLL_2}
		\ln\big( \frac{1}{\ISLL_1}+\frac{1}{\ISLL_2}+\LinkScal^2 \big)
		-e^{-\LinkDens \LocaPart}\frac{d}{2}
		\sum_{m=1}^{\infty}\frac{(\LinkDens \LocaPart)^m}{m!}
		\int_{\ISLL_1,\ldots,\ISLL_m} \ln
		\Big( \frac{\tISLL_1\cdots\tISLL_m}{\tISLL_1 +\cdots +\tISLL_m} \Big)
	\Bigg\rbrace .
\end{eqnarray}
\end{widetext}
Here, the variable $\tISLL$ is defined as $\tISLL\equiv\big(\frac{1}{\ISLL}+\LinkScal^2\big)^{-1}$, where $\ISLL$ is the inverse square localization length $\ISLL\equiv 1/\LocaLeng^2$, and we have introduced the shorthand for the integrals $\int_{\ISLL}\equiv \int d\ISLL \DistISLL(\ISLL)$.
The dimensionless factor $\UnivPara$ in Eq.~(\ref{EQ:HSP}) is given by
\begin{eqnarray}\label{EQ:theta}
	\UnivPara\equiv -\frac{\LinkDens \LocaPart^2}{2}+\LinkDens \LocaPart
	-1+e^{-\LinkDens \LocaPart}.
\end{eqnarray}
We shall see in Section \ref{SEC:Heterogeneity} that $\UnivPara$ also controls the mean value of the shear modulus, as well as the amplitude of the disorder correlators that involve the residual stress fields.

To obtain the disorder-averaged free energy, we shall make the stationary-point approximation:
\begin{eqnarray}
	Z_{1+n}\simeq e^{-\frac{\HVOPSP}{\BoltCons T}} .
\end{eqnarray}
Thus, we can obtain the Helmholtz free energy using Eq.~(\ref{EQ:HFEDisoAver}), arriving at the result
\begin{widetext}
\begin{eqnarray}
	\lda \HelmFreeEner_{\textrm{SP}} \rda \!&=&\! -\BoltCons T  \lim _{n \to 0} \frac{1}{n}
		\Big( \frac{Z_{1+n}}{Z_1}-1 \Big) \nonumber\\
	\!&=&\! -\PartNumb \BoltCons T \Big(1-\frac{\LinkDens}{2}+\UnivPara \Big) \ln\Volu
			-\PartNumb \BoltCons T \frac{\UnivPara d}{2}\big(\ln(2\pi)+\Contraction^2 \big)
			+\frac{\ExclVolu\PartNumb^2}{2\Volu}
		-\frac{\LinkDens \PartNumb \BoltCons}{2} \ln \LinkPote_0 \nonumber\\
	&&+ \PartNumb \BoltCons T \frac{\LinkDens\LocaPart^2}{2}\frac{d}{2}
			\int_{\ISLL_1,\ISLL_2}
			\!\ln\Big( \frac{1}{\ISLL_1}+\frac{1}{\ISLL_2}+\frac{\LinkScal^2}{\BoltCons T} \Big)\!
			\!+\!\PartNumb \BoltCons T \frac{d}{2} e^{-\LinkDens \LocaPart}
		\sum_{m=1}^{\infty}\frac{(\LinkDens \LocaPart)^m}{m!}
		\!\int_{\ISLL_1,\ldots,\ISLL_m}\! \!\!\ln
		\Big( \frac{\tISLL_1\cdots\tISLL_m}{\tISLL_1 +\cdots +\tISLL_m} \Big) ,
\end{eqnarray}
where we have used the mean-field value of $Z_1$, from Eq.~(\ref{EQ:Z1SP}), and we have also made an expansion for small $n$ of the renormalized excluded-volume parameter $\ExclVoluRN$, using
\begin{eqnarray}
	\LinkPoteRepl_{0}=(\LinkPote_{0})^{1+n}
	=\LinkPote_{0}\big(1+n\ln \LinkPote_{0} +O(n^2)\big).
\end{eqnarray}
In order to study elasticity, we shall need to know the disorder-averaged Gibbs free energy $\lda\GibbFreeEner\rda$, which is given by a Legendre transformation, Eq.~(\ref{EQ:LegeTran}):
\begin{eqnarray}
	 \lda\GibbFreeEner_{\textrm{SP}}\rda=\lda\HelmFreeEner_{\textrm{SP}}\rda+\Pres\Volu .
\end{eqnarray}
{\colorred{We can insert the pressure $p$, given by Eq.~(\ref{EQ:AverPres}), and drop the slowly-varying $\ln\Volu$ term given that $\UnivPara\simeq\LinkDens/2-1$ provided the system is not close to the critical point $\LinkDens_C=1$.  In the limit $\frac{\ExclVolu N}{\BoltCons T \Volu}\gg \LinkDens$ we arrive at}}
\begin{eqnarray}\label{EQ:SPGIBBS}
	\lda \GibbFreeEner_{\textrm{SP}} \rda
	\!&\simeq&\! \frac{\ExclVolu\PartNumb^2}{2\Volu_0}
		\Big\lbrack 2+\Big(\frac{\Volu}{\Volu_0}-1\Big)^2\Big\rbrack
			-\PartNumb \BoltCons T \frac{\UnivPara d}{2}\big(\ln(2\pi)+\Contraction^2 \big)
			-\frac{\LinkDens \PartNumb \BoltCons}{2} \ln \LinkPote_0 \nonumber\\
	&&+ \PartNumb \BoltCons T \frac{\LinkDens\LocaPart^2}{2}\frac{d}{2}
			\int_{\ISLL_1,\ISLL_2}\!\!
			\ln\Big( \frac{1}{\ISLL_1}+\frac{1}{\ISLL_2}+\frac{\LinkScal^2}{\BoltCons T} \Big)
			\!+\!\PartNumb \BoltCons T \frac{d}{2} e^{-\LinkDens \LocaPart}
		\!\sum_{m=1}^{\infty}\frac{(\LinkDens \LocaPart)^m}{m!}
		\!\int_{\ISLL_1,\ldots,\ISLL_2} \!\!\!\ln
		\Big( \frac{\tISLL_1\cdots\tISLL_m}{\tISLL_1 +\cdots +\tISLL_m} \Big).\,\,
\end{eqnarray}
\end{widetext}
Using the relation~(\ref{EQ:VContraction}), we can obtain the stationary values of the contraction $\Contraction$ that minimizes the disorder-averaged Gibbs free energy by solving
\begin{align}
	\frac{\delta \lda \GibbFreeEner_{\textrm{SP}} \rda }{ \delta \Contraction}=0 .
\end{align}
{\colorred{In the limit $\frac{\ExclVolu N}{\BoltCons T \Volu}\gg \LinkDens$, the solution is
\begin{eqnarray}\label{EQ:ContRLPM}
	\Contraction \simeq 1-\frac{ \UnivPara\Volu_0 \BoltCons T}{\ExclVolu \PartNumb d}.
\end{eqnarray}
The limit $\frac{\ExclVolu N}{\BoltCons T \Volu}\gg \LinkDens$ is the same as the limit taken below Eq.~(\ref{EQ:ExclVoluRN}), indicating that the excluded-volume repulsion is much stronger than the attractive effects of the links.}}
This contraction of the volume due to the introduction of links at a given pressure is a result of both the reduction of the total number of translational degrees of freedom, i.e., the change of the \lq\lq osmotic pressure\rlap,\rq\rq\thinspace\ and the attractive interactions induced by the links. We shall see later that this contraction is consistent with a particular phenomenological model of a disordered elastic medium that we shall introduce.

\subsection{Goldstone fluctuations in the RLPM}

\subsubsection{Spontaneous symmetry breaking}
\label{SEC:SSB}
To characterize the Goldstone modes of fluctuations associated with the random solid state, we shall first look at the pattern of symmetry breaking accompanying the transition to this state.

The Hamiltonian~(\ref{EQ:HVOPHRS}) for the liquid-to-soft-random-solid transition has the symmetry of independent translations and rotations of each replica. The translational invariance of the Hamiltonian can be readily verified by making the transformation
\begin{eqnarray}
	\hat{x} &\to& \hat{x}'=\hat{x}+\hat{a} , \nonumber\\
	\VOP(\hat{x}) &\to& \VOP'(\hat{x}')=\VOP(\hat{x}) = \VOP(\hat{x}'-\hat{a}) ,
\end{eqnarray}
where $\hat{a}\equiv(a^0,a^1,\ldots,a^n)$ represents a replicated translation. In momentum space this transformation reads
\begin{eqnarray}
	\VOP_{\hat{p}} &\to & \VOP'_{\hat{p}}=e^{i\hat{p}\cdot\hat{a}}\, \VOP_{\hat{p}}\, .
\end{eqnarray}
It is easy to check that, by inserting this transformed order parameter back into the Hamiltonian~(\ref{EQ:HVOPHRS}) and making a change of variables, the same Hamiltonian but for the field $\VOP '$ is recovered. Similarly, one can verify invariance under independent rotations $\hat{\Tens{O}}\equiv(\Tens{O}^{0},\Tens{O}^{1},\ldots,\Tens{O}^{n})$ with
\begin{eqnarray}
	\VOP_{\hat{p}} \to \VOP'_{\hat{p}} = \VOP_{\hat{\Tens{O}}^{-1}\cdot\hat{p}} \, .
\end{eqnarray}

The order parameter in the liquid state (i.e., $\VOP=0$) has the full symmetry of the Hamiltonian, and at the transition to the soft random solid state it is the symmetry of \emph{relative} translations and rotations between different replicas that is spontaneously broken. However, the symmetry of \emph{common} translations and rotations of all replicas are preserved, and this reflects the important notion that from the macroscopic perspective the system remains translationally and rotationally invariant, even in the random solid state. This entire pattern of symmetry breaking amounts to an unfamiliar but essentially conventional example of the Landau paradigm.

This broken symmetry of relative translations and rotations between different replicas can be understood as a result of particle localization. Because a delocalized liquid particle can explore the whole volume via its thermal fluctuations, and in thermal equilibrium its positions in different replicas are uncorrelated, the liquid state is invariant, under separate translation and rotation of individual replica. On the contrary, for a localized particle, its positions in the various replicas are strongly correlated, and therefore the symmetries of relative translations and rotations are broken.

It is straightforward to verify that the form of the random-solid-state order parameter,  Eq.~(\ref{EQ:VOPAnsatzR}), correctly implements this pattern of symmetry breaking.
To see this, we can use the complete orthonormal basis in replica space defined in Section \ref{SEC:FTMF}, and define an alternative basis involving a \lq\lq replica (almost) body-diagonal\rq\rq\ unit vector
\begin{eqnarray}
	\BASEl \equiv \frac{1}{\sqrt{1+n\Contraction^2}} \big(\BASE^{0}+\Contraction\ReplSumOne\BASE\REPa\big) .
\end{eqnarray}
Relative to $\BASEl$, we may decompose a $(1+n)d$ dimensional vector $\hat{x}$ into its longitudinal ($\lambda$) and transverse ($\tau$) components:
\begin{eqnarray}
	\hat{x}=\hat{x}_{\lambda}+\hat{x}_{\tau}, \quad\,
	\hat{x}_{\lambda} = (\hat{x}\cdot\BASEl)\BASEl, \quad\,
	\hat{x}_{\tau} = \hat{x}-\hat{x}_{\lambda} \, .
\end{eqnarray}
Note that $\hat{x}_{\lambda}$ and $\hat{x}_{\tau}$ are both $(1+n)d$-dimensional vectors, but $\hat{x}_{\lambda}$ has only $d$-degrees of freedom (given by $\hat{x}\cdot\BASEl$), and $\hat{x}_{\tau}$ has only $nd$-degrees of freedom.

\begin{figure}[htbp]
	\includegraphics[width=.48\textwidth]{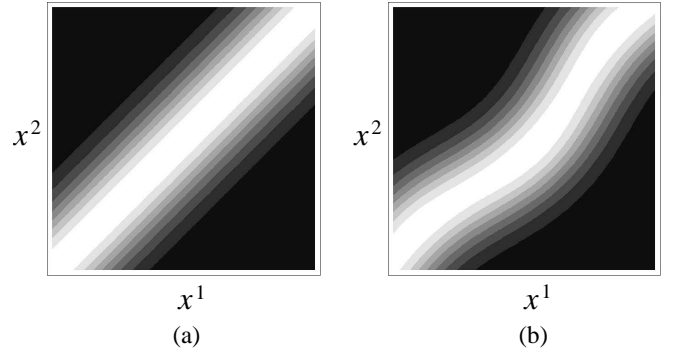}
	\caption{(a)~Schematic plot of the value of the order parameter (brightness) at the stationary point, for the illustrative two replicas, labeled by $x^1$ and $x^2$. (b)~Schematic plot of the value of the order parameter (brightness) for a Goldstone deformation of the stationary point for two replicas.}
	\label{FIG:OPab}
\end{figure}

By this decomposition, the vector $\hat{z}=(z,\Contraction z, \Contraction z, \ldots )$, which characterizes the mean positions of a particle in the stationary-point state, can be written as
\begin{eqnarray}
	\hat{z} = \sqrt{1+n\Contraction^2} \, z\, \BASEl ,
\end{eqnarray}
which points purely in the $\BASEl$ direction. As a result, the stationary order parameter, Eq.~(\ref{EQ:VOPAnsatzR}), can be written as
\begin{eqnarray}\label{EQ:OPInv}
	\VOPSP(\hat{x}) &=& \LocaPart \int \frac{dz}{\Volu_0}
		\int d\ISLL \DistISLL(\ISLL)
		\Big(\frac{\ISLL}{2\pi}\Big)^{\frac{(1+n)\Dime}{2}} \nonumber\\
		&& \quad\times e^{-\frac{\ISLL}{2} \vert \hat{x}_{\lambda}-\hat{z}_{\lambda}\vert^2
			-\frac{\ISLL}{2} \vert \hat{x}_{\tau}\vert^2} - \frac{\LocaPart}{\Volu_0\Volu^n} \nonumber\\
		&=& \LocaPart \int d\ISLL \DistISLL(\ISLL)
				\Big(\frac{\ISLL}{2\pi}\Big)^{\frac{(1+n)\Dime}{2}}
				\Big(\frac{2\pi}{\ISLL(1+n\Contraction^2)}\Big)^{\frac{d}{2}} \nonumber\\
		&& \quad\times	 e^{-\frac{\ISLL}{2} \vert \hat{x}_{\tau}\vert^2} - \frac{\LocaPart}{\Volu_0\Volu^n} ,
\end{eqnarray}
where in the last line we have integrated out the $d$-dimensional vector $z$. It is evident that this value of order parameter does not depend on $\hat{x}_{\lambda}$, which means that it is \emph{invariant under translations in the $\BASEl$ direction, corresponding to common translations and rotations of all replicas (albeit appropriately contracted by $\Contraction$ in replicas $1$ through $n$)}.
This stationary order parameter is shown schematically in Fig.~\ref{FIG:OPab}(a) for two replicas.
The Gaussian-like form in the $\hat{x}_{\tau}$ direction indicates a \lq\lq condensation\rq\rq~between different replicas. This is called a \emph{molecular bound state} in Ref.~\cite{Goldbart2004}.

\subsubsection{Goldstone fluctuations}\label{SEC:GSFT}
With the pattern of continuous symmetry breaking just outlined, we can write down the form that the order parameter takes when it is subject to
\lq\lq Goldstone fluctuations\rq\rq :
\begin{eqnarray}\label{EQ:VOPGSR}
	\VOPGS(\REPX)&=&\LocaPart \int \frac{dz}{\Volu_0}\int d\ISLL\, \DistISLL(\ISLL)
	\Big(\frac{\ISLL}{2\pi}\Big)^{\frac{(1+n)\Dime}{2}}
	\nonumber\\ \noalign{\medskip}
	&& \,\times e^{-\frac{\ISLL}{2}
	\{ \ReplSum\vert x\REPa-\DefoPosi\REPa(z) \vert^2 \}}
	-\frac{\LocaPart}{\Volu_0 \Volu^n} ,
\end{eqnarray}
where $\DefoPosi^{0}(z)=z$. Therefore the Goldstone deformation of the order parameter is parameterized by the $n$ independent functions $\{\DefoPosi^{1}(z),\ldots,\DefoPosi^{n}(z)\}$.
The stationary form of the order parameter, Eq.~(\ref{EQ:VOPAnsatzR}), describes a system in which the mean positions of the replicas of the thermally fluctuating particles are located at $(x^0,x^1,\ldots,x^n)=(z,\Contraction z,\ldots,\Contraction z)$. We shall refer to these positions as the \lq\lq \emph{centers of the thermal cloud}\rlap.\rq\rq\ By comparing the undeformed order parameter~(\ref{EQ:VOPAnsatzR}) and the \lq\lq Goldstone-deformed\rq\rq~one, Eq.~(\ref{EQ:VOPGSR}), we see that the Goldstone-deformed order parameter describes a system in which the mean positions of the replicas of the fluctuating particles are displaced from $(z,\Contraction z,\ldots,\Contraction z)$ to $(z, \DefoPosi^{1}(z),\ldots,\DefoPosi^{n}(z))$. Thus, $\DefoPosi\REPa(z)$ ($\alpha=1,2,\ldots,n$) represent the deformed mean positions in the measurement replicas.

We require that the deformations $\Contraction z \to \DefoPosi\REPa(z)$ be \emph{pure shear} deformations. This constraint can be expressed as $\textrm{det}\big(\partial \DefoPosi_i\REPa(z)/\partial (\Contraction z)_j\big)=1$; it guarantees that the Goldstone fluctuation does not excite the LRS (i.e., each replica still has homogeneous density), which would be extremely energetically costly, owing to the large excluded-volume interaction. The 0RS has already been removed from the theory, and one can easily check that it remains zero in this Goldstone-deformed order parameter. The vanishing of the order parameter in the 1RS can be verified by taking the momentum-space Goldstone deformation and making a change of variables:
\begin{widetext}
\begin{eqnarray}
	(\VOPGS)_{\hat{q}}\big\vert_{\hat{q}=p\BASE\REPa}
	&=& \LocaPart \int \frac{dz}{\Volu_0}\int d\ISLL \DistISLL(\ISLL)
		 e^{-\frac{\vert p \vert^2}{2\ISLL}
		-ip \, \DefoPosi\REPa(z)}
		-\LocaPart\delta^{((1+n)d)}_{\REPP,0}	\nonumber\\
	&=&\LocaPart \int \frac{d\DefoPosi\REPa}{\Volu} \frac{\Volu}{\Volu_0}
		\Big\vert\frac{\partial z}{\partial \DefoPosi\REPa}\Big\vert
		\int d\ISLL \DistISLL(\ISLL)
		 e^{-\frac{\vert p \vert^2}{2\ISLL}
		-ip \, \DefoPosi\REPa(z)}-\LocaPart\delta^{((1+n)d)}_{\REPP,0}=0.
\end{eqnarray}
\end{widetext}
This result indicates that the deformed state has the same local density as the (undeformed) stationary-point state.

\begin{figure}[htbp]
	\centering
		\includegraphics[width=.45\textwidth]{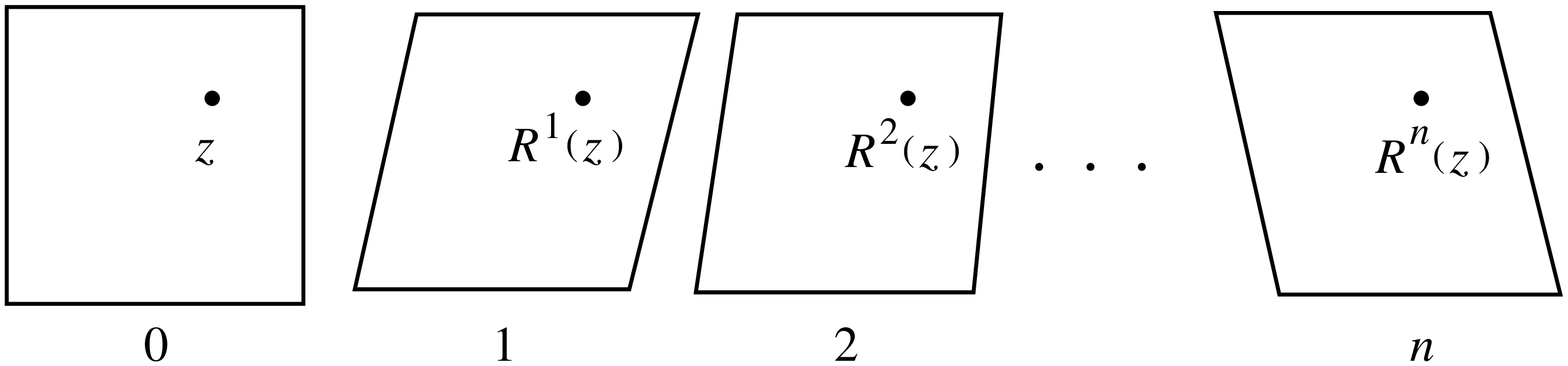}
	\caption{Example of a Goldstone-deformed state. The system of replicas $0$ through $n$ are shown. The mean positions of the replicas of a thermally fluctuating particle are displaced to $(z, \DefoPosi^{1}(z),\ldots,\DefoPosi^{n}(z))$ in this Goldstone-deformed state, which characterizes an $n$-fold replicated deformation field. Here, for simplicity, we only show spatially homogeneous deformations, and it is worth noticing that the volumes of the measurement replicas, i.e., replicas $1$ through $n$, are contracted by a factor $\Contraction^{\Dime}$.}
	\label{FIG:ReplGoldstone}
\end{figure}

We ought to clarify the following point about this Goldstone deformation.  As we have already mentioned, the symmetry that is broken at the transition is that of relative translations and rotations of the various replicas, the symmetry of common translations and rotations remaining intact.
As a result, the Goldstone deformation should be constructed via $z$-dependent translations of the order parameter in the $\hat{x}_{\tau}$ direction, i.e., \emph{the broken symmetry direction}. However, if we look at the deformation field defined by $\hat{\GSDF}\equiv\hat{\DefoPosi}-\hat{z}$, we find that $\hat{\GSDF}=(0,\DefoPosi^{1}(z)-\Contraction z, \DefoPosi^{2}(z)-\Contraction z , \ldots)$ is in  fact \emph{not} in the broken symmetry direction $\hat{x}_{\tau}$, because it has an $\hat{x}_{\lambda}$ component, viz.,
\begin{eqnarray}
	\hat{\GSDF}_{\lambda}=(\hat{\GSDF}\cdot\BASEl )\,\, \BASEl
	= \frac{\Contraction}{\sqrt{1+n\Contraction^2}} \ReplSumOne \GSDF\REPa(z) \,\, \BASEl .
\end{eqnarray}
This $\hat{\GSDF}_{\lambda}$ component is actually redundant. This can be seen by decomposing the quadratic form $\vert \hat{x}-\hat{z}-\hat{\GSDF} \vert^2$ as
\begin{eqnarray}
	\vert \hat{x}-\hat{z}-\hat{\GSDF} \vert^2 = \vert \hat{x}_{\lambda}- \hat{z}_{\lambda}-\hat{\GSDF}_{\lambda} \vert^2
	+ \vert \hat{x}_{\tau}-\hat{\GSDF}_{\tau} \vert^2 ,
\end{eqnarray}
and noting that in the form of the order parameter~(\ref{EQ:VOPGSR}) one can change the integration variable (which is a $d$-dimensional vector) from $z$ to
\begin{eqnarray}
	y\equiv z+\frac{1}{\sqrt{1+n\Contraction^2}} \,\hat{\GSDF}\cdot\BASEl  ,
\end{eqnarray}
so the longitudinal component of the $(1+n)\Dime$-dimensional vector $\hat{y}\equiv(y,\Contraction y,\ldots,\Contraction y)$ is $\hat{y}_{\lambda}=\hat{z}_{\lambda}+\hat{\GSDF}_{\lambda}$.
The Jacobian of this change of variables is unity, provided that each deformation $z\to \GSDF\REPa(z)$ is a pure shear deformation~\footnote{This relation is correct to linear order in $\GSDF$ or linear order in $n$.}. With this change of variables the Goldstone-deformed order parameter attains the form
\begin{widetext}
\begin{eqnarray}\label{EQ:VOPGSRTWO}
	\VOPGS(\hat{x}) &=& \LocaPart \int \frac{dy}{\Volu_0} \int d\ISLL \DistISLL(\ISLL)
			\Big(\frac{\ISLL}{2\pi}\Big)^{\frac{(1+n)\Dime}{2}}  
	 e^{-\frac{\ISLL}{2} \vert \hat{x}_{\lambda} - \hat{y}_{\lambda}\vert^2
			-\frac{\ISLL}{2} \vert \hat{x}_{\tau} - \hat{\GSDF}'_{\tau}(y)\vert^2}
		- \frac{\LocaPart}{\Volu_0\Volu^n} \nonumber\\
	&=& \LocaPart \int \frac{dy}{\Volu_0} \int d\ISLL \DistISLL(\ISLL)
			\Big(\frac{\ISLL}{2\pi}\Big)^{\frac{(1+n)\Dime}{2}} 
	e^{-\frac{\ISLL}{2} \vert \hat{x} - \hat{y}-\hat{\GSDF}'_{\tau}(y) \vert^2}
		- \frac{\LocaPart}{\Volu_0\Volu^n} \nonumber\\
	&\equiv& \VOPGS'(\hat{y}) ,
\end{eqnarray}
\end{widetext}
where the transformed deformation field $\hat{\GSDF}'$ is defined via $\hat{\GSDF}'_{\tau}(y)=\hat{\GSDF}_{\tau}(z)$.
With this change of variables, the order parameter field, $\VOPGS(\hat{x})$ of $\hat{x}$, is transformed to a new field, $\VOPGS'(\hat{y})$ of $\hat{y}$, which can be viewed as being the stationary point $\VOP_{SP}(\hat{y})$ but locally translated purely in the $\hat{y}_{\tau}$ directions,
because the deformation is $\hat{\DefoPosi}'(y)=\hat{y}+\hat{\GSDF}'_{\tau}(y)$. Comparing with the deformation before the change of variables, $\hat{\DefoPosi}(z)=\hat{z}+\hat{\GSDF}(z)$, it is evident that the $\GSDF_{\lambda}$ component is actually removed, and the deformation field $\hat{\GSDF}$ only affects the $\hat{x}_{\tau}$ direction. Therefore, $\GSDF_{\lambda}$ is a redundant component in this field-theoretic description of the Goldstone fluctuation. Note that in these two representations of the Goldstone fluctuation, Eqs.~(\ref{EQ:VOPGSR}) and (\ref{EQ:VOPGSRTWO}), the number of degrees of freedom of the deformation field, $\GSDF(z)$ or $\hat{\GSDF}'_{\tau}(y)$, is $nd$, because in $\GSDF(z)$ one has the constraint $\GSDF^{0}=0$.

The reason we choose to adopt the form of Goldstone fluctuation given in Eq.~(\ref{EQ:VOPGSR}) is that in the true physical system that we are intending to describe, the preparation state (replica $0$) is not deformed. Although in the field theory the $1+n$ replicas feature symmetrically (apart from the contraction $\Contraction$), physically, one should only have Goldstone fluctuations that deform replicas $1$ through $n$, as these are the replicas associated with the measurement states, on which deformations are actually performed. Therefore, although Eqs.~(\ref{EQ:VOPGSR}) and (\ref{EQ:VOPGSRTWO}) are mathematically equivalent, Eq.~(\ref{EQ:VOPGSR}) provides a better physical description of the Goldstone fluctuations~\footnote{
In earlier work (see Refs.~\cite{Goldbart2004, Mukhopadhyay2004}),
the form of the Goldstone fluctuation was taken to be
$	\VOPGS(\hat{x}) = \LocaPart \int d\ISLL \DistISLL(\ISLL)
			\big(\ISLL/2\pi\big)^{(1+n)\Dime/2}
			 e^{-\frac{\ISLL}{2} \vert \hat{x}_{\tau} - \hat{\GSDF}_{\tau}(x_{\lambda})\vert^2}
			 - \LocaPart/\big(\Volu_0\Volu^n\big)$.
This form
(which we shall call the \lq\lq old Goldstone deformation\rq\rq)
differs from the Goldstone deformation that we are currently using
(the \lq\lq new Goldstone deformation\rq\rq) in two ways:
(i)~In the old Goldstone deformation, the deformation field was taken to be a function of $x_{\lambda}$
[and, as a result, $z$ can be integrated out, as in the stationary-point form (\ref{EQ:OPInv})].
However, in the new Goldstone deformation, the deformation is instead taken to be function of $z$.
The new Goldstone deformation is more physical, in the sense that the deformation field should be defined in terms of the the \emph{mean} positions $z$ during thermal fluctuations, not the \emph{instantaneous} positions of the particles $\hat{x}$.
In the new Goldstone deformation, it is clear that it is the mean positions that are deformed, $\hat{z} \to \hat{\DefoPosi}(z)$, but the shape of the thermal cloud, which corresponds to a \lq\lq massive\rq\rq~mode, is left untouched.  (The point was already made in Ref.~\cite{Ulrich2006}, but we make it again, here, for the sake of completeness.)\thinspace\
(ii)~The deformation field in the old Goldstone deformation lies in the $\hat{x}_{\tau}$ direction, whereas the new Goldstone deformation has a deformation field in replicas $1$ through $n$.
This issue has already been discussed in Sec.~\ref{SEC:GSFT}, as has been the point that these two representations are related by a change of variables; see Eq.~(\ref{EQ:VOPGSRTWO}).
The new Goldstone structure is more physical, in the sense that the preparation state (replica $0$) cannot be deformed once the sample has been made.
}.

\subsubsection{Energetics of Goldstone deformations}
\label{SEC:EnerGoldSton}
\begin{figure}[htbp]
	\centering
		\includegraphics[width=.4\textwidth]{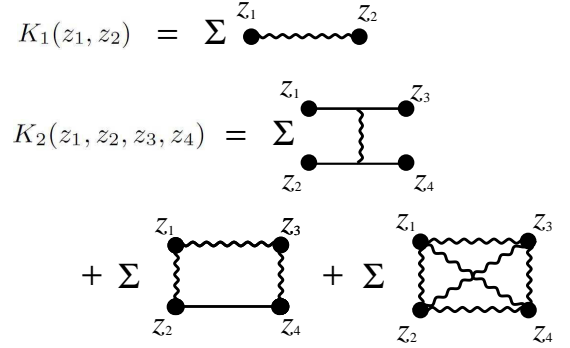}
	\caption{Diagrams for the functions $\KernOne(z_1,z_2)$ and $\KernTwo(z_1,z_2,z_3,z_4)$. In these diagrams, the straight lines represent delta functions, and the wavy lines represent Gaussian potentials between pairs of points, averaged over the distribution of localization lengths, as indicated schematically by the summation signs, which also indicate symmetrization over the arguments. The typical localization length provides the characteristic lengthscale for $\KernOne$ and $\KernTwo$. The full expression is listed in Appendix~\ref{APP:HGS}.}
	\label{FIG:KK}
\end{figure}

To obtain the energy of a Goldstone deformed state, we take the momentum-space version of the Goldstone-deformed order parameter, Eq.~(\ref{EQ:VOPGSR}),
\begin{eqnarray}\label{EQ:VOPGSM}
	(\VOPGS)_{\REPP}&=&\LocaPart \int \frac{dz}{\Volu_0}\int d\ISLL \DistISLL(\ISLL)
	 \nonumber\\
	&& \quad\times e^{-\frac{\vert\REPP\vert^2}{2\ISLL}
	-i\REPP\cdot\hat{\DefoPosi}(z)}
	-\LocaPart \delta^{(1+n)d}_{\REPP,0} ,
\end{eqnarray}
and insert it into the Hamiltonian~(\ref{EQ:HVOPHRS}). After a lengthy calculation (see Appendix~\ref{APP:HGS}), similar to the one for the stationary-point free energy, we arrive at the energy of the Goldstone deformed state:
\begin{eqnarray}\label{EQ:HGS}
	\HVOPGF= \HVOPSP+H_{\VOP}^{\DefoScalPsi}.
\end{eqnarray}
Here, we use the short-hand
\begin{eqnarray}
	 \!\!\!\!\DefoScalPsi(z_1,z_2)\!\!&\equiv&\!\!(\hat{\DefoPosi}(z_1)\!-\!\hat{\DefoPosi}(z_2))^2\!-\!(1+n)(z_1\!-\!z_2)^2 \nonumber\\
	 \!\!&=&\!\!\!\! \ReplSumOne\! \big((\DefoPosi\REPa(z_1)\!\!\!-\!\!\DefoPosi\REPa(z_2))^2\!-\!(z_1\!-\!z_2)^2\!\big)
\end{eqnarray}
to denote the deformation. The term $\HVOPSP$ is the Hamiltonian at the stationary point, as given in Eq.~(\ref{EQ:HSP}), and the term $H_{\VOP}^{\DefoScalPsi}$ accounts for the increase in the energy due to Goldstone deformation, and is given by
\begin{eqnarray}\label{EQ:GibbEnerGoldSton}
	\!\!\!\! H_{\VOP}^{\DefoScalPsi}
	\!\!\!\!&=&\!\!\frac{1}{2}\int dz_1 dz_2 \KernOne(z_1,z_2)\DefoScalPsi(z_1,z_2)
		\nonumber\\
	&&\!\!-\frac{1}{8\BoltCons T}\int dz_1 dz_2 dz_3 dz_4 \KernTwo(z_1,z_2,z_3,z_4)
		\nonumber\\
	&& \! \times \DefoScalPsi(z_1,z_2)\DefoScalPsi(z_3,z_4)
		\!-\!\!n\PartNumb \BoltCons T \frac{\UnivPara d}{2}(\Contraction^2\!-\!1\!) , \,
\end{eqnarray}
{{\colorred
where we have kept terms to linear order in $n$ and quadratic order in $\DefoScalPsi^2$ in this expansion.}}
The functions $\KernOne(z_1,z_2)$ and $\KernTwo(z_1,z_2,z_3,z_4)$ are bell-shaped functions of the distances $(z_1-z_2)$ and $(z_1-z_2)$, $(z_2-z_3)$, $(z_3-z_4)$, as shown in Fig.~\ref{FIG:KK}. They are independent of the centers of mass, $(z_1+z_2)/2$ for $\KernOne(z_1,z_2)$, and $(z_1+z_2+z_3+z_4)/4$ for $\KernTwo(z_1,z_2,z_3,z_4)$. The forms of $\KernOne(z_1,z_2)$ and $\KernTwo(z_1,z_2,z_3,z_4)$ are given in Appendix~\ref{APP:HGS}. In specific, $\KernOne(z_1,z_2)$ has the following schematic functional form
\begin{eqnarray}
	\KernOne(z_1,z_2) = \int dr \, C(r) e^{-\frac{\vert z_1-z_2\vert^2}{2r^2}} ,
\end{eqnarray}
which is a superposition of Gaussian distributions on the scale of $r$, where $r$ is a certain combination of localization lengths $\LocaLeng$, weighted by the distribution of those localization lengths.  The kernel $\KernTwo(z_1,z_2,z_3,z_4)$ has a similar structure, except that it also contains delta-function factors such as $\delta(z_1-z_3)$, which should be associated with the scale of the short-distance cutoff of the theory, which is also on the scale of typical localization length.

The form of the energy increase due to Goldstone fluctuations, Eq.~(\ref{EQ:GibbEnerGoldSton}), can be understood intuitively as follows.
This energy actually describes the elastic energy of replicated shear deformations of the system, as we have explained in Section~\ref{SEC:GSFT} (i.e., the Goldstone fluctuations in the RLPM are replicated shear deformations).
Thus, it is evident that the first term in Eq.~(\ref{EQ:GibbEnerGoldSton}) represents a coupling of the deformations $\hat{\DefoPosi}(z_1)$ and $\hat{\DefoPosi}(z_2)$ at the points $z_1$ and $z_2$.
The coupling function, $\KernOne(z_1,z_2)$, given in Eq.~(\ref{EQ:KoneApp}), has a magnitude controlled by the probability $\LocaPart$ for a particle to be localized, and a lengthscale controlled by the typical localization length; these two quantities are, in turn, determined by the link density parameter $\LinkDens$, which is the control parameter for the RLPM.
In particular, the first term in $\KernOne(z_1,z_2)$ carries a factor $\LocaPart^2$, which is the probability for both of the two mass-points, $z_1$ and $z_2$, to be in the infinite cluster, and therefore to have an elastic interaction between their deformations.
The second term in $\KernOne(z_1,z_2)$, which involves a summation over $m$ from $2$ to $\infty$, takes into account the interactions between these two points that are mediated via other mass points.
The four-point term, with coupling function $\KernTwo(z_1,z_2,z_3,z_4)$, is a bit more complicated.
It couples replicas of the shear deformation to one another and to themselves, also on the scale of the localization length.
These two terms in the elastic energy embody the statistics of the elastic free energy of the soft random solid in the language of replicas, and thus encode information about the statistics of the quenched random elastic parameters.
In Section~\ref{SEC:Phen} we shall uncover this statistical content via the introduction of a phenomenological model inspired by this elastic energy.

In order to compare with the phenomenological elastic free energy that we shall discuss in Section~\ref{SEC:Phen}, it is useful to make an alternative decomposition of the Hamiltonian to the one made in Eq.~(\ref{EQ:HGS}).  That decomposition (\ref{EQ:HGS}) was in terms of the stationary-point part and the part due to fluctuations.  The alternative decomposition reads
\begin{eqnarray}\label{EQ:HSepa}
	\HVOPGF= H_{\VOP}^{(0)}+H_{\VOP}^{(\DefoPosi)}.
\end{eqnarray}
The relation between the two decompositions is given by
\begin{eqnarray}
	H_{\VOP}^{(0)} &=& \HVOPSP-\ContEner(\Contraction) , \nonumber\\
	H_{\VOP}^{(\DefoPosi)} &=& H_{\VOP}^{\DefoScalPsi} \,\,\,\, + \ContEner(\Contraction) ,
\end{eqnarray}
where $\ContEner(\Contraction)$ accounts for the energy of the stationary point, measured with respect to the state right after linking, which is actually the elastic energy of the contraction $\Contraction$, i.e.,
\begin{eqnarray}\label{EQ:CE}
	\ContEner(\Contraction)
	= \frac{\ExclVolu \PartNumb^2}{2\Volu_0} \Big( \frac{\Volu}{\Volu_0}-1 \Big)^2
		+\PartNumb \BoltCons T \frac{\UnivPara d}{2}(\Contraction^2-1).
\end{eqnarray}
In Eq.~(\ref{EQ:HSepa}), we are separating the Hamiltonian into two parts. $H_{\VOP}^{(0)}$ gives energy of \lq\lq the state right after linking\rlap,\rq\rq\ which is a state that has not been allowed to contract after the links were made and thus has the same volume and shape as the liquid state. In the state right after linking, the mean positions of the replicas of the thermally fluctuating particle (i.e., the centers of the thermal cloud) are located at the positions $(x^0,x^1,\ldots,x^n)=(z,z,\ldots,z)$.
The other part of the Hamiltonian is the elastic energy of the deformation away from this state, $H_{\VOP}^{(\DefoPosi)}$. This is different from the separation made in Eq.~(\ref{EQ:HGS}), in which one has the stationary point energy $\HVOPSP$ and the energy-increase due to Goldstone deformations $H_{\VOP}^{\DefoScalPsi}$.

\section{Phenomenological approach to the elasticity of soft random solids}\label{SEC:Phen}

\subsection{Phenomenological nonlocal elastic free energy}\label{SEC:PhenFE}
As discussed in Section~\ref{SEC:Intr}, in the classical theory of rubber elasticity~\cite{Treloar1975}, rubbery materials are modeled as incompressible networks of entropic Gaussian chains, and the resulting elastic free energy density is given by
\begin{eqnarray}
	f = \frac{\SheaModu}{2} \, \textrm{Tr} \,\, \DefoGrad^{\textrm{T}} \DefoGrad
\end{eqnarray}
for spatially uniform deformations $r\to \DefoGrad \cdot r$. Incompressibility is incorporated via the constraint $\textrm{det} \DefoGrad = 1$. For the shear modulus $\SheaModu$, the classical theory gives the result $n_c \, \BoltCons \, T$, where $n_c$ is the density of effective chains in the network.

The phenomenological model that we now discuss is in the same spirit as the classical theory of rubber elasticity. However, to account for the heterogeneity of the medium we need to introduce the additional feature of quenched randomness into the model, and thus the entropic Gaussian chains are allowed to be of heterogeneous length and density. Furthermore, the classical theory is a local elasticity theory, which is valid at length scales that are much longer than the effective chain length of the polymers. By contrast, our phenomenological model is a nonlocal theory, which explicitly takes into account the finite length of polymer chains, as well as their variations.

Inspired by the form of the energy of Goldstone fluctuations determined from the RLPM in Section~\ref{SEC:EnerGoldSton}, we choose the following elastic free energy $\FreeEnerPhen_{\NonLocaKern}$, associated with a deformation of the soft random solid state that maps the mass point at $z$ to the new location $\DefoPosi(z)$:
\begin{eqnarray}\label{EQ:phenom_model}
	\FreeEnerPhen_{\NonLocaKern} \!\!&=& \! \frac{1}{2} \!\int\! dz_1\, dz_2\, \NonLocaKern (z_1,z_2)
	\big(\vert \DefoPosi(z_1)-\DefoPosi(z_2) \vert^2 \! -\vert z_1-z_2 \vert^2\big) \nonumber\\
	\noalign{\medskip}
	&& +\frac{\BulkModuZero}{2}\int dz
	\Big\lbrace\textrm{det}\Big(\frac{\partial\DefoPosi_i(z)}{\partial z_j}\Big)-1 \Big\rbrace^2 ,
\end{eqnarray}
where $\NonLocaKern (z_1,z_2)$ is a \emph{nonlocal harmonic attraction} that serves to pull the two \lq\lq mass points\rq\rq\ (i.e., coarse-grained volume-elements) at $z_1$ and $z_2$ towards one another. The kernel $\NonLocaKern (z_1,z_2)$ originates in the entropy of the molecular chains of the heterogeneous network, and we model it as \lq\lq zero rest-length\rq\rq~springs having random spring coefficient. Notice that $\NonLocaKern (z_1,z_2)$ is a \emph{coarse grained} consequence of many molecular chains and, more importantly, is an \emph{entropic} effect and does not depend on the choice of precise form of microscopic attractive interactions.

We take $\NonLocaKern (z_1,z_2)$ to be a quenched random function of the two positions, $z_1$ and $z_2$, symmetric under $z_1 \leftrightarrow z_2$. We assume that the disorder average of $\NonLocaKern (z_1,z_2)$ is $\NonLocaKernZero(z_{1}-z_{2})\equiv [\NonLocaKern(z_{1},z_{2})]$, i.e., is translationally invariant. Furthermore, we define the fluctuation part of $\NonLocaKern (z_1,z_2)$ to be $\NonLocaKernOne (z_1,z_2)\equiv \NonLocaKern (z_1,z_2) - \NonLocaKernZero(z_{1}-z_{2})$. In the following analysis, we assume that $\NonLocaKernOne \ll \NonLocaKernZero$ in order to make a necessary perturbative expansion.

In the second term in Eq.~(\ref{EQ:phenom_model}), the determinant of the deformation gradient tensor $\DefoGrad_{ij} (z)$[$\equiv \partial \DefoPosi_i/\partial z_j$] captures the change of the volume and, correspondingly, the parameter $\BulkModuZero$, which we take to be large, heavily penalizes density variations. This large $\BulkModuZero$ results from a competition between (i) repulsions (either direct or mediated via a solvent, e.g., excluded-volume), and (ii)~intermolecular attractions and external pressure.

In discussion of elasticity that follows, we exploit the notions of a \lq\lq reference space\rq\rq\ and a \lq\lq target space\rq\rq~for any deformation $\DefoPosi(z)$. The reference space, labeled by the $\Dime$-dimensional vector $z$, is the space \emph{before} the deformation, whereas the target space, labeled by the $\Dime$-dimensional vector $\DefoPosi(z)$, is the space \emph{after} the deformation.

\subsection{Disorder average of the phenomenological model via the replica method}
To make a comparison with the RLPM, and thus to obtain information about the statistics of the nonlocal kernel $\NonLocaKern$ that characterizes the disorder present in the phenomenological model, we shall use the replica method to average the elastic free energy~(\ref{EQ:phenom_model}) over the quenched disorder, whose statistics will be specified below.

We follow a recipe similar to the one used in Section~\ref{SEC:RLPMReplica} [see Eq.~(\ref{EQ:HFReplica})]. The elastic free energy,
Eq.~(\ref{EQ:phenom_model}), contains the random ingredient $\NonLocaKern$. As with the RLPM,
the physical quantity to be disorder-averaged is the free energy at a given pressure, but now with the deformations $\DefoPosi(z)$ as thermally fluctuating field. Therefore, we need to take $\FreeEnerPhen$, defined in Eq.~(\ref{EQ:phenom_model}), as the effective Hamiltonian, because it is the elastic energy for a given deformation field specified by $\DefoPosi(z)$, and then calculate the free energy at a given temperature, via the partition function
\begin{eqnarray}
	Z_{\NonLocaKern}
	=\int \mathcal{D} \DefoPosi
		e^{-\FreeEnerPhen_{\NonLocaKern}(\DefoPosi(z))/\BoltCons T} ,
\end{eqnarray}
with $\FreeEnerPhen$ depending on the quenched randomness through its kernel $\NonLocaKern$. The Gibbs free energy is related to this partition function via $\GibbFreeEner = -\BoltCons T \ln Z$, and $\GibbFreeEner$ is the quantity that should be averaged over the quenched disorder. Note that it is the Gibbs free energy, instead of the Helmholtz free energy, that is related to this partition function $Z$, because in the elastic energy $\FreeEnerPhen_{\NonLocaKern}$ one has a fixed pressure, which is accounted for by the $\BulkModuZero$ term, the volume being allowed to fluctuated.
The disorder average of the Gibbs free energy can be computed using the replica technique:
\begin{eqnarray}
	\lda \GibbFreeEner \rda
	&=& -\BoltCons T \int \mathcal{D} \NonLocaKern\, \ProbG
			\ln Z_{\NonLocaKern} \nonumber\\
	&=& -\BoltCons T\int \mathcal{D} \NonLocaKern\, \ProbG  \lim_{n\to 0}
			\frac{Z_{\NonLocaKern}^n-1}{n} \nonumber\\
	&=& -\BoltCons T \lim_{n\to 0} \frac{\partial}{\partial n}\Big\vert_{n \to 0}
			\lda Z_{\NonLocaKern}^n \rda ,
\end{eqnarray}
where we have used $\lda\ldots\rda\equiv
\int \mathcal{D} \NonLocaKern \, \ProbG \ldots$ once again to denote a disorder average, but this time over the values of the quenched random kernel $\NonLocaKern$, weighted by an as-yet unknown distribution $\ProbG$.
In the present setting, we do not have a \lq\lq zeroth replica\rlap,\rq\rq\ as such a replica arises from the Deam-Edwards distribution of the links, and this is not the type of quenched disorder that we have in mind. Rather, in the present setting we regard the distribution of disorder $\ProbG$ as a physical quantity that is unknown but is to be determined through a comparison with the analysis of the Goldstone fluctuations of the RLPM. The replica partition function is then given by
\begin{eqnarray}\label{EG:PhenZn}
	\mathbb{Z}_{n}&\equiv& \lda Z_{\NonLocaKern}^n \rda
	= \int \mathcal{D} \NonLocaKern \,\ProbG\,  Z_{\NonLocaKern}^n \nonumber\\
	&=& \int \mathcal{D} \NonLocaKern \,\ProbG
			\int \ReplProdOne \mathcal{D} \DefoPosi\REPa
				 e^{-\ReplSumOne\FreeEnerPhen_{\NonLocaKern}(\DefoPosi\REPa(z))/\BoltCons T}
				 \nonumber\\
	&=& \int \ReplProdOne \mathcal{D} \DefoPosi\REPa\,
			e^{-\FreeEnerPhen_n/\BoltCons T} ,
\end{eqnarray}
in which we functionally integrate over the configurations of the $n$-fold replicated displacement fields $\DefoPosi\REPa$.  We have also introduced the effective pure Hamiltonian $\FreeEnerPhen_n$ that governs the replicated deformation fields:
\begin{eqnarray}\label{EQ:PhenReplPartFunc}
	\FreeEnerPhen_n &\equiv& -\BoltCons T \ln \big\lda
			e^{-\ReplSumOne\FreeEnerPhen_{\NonLocaKern}(\DefoPosi\REPa(z))/\BoltCons T}
		\big\rda .
\end{eqnarray}
The exponential and the logarithm in Eq.~(\ref{EQ:PhenReplPartFunc}) can jointly be expanded in terms of cumulants, and thus we arrive at the form
\begin{widetext}
\begin{eqnarray}
	\FreeEnerPhen_n &=& -\BoltCons T \Big\lbrace
			-\Big\lda \ReplSumOne\FreeEnerPhen_{\NonLocaKern}(\DefoPosi\REPa(z))/\BoltCons T \Big\rda_c
			+\frac{1}{2} \Big\lda
					\sum_{\alpha,\beta=1}^{n}
					\FreeEnerPhen_{\NonLocaKern}(\DefoPosi\REPa(z))
					\FreeEnerPhen_{\NonLocaKern}(\DefoPosi\REPb(z))/\BoltCons T
				\Big\rda_c
			- \cdots
		\Big\rbrace ,
\end{eqnarray}
where $\lda \ldots \rda_c$ are connected statistical moments (i.e., cumulants) associated with the probability distribution of the disorder $\ProbG$, and the omitted terms are $O[(\FreeEnerPhen/\BoltCons T)^3]$.
The elastic energy $\FreeEnerPhen_{\NonLocaKern}$ for a given realization of disorder $\NonLocaKern$ and a given deformation field $\DefoPosi(z)$ is given in Eq.~(\ref{EQ:phenom_model}); inserting this form for $\FreeEnerPhen_{\NonLocaKern}$ we obtain
\begin{eqnarray}\label{EQ:FEPn}
	\FreeEnerPhen_n &=& \frac{\BulkModuZero}{2} \int _{z} \ReplSumOne
			\big(\vert\partial \DefoPosi\REPa \vert-1\big)^2
		+\frac{1}{2}\int_{z_1,z_2} \lda \NonLocaKern(z_1,z_2) \rda_c \DefoScalPsi (z_1,z_2)
		\nonumber\\
		\noalign{\medskip}
	&&	-\frac{1}{8\BoltCons T} \int_{z_1,z_2,z_3,z_4}
			\lda \NonLocaKern(z_1,z_2) \NonLocaKern(z_3,z_4) \rda_c
			\DefoScalPsi (z_1,z_2)\DefoScalPsi (z_3,z_4)
		+ O(\DefoScalPsi^3) ,
\end{eqnarray}
\end{widetext}
and we remind the reader of the definition of $\DefoScalPsi$, first given in Section~\ref{SEC:EnerGoldSton}:
\begin{align}
	 \DefoScalPsi\equiv\big(\hat{\DefoPosi}(z_1)-\hat{\DefoPosi}(z_2)\big)^2-(1+n)(z_1-z_2)^2 .
\end{align}
Up to quadratic order in $\DefoScalPsi$, the effective pure Hamiltonian $\FreeEnerPhen_n$ of Eq.~(\ref{EQ:FEPn}) has \emph{precisely the same form} as the energy of the Goldstone fluctuations~(\ref{EQ:GibbEnerGoldSton}), derived microscopically from the RLPM. Thus, the RLPM actually provides a \emph{derivation} of the phenomenological model we proposed in Section~\ref{SEC:PhenFE}, and \emph{justifies}, from a microscopic perspective, the phenomenological elastic free energy~(\ref{EQ:phenom_model}) with its quenched randomness. Therefore, the probability distribution $\ProbG$ of the quenched randomness in Eq.~(\ref{EQ:phenom_model}) is contained in the RLPM. By comparing the two schemes, i.e., Eqs.~(\ref{EQ:GibbEnerGoldSton}) and (\ref{EQ:FEPn}), we arrive at a statistical description of the quenched random kernel $\NonLocaKern$ in the phenomenological model~(\ref{EQ:phenom_model}), as we shall now show.

\subsection{Comparing the Gibbs free energies of the RLPM and the phenomenological model}
The RLPM is a semi-microscopic random network model, our analysis of which led to the disorder-averaged Gibbs free energy~(\ref{EQ:HFEDisoAverDeri},\ref{EQ:LegeTran}):
\begin{eqnarray}\label{EQ:GRLPM}
	\lda \GibbFreeEner \rda = -\BoltCons T \lim _{n \to 0}
		\frac{\partial}{\partial n} \ln Z_{1+n} + \Pres\Volu ,
\end{eqnarray}
with
\begin{eqnarray}\label{EQ:ZRLPMVOP}
	Z_{1+n}=\int \mathcal{D} \VOP \, e^{-H_{\VOP}/\BoltCons T} .
\end{eqnarray}
{\colorred{
The functional integration here is over all possible configurations of the order-parameter field $\VOP$.  By contrast, in the phenomenological model the replicated partition function $\mathbb{Z}_{n}$ involves a functional integration over the $n$-fold replicated deformation field $\DefoPosi\REPa$, as in Eq.~(\ref{EG:PhenZn}).  To obtain the equivalence between the RLPM and the phenomenological model it is useful to proceed in two steps.  First, we note that in the random solid state the Boltzmann weight in Eq.~(\ref{EQ:ZRLPMVOP}) [i.e., $\exp\left(-H_{\VOP}/\BoltCons T\right)$]
is heavily concentrated near the stationary point $\VOP_{SP}$ and the Goldstone-deformed states $\VOPGS$, and decreases steeply for other sectors of fluctuations.  This suggests that we parametrize the fluctuating field $\VOP$ in terms of an \lq\lq amplitude\rlap,\rq\rq\ which we take to be $\VOP_{SP}$, plus \lq\lq radial\rq\rq\ fluctuations around it, together with an appropriate set of generalized \lq\lq angular\rq\rq\ {\colorgreen Goldstone} variables, which are the $n$ independent $\Dime$-vector fields in $\hat{\DefoPosi}$, as, e.g., in Eq.~(\ref{EQ:VOPGSM}).  We then make an exact change of functional integration variables, from $\VOP$ to these radial and angular variables, and this introduces a corresponding Jacobian factor.  The second step is to recognize that the radial fluctuations are \lq\lq massive\rq\rq\ (i.e., they have restoring forces, in contrast with the angular fluctuations, which are \lq\lq massless\rq\rq).  Thus, if we were to integrate these radial fluctuations we would obtain small corrections to the terms of the remaining, effective, angular-variable theory.  We therefore elect to treat the radial variables at the classical level, which amounts to neglecting the radial fluctuations.  Under this condition, the aforementioned Jacobian factor does not depend on the angular variables, and therefore it contributes only a constant multiplicative factor to the functional integral, which can be safely omitted.  This procedure enables us to arrive at the following approximate form for the replica partition function of the RLPM, Eq.~(\ref{EQ:ZRLPMVOP}):
}}
\begin{eqnarray}
	Z_{1+n}\approx e^{-\HVOPSP/\BoltCons T}
		\int \ReplProdOne \mathcal{D} \DefoPosi\REPa\, e^{-H_{\VOP}^{\DefoScalPsi}/\BoltCons T},
\end{eqnarray}
with $\HVOPSP$ and $H_{\VOP}^{\DefoScalPsi}$ given in Eqs.~(\ref{EQ:HSP}) and (\ref{EQ:GibbEnerGoldSton}).

In our phenomenological model, introduced in Section~\ref{SEC:PhenFE}, the disorder-averaged Gibbs free energy is given by
\begin{eqnarray}\label{EQ:GPhen}
	\lda \GibbFreeEner \rda =
			-\BoltCons T \lim_{n\to 0} \frac{\partial}{\partial n}\Big\vert_{n \to 0}
			\mathbb{Z}_{n}\, ,
\end{eqnarray}
with
\begin{eqnarray}
	\mathbb{Z}_{n}=\int \ReplProdOne \mathcal{D} \DefoPosi\REPa\, e^{-\FreeEnerPhen_n/\BoltCons T},
\end{eqnarray}
where $\FreeEnerPhen_n$ is given in Eq.~(\ref{EQ:FEPn}). 

The Gibbs free energies---for the RLPM and for the phenomenological model---are supposed to be equal, up to an additive constant, because they both capture the Gibbs free energy of a soft random solid system having elastic deformations. It is this equality that we shall now exploit to characterize, via the RLPM, the distribution of quenched disorder $\ProbG$ in the phenomenological model. Actually, we can directly identify the Hamiltonian $\FreeEnerPhen_n$ with $H_{\VOP}^{(\DefoPosi)}$, because the functional integration over the replicated deformation field $\DefoPosi\REPa$ is common to both the RLPM and the phenomenological model, in the sense that the deformation $z\to\DefoPosi\REPa(z)$, in both the RLPM and the phenomenological model, takes the \emph{state right after linking} as the reference state. Therefore, we have the relation
\begin{eqnarray}\label{EQ:Comp}
	\FreeEnerPhen_n = H_{\VOP}^{(\DefoPosi)} .
\end{eqnarray}
Notice that, here, the RLPM Hamiltonian is $H_{\VOP}^{(\DefoPosi)}$, not $H_{\VOP}^{\DefoScalPsi}$, because it is $H_{\VOP}^{(\DefoPosi)}$ that is the energy measured from the \emph{state right after linking}, which matches the definition of reference state in the phenomenological theory, whereas $H_{\VOP}^{\DefoScalPsi}$ is the energy measured from the \emph{stationary point}, which differs from the \emph{state right after linking} by the energy associated with the contraction $\ContEner(\Contraction)$, given in Eq.~(\ref{EQ:CE}).

By the comparison stated Eq.~(\ref{EQ:Comp}), we arrive at the following determination of the quenched-disorder characteristics of the phenomenological model (LHS) in terms of the elastic properties of the RLPM (RHS):
\begin{subequations}\label{EQ:CompDA}
\begin{eqnarray}
	\lda \NonLocaKern(z_1,z_2) \rda_c &=& \KernOne(z_1,z_2),
	\label{EQ:KOne}\\
	\lda \NonLocaKern(z_1,z_2) \, \NonLocaKern(z_3,z_4) \rda_c &=& \KernTwo(z_1,z_2,z_3,z_4),
	\label{EQ:KTwo}\\
	\BulkModuZero &=& \ExclVolu \PartDens^2 , \label{EQ:BMZ}
\end{eqnarray}
\end{subequations}
where $\PartDens \equiv \PartNumb / \Volu_0$ is the number-density of the particles in the preparation state.
The functions $\KernOne(z_1,z_2)$ and $\KernTwo(z_1,z_2,z_3,z_4)$ are defined in Appendix~\ref{APP:HGS}, and have been discussed in Section~\ref{SEC:EnerGoldSton}

\section{Phenomenological model at fixed disorder: Relaxation, excitation, and deformation}\label{SEC:relaxation}
\subsection{Relaxation to the stable state at fixed disorder}\label{SEC:subrelaxation}
\begin{figure}[htbp]
	\centering
		\includegraphics[width=.45\textwidth]{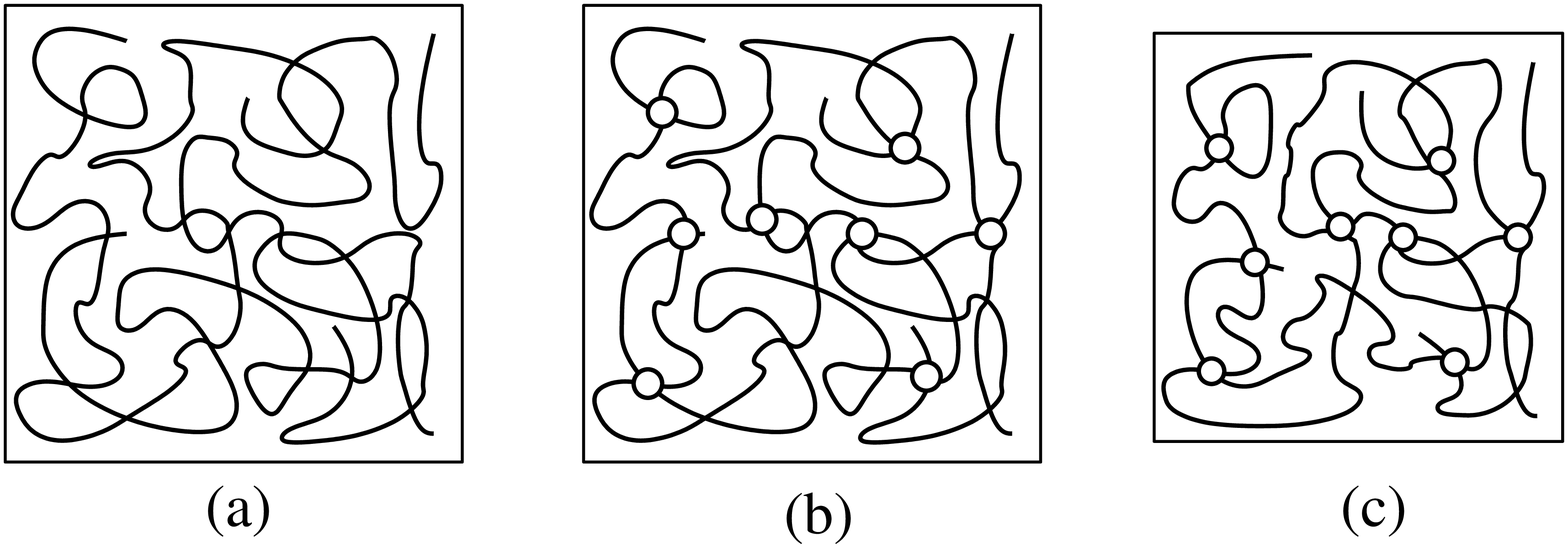}
	\caption[Schematic plot of the relaxation process under a fixed pressure.]{Schematic plot of the relaxation process under a fixed pressure. (a) The liquid state with no linking. (b) The \emph{state right after linking}. cross-links are added to the system, and an infinite cluster is formed. This state is not stable, because of the inward pressure and local stresses. (c) The \emph{relaxed state}. The system undergoes a uniform contraction and random local deformations that release the unbalanced stress introduced by cross-linking.}
	\label{FIG:Rela}
\end{figure}

The free energy $\FreeEnerPhen$ provides a natural description of the heterogeneous elasticity of soft random solids. However, its stable state is not $\DefoPosi(z)=z$ (i.e., the state $\DefoPosi(z)=z$ does not satisfy the stationarity condition $\delta \FreeEnerPhen/\delta \DefoPosi(z)=0$). There are two reasons for this instability.
First, the attraction $\NonLocaKern$ causes a small, spatially uniform, contraction [the fractional volume change being $O(1/\BulkModuZero)$].
Second, the randomness of $\NonLocaKern$ additionally destabilizes this contracted state, causing the adoption of a randomly deformed stable state.  We denote this relaxation as
\begin{eqnarray}\label{EQ:Defiv}
	z\to\tz\equiv\Contraction z + \RelaRand(z) ,
\end{eqnarray}
in which $\Contraction$ describes the uniform contraction and $\RelaRand(z)$ describes the random local deformation.
This relaxation process can be understood in the setting of the preparation of a sample of rubber via a hypothetic instantaneous cross-linking: cross-linking not only drives the liquid-to-random-solid transition but it also generates a uniform inward pressure, as well as introducing random stresses, as shown in Fig.~\ref{FIG:Rela}
As a result, immediately after cross-linking the state is not stable, but relaxes to a new stable state, determined by the particular realization of randomness created by the cross-linking. In the following discussion, we shall use the following nomenclature:
\begin{subequations}
\begin{eqnarray}
	\DefoPosi(z)\!\!&=&\!\!z  \!\iff\! \textrm{the state right after linking,} \\
	\DefoPosi(z)\!\!&=&\!\!\tz\equiv\Contraction z + \RelaRand(z) \!\iff\! \textrm{the relaxed state.}\,
\end{eqnarray}
\end{subequations}
The state right after linking, here, is the same as the one just defined, following Eq.~(\ref{EQ:HSepa}), which has energy $H_{\VOP}^{(0)}$ in the RLPM, because they both describe the state that has undergone no deformation since being linked.

By writing the relaxation as $z\to\tz \,[\,\equiv\Contraction z + \RelaRand(z)]$ we are making the approximations that the contraction $\Contraction$ is homogeneous and that the random deformations $\RelaRand(z)$ are pure shear, which means that any randomness in the \emph{bulk} deformation is ignored. This can be understood by looking at the orders of magnitude of the deformations. The uniform contraction is of order $O(\NonLocaKernZero/\BulkModuZero)$, and the random local shear deformations are of order $O(\NonLocaKernOne/\NonLocaKernZero)$. The random local bulk deformations is, however, of order $O(\NonLocaKernOne/\BulkModuZero)$, and is thus much smaller than the other two deformations, given the assumptions that (i) the fluctuations of the shear modulus are much smaller than the mean value (the shear modulus corresponds to $\NonLocaKern$, as we shall see later), and (ii) the shear modulus is much smaller than the bulk modulus.

With these assumptions, we can insert the form $\DefoPosi(z)=\tz(z)$ into the stationarity condition, and solve for the relaxed state, which is characterized by $\Contraction$ and $\RelaRand(z)$.  As we have just discussed, for the contraction, only the homogeneous part is admitted, so the variational equation for $\Contraction$ assumes $\NonLocaKernOne=0$ and thus $\RelaRand(z)=0$ and, stationarity requires
\begin{eqnarray}\label{EQ:StabGlob}
	\frac{\partial \FreeEnerPhen}{\partial \Contraction} = 0 .
\end{eqnarray}
Thus, for the present model, Eq.~(\ref{EQ:phenom_model}), we have
\begin{eqnarray}\label{EQ:StabCont}
	\!\!0\!\!&=&\!\!\frac{\partial }{\partial \Contraction}
		\Big\lbrace
			\frac{1}{2} \int dz_1 \, dz_2 \, \NonLocaKernZero (z_1,z_2) \,
				(\Contraction^2-1)\, \vert z_1-z_2 \vert ^2 \nonumber\\
	&&\quad\quad	+\frac{\BulkModuZero}{2}\int dz \big(\Contraction^{\Dime}-1 \big)^2
		\Big\rbrace\nonumber\\ \noalign{\medskip}
	&=& \Volu_0 \big(
			\Dime \, \MeanSheaModu \, \Contraction
			+ \BulkModuZero \, \big(\Contraction^{\Dime}-1 \big) \Dime \Contraction^{\Dime-1}
		\big) .
\end{eqnarray}
By solving this equation to leading order in $\MeanSheaModu/\BulkModuZero$, we obtain
\begin{eqnarray}\label{EQ:SoluCont}
	\Contraction \approx 1-({\MeanSheaModu}/{d\BulkModuZero}) ,
\end{eqnarray}
where
\begin{equation}\label{EQ:MSheaModu}
\MeanSheaModu\equiv
\frac{1}{\Dime}
\int dz_2\,(z_1-z_2)^2\,\NonLocaKernZero(z_1-z_2).
\end{equation}
As we shall see below, $\MeanSheaModu$ is actually the mean shear modulus.

The stationarity condition for the random local deformation $\RelaRand(z)$ reads
\begin{eqnarray}\label{EQ:StabLoca}
	\frac{\delta \FreeEnerPhen}{\delta \RelaRand_a(z)} = 0 ,
\end{eqnarray}
and for the present model, Eq.~(\ref{EQ:phenom_model}), this condition becomes
\begin{eqnarray}\label{EQ:RelaV}
	&& 2(\Contraction z_a + \RelaRand_a(z))\int dz_2 \NonLocaKern(z,z_2) \nonumber\\
	&&	-2\!\int dz_2\NonLocaKern(z,z_2)\big(\Contraction z_{2,a} + \RelaRand_a(z_2)\big) \nonumber\\
	&&		-\BulkModuZeroP \partial_{a} \big(\partial_i \RelaRand_i(z)\big) =0 .
\end{eqnarray}
Here, the last term, $\BulkModuZeroP \partial_{a} \big(\partial_i \RelaRand_i(z)\big)$, is associated with density variations, and arises from the variation of the second term in the elastic free energy~(\ref{EQ:phenom_model}), which is $\frac{\BulkModuZero}{2}\int dz
	\big\lbrack\textrm{det}\big(\partial\DefoPosi_i(z)/\partial z_j\big)-1 \big\rbrack^2$. In the following discussion we shall call this the \lq\lq bulk term\rq\rq\ in the elastic free energy, and we have made the definition $\BulkModuZeroP\equiv\Contraction^{2d-2}\BulkModuZero$; see Appendix~\ref{APP:Relaxation} for the expansion.

The stationarity equation~(\ref{EQ:RelaV}) for $\RelaRand(z)$ can be solved perturbatively, by assuming that $\NonLocaKernZero$ is of zeroth order and that $\NonLocaKernOne$ and $\RelaRand(z)$ are of first order;
see Appendix~\ref{APP:Relaxation} for the explicit calculation. In momentum space, the result is
\begin{eqnarray}\label{EQ:SoluV}
	\vec{\RelaRand}_{p}
	&=& \frac{\PPerpT \cdot \vec{\RandForc}_{p}}
				{2\DiffG_p}
			+\frac{\PLongT \cdot \vec{\RandForc}_{p}}
				{\BulkModuZeroP+ 2\DiffG_p} ,
\end{eqnarray}
where $p$ is a $\Dime$-dimensional momentum vector. The notation $\vec{\RandForc}_{p}$ and $\DiffG_p$ are defined as
\begin{eqnarray}
	\RandForc_{a,p}&\equiv& -2\Contraction \Big(
		i\frac{\partial}{\partial p_{a}} \NonLocaKernOne_{p,0}
		-i\frac{\partial}{\partial p'_{a}}\Big\vert_{p'=0}  \NonLocaKernOne_{p,p'}
	\Big), \nonumber\\
	\DiffG_p &\equiv& \NonLocaKernZero_{0}-\NonLocaKernZero_{p} .
\end{eqnarray}
Notice that $\RandForc_{a,p}$ is actually the random force in the state that is contracted but not yet equilibrated for randomness.
The definitions of the projection operators $\PLongT$ and $\PPerpT$ are respectively,
\begin{eqnarray}\label{EQ:DefiProjText}
	\PLong_{ij} &\equiv& p_i p_i /p^2 , \nonumber\\
	\PPerp_{ij} &\equiv& \delta_{ij}-p_i p_i /p^2 .
\end{eqnarray}
We use bold letters to denote $\Dime$-dimensional rank-$2$ tensors, and add an overhead arrow [such as $\vec{\RelaRand}$\,] to denote vectors, when needed.

{{\colorred
In the solution~(\ref{EQ:SoluV}), the second term is much smaller than the first term, due to the large bulk modulus $\BulkModuZeroP$. In the incompressible limit (i.e., $\BulkModuZero \to \infty$), we have
\begin{eqnarray}
	\vec{\RelaRand}_{p}=\frac{\PPerpT \cdot \vec{\RandForc}_{p}}
				{2\DiffG_p} ,
\end{eqnarray}
which is a purely transverse field, meaning that it satisfies $p_i\,\RelaRand_{i,p}=0$ or, equivalently, that $\partial_i\,\RelaRand_{i}(x)=0$, which is the only type of deformation that can occur in an incompressible medium.}}


\subsection{Excitation around the relaxed state at fixed disorder}\label{SEC:EFERS}
In order to obtain a description of the elasticity of the relaxed state $z\to\tz$, which is a stable state and thus relevant for experimental observations, we re-expand the phenomenological elastic free energy~(\ref{EQ:phenom_model}) around the relaxed state. This amounts to taking the relaxed state $\tz(z)\,[\,\equiv\Contraction z+\RelaRand(z)]$ as the new reference state, and deriving the elastic free energy for deformations
(i.e., excitations) relative to this state.

To do this, we study the free energy for the following elastic deformation:
\begin{eqnarray}
	z\to \DefoPosi(z)=\tz(z) + \deformation(z) ,
\end{eqnarray}
where $\deformation(z)$ is a deformation away from the relaxed reference state. It will be convenient to make the following change of the independent variables:
\begin{eqnarray}
	z\to\tz(z)\equiv\Contraction z+\RelaRand(z) ,
\end{eqnarray}
which has the Jacobian determinant
\begin{eqnarray}\label{EQ:zJaco}
	\Jaco (z) \equiv \Big\vert \frac{\partial \tz_i}{\partial z_j}\Big\vert
	\approx \Contraction^{\Dime} \big(1+\Contraction^{-1}\partial_j\RelaRand_j(z)\big).
\end{eqnarray}
With this change of variables, the phenomenological elastic free energy is expressed as
\begin{widetext}
\begin{eqnarray}\label{EQ:REExpaEner}
	\FreeEnerPhen &=& \FreeEnerPhen_0 +
		\frac{1}{2} \int d\tz_1\, d\tz_2\, \Jaco(z_1)^{-1}\,\Jaco(z_2)^{-1}
			\tNonLocaKern (\tz_1,\tz_2)\Big(
				\big\vert \tz_1+\tu(\tz_1)-\tz_2-\tu(\tz_2)\big\vert^2
				- \big\vert \tz_1-\tz_2\big\vert^2
			\Big) \nonumber\\
			\noalign{\medskip}
	&&+ \frac{\BulkModuZero}{2} \int d\tz\, \Jaco(z)^{-1}
			\Big\lbrace
				\Jaco(z) \textrm{det} \Big( \frac{\partial \tz_i+\tu_i(\tz)}{\partial \tz_j}
					\Big)-1
			\Big\rbrace^2 ,
\end{eqnarray}
\end{widetext}
where we have made the definitions
\begin{subequations}
\begin{eqnarray}\label{EQ:tGtu}
	\tNonLocaKern(\tz_1,\tz_2)&\equiv&\NonLocaKern(z(\tz_1),z(\tz_2)), \\
	\tu(\tz)&\equiv&\deformation(z(\tz)), \\
	\tDefoPosi(\tz)&\equiv& \tz+\tu(\tz),
\end{eqnarray}
\end{subequations}
with $z(\tz)$ denoting the mapping of the mass point $\tz$ in relaxed state back to the mass point $z$ in the state right after linking, i.e., the inverse of the $\tz(z)$ mapping. The change of the free energy due to choosing a different reference state is defined as
\begin{eqnarray}
	\!\!\FreeEnerPhen_0\!\equiv\frac{1}{2} \!\int\!\! dz_1\, dz_2 \,
	\NonLocaKern (z_1,z_2)\big(\vert \tz_1-\tz_2\vert^2 \!\!-\! \vert z_1-z_2\vert^2 \big) ,
\end{eqnarray}
which is a constant for any given realization of the randomness.

In order to obtain a direct description of the elastic energy relative to the relaxed state, we expand the quenched random nonlocal kernel in the relaxed state, $\tNonLocaKern(\tz_1,\tz_2)$, defined in Eq.~(\ref{EQ:tGtu}) as  \begin{eqnarray}\label{EQ:tNonLocaKernText}
	\!\!\!\! \tNonLocaKern_{\tp_1,\tp_2} \!\!
	&\approx& \!\! \NonLocaKernZero_{\tp_1,\tp_2}+\NonLocaKernOne_{\tp_1,\tp_2}
	\nonumber\\
	&&\!\!		-i \big(
				\tp_1\cdot\!\vec{\RelaRand}_{(\tp_1+\tp_2)} \NonLocaKernZero_{\tp_2}
				\!+\!\tp_2\cdot\!\vec{\RelaRand}_{(\tp_1+\tp_2)} \NonLocaKernZero_{\tp_1}
			\big),\,
\end{eqnarray}
where $\vec{\RelaRand}$ is the random local deformation field defined in Eq.~(\ref{EQ:Defiv}).  We then expand the elastic free energy (\ref{EQ:REExpaEner}) as a power series in the small deformation $\tu(\tz)$ away from the relaxed state $\tz$. The computation of this expansion is given in Appendix~\ref{APP:ReExpandFreeEner}.

As we shall show in Section~\ref{SEC:Heterogeneity}, the statistics of the quenched randomness present in this phenomenological theory can be determined via a comparison with the RLPM. Through this comparison, we find that, the lengthscale of the nonlocal kernel $\NonLocaKern$ is actually the typical localization length, which is small compared to the lengthscales on which our theory of elasticity applies, because the deformations in this theory are associated with Goldstone fluctuations in the RLPM, which feature lengthscales larger than the typical localization length.

Thus, it is reasonable to make a \emph{local expansion} of the elastic energy (\ref{EQ:REExpaEner}) relative to the relaxed state in terms of the strain tensor $\StraTensT$. The resulting form, which we shall call \lq\lq the local form of the elastic energy relative to the relaxed state\rlap,\rq\rq is in the form of Lagrangian elasticity.  As will be seen in Section~\ref{SEC:Corr}, the advantage of this local form of the elastic energy is that one can extract from it \emph{large-distance behavior} of the disorder correlators of the elastic parameters, which turn out to be \emph{universal}.  The calculation for this local expansion is given in Appendix~\ref{APP:ReExpandFreeEner}. The resulting local form of elastic energy is
\begin{eqnarray}\label{EQ:FELagr}
	\FreeEnerPhen
	&=& \int d\tz \big\lbrace
				\textrm{Tr}(\StreT(\tz)\cdot\tStraTensT(\tz)) \nonumber\\
	&&		+ \SheaModu(\tz) \textrm{Tr} \tStraTensT (\tz)^2
				+ \frac{\BulkModu(\tz)}{2} (\textrm{Tr} \tStraTensT (\tz))^2
			\big\rbrace ,
\end{eqnarray}
where the strain tensor relative to the relaxed state is defined as
\begin{eqnarray}
	\tStraTens_{ij} (\tz) \equiv \frac{1}{2}\Big(\frac{\partial \tu_j}{\partial \tz_i}+\frac{\partial \tu_i}{\partial \tz_j} +\frac{\partial \tu_l}{\partial \tz_i}\frac{\partial \tu_l}{\partial \tz_j}\Big) ,
\end{eqnarray}
and the heterogeneous elastic parameters, viz., the residual stress $\StreT$, the shear modulus $\SheaModu$, and the bulk modulus $\BulkModu$, are given in momentum space, by
\begin{subequations}
\label{EQ:RelaStre}
\begin{eqnarray}
	\Stre_{ij,\tp}
	&=& -\frac{\partial^2}{\partial \tq_i \partial \tq_j}
		\Big\vert_{q=0} \NonLocaKernOne_{\tp-\tq,\tp}
			+ i \delta_{ij} \frac{i\tp\cdot\vec{\RandForc}_{\tp}}{\vert \tp \vert ^2}\label{EQ:Stress}\\
	&& \quad -\frac{\RandForc_{a,\tp}}{\vert \tp \vert ^2}\big(
				\tp_{i} \PPerp_{ja,\tp} + \tp_{j} \PPerp_{ia,\tp}
			\big) , \nonumber\\
	\SheaModu_{\tp}
	&=& \MeanSheaModu \Volu_0 \delta_{\tp} -
			\frac{i\tp\cdot\vec{\RandForc}_{\tp}}{\vert \tp \vert ^2} , \label{EQ:Shear}\\
	\BulkModu_{\tp}
	&=& \BulkModuZero \Volu_0 \delta_{\tp} + 2\Big\lbrace
										\frac{i\tp\cdot\vec{\RandForc}_{\tp}}{\vert \tp \vert ^2}
										-\MeanSheaModu \Volu_0 \delta_{\tp}
								\Big\rbrace . \label{EQ:Bulk}
\end{eqnarray}
\label{Eq:ThreeResults}
\end{subequations}
In the expression for $\Stre_{ij,\tp}$ we have kept terms only to leading order in the momentum $\tp$ (see Appendix~\ref{APP:ReExpandFreeEner} for the derivation)~\footnote{To be consistent with the RLPM, we have used finite-volume versions of the Fourier transform and Kronecker delta function in momentum space; we shall take the continuum limit later on, in the final results in Section~\ref{SEC:Local}.  Strictly speaking, the differentiations in Eq.~(\ref{EQ:Stress}) should be understood as the corresponding difference quotients.}. It is worth mentioning that, to leading order in the momentum $\tp$, this residual stress satisfies the stability condition $\tp_i\,\Stre_{ij,\tp}=0$, because the reference state of this elastic free energy, the relaxed state, is a stable state. This will also be shown more directly in the final results in Section~\ref{SEC:Corr}.

\subsection{Nonaffine deformation at fixed disorder}\label{SEC:NAD}
Because of the quenched disorder present in the elastic parameter $\NonLocaKern$ of our phenomenological model, Eq.~(\ref{EQ:phenom_model}), upon the application of external stress, the system will respond by adopting a strain field that is nonaffine. This means that the strain tensor will be spatially inhomogeneous even though the applied stress is homogeneous.
Such nonaffine deformations reflect the quenched randomness of the elasticity, and can be derived for a given realization of the disorder and a given macroscopic deformation by external stress. Because the deformation is the quantity that is directly measurable in experiments, it is useful to derive the relationship between the nonaffine deformation and the quenched randomness in the elastic parameters. Then, by comparing with the RLPM, we shall obtain a statistical description of the nonaffine deformations, as we shall discuss in Section~\ref{SEC:SNAD}.

\begin{figure}[htbp]
	\centering
		\includegraphics[width=.45\textwidth]{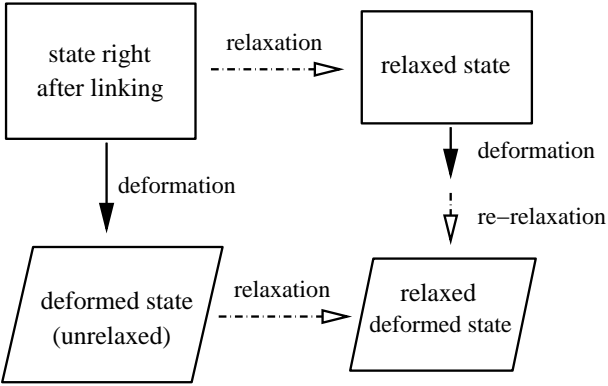}
	\caption[Illustration of the relaxed deformed state.]{Illustration of the \emph{relaxed deformed state}. Theoretically, the relaxed deformed state can be reached via two routes: in the first, the upper route in this plot, one applies the deformation $\DefoGrad$ on the already-relaxed state, and lets the system re-relax while keeping the external deformation $\DefoGrad$ to the relaxed deformed state; the second route, the lower route in this plot, one deforms the system before relaxation is allowed, and then lets the system relax while maintaining the deformation $\DefoGrad$. These two routes reach the same final state, the relaxed deformed state, which characterizes the \emph{nonaffine deformations} that the system undergoes under external deformation. For convenience of calculation, we use the lower route to determine the nonaffine deformations.}
	\label{FIG:RelaDefo}
\end{figure}

To study nonaffine deformations, it is convenient to take the \lq\lq state right after linking\rq\rq~[i.e., the state $\DefoPosi(z)=z$] as the reference state, and to re-derive the relaxation in the presence of \emph{a given deformation} $\DefoGrad$. This is equivalent to applying the deformation $\DefoGrad$ to the relaxed state, and then letting the system further relax for this given deformation, as shown in Fig.~\ref{FIG:RelaDefo}.
The relaxed state for this given deformation $\DefoGrad$, which we term the \lq\lq relaxed deformed state\rlap,\rq\rq\ is described by the deformation $z\to\tzL (z) $. We suppose that
\begin{eqnarray}\label{EQ:tzl}
	\tzL (z) = \Contraction \DefoGrad \cdot z + \RelaRandL(z) .
\end{eqnarray}
For simplicity, we assume that the deformation $\DefoGrad$ is pure shear (i.e., $\textrm{det}\,\DefoGrad=1$).

Next, we use the two stationarity conditions, Eqs.~(\ref{EQ:StabGlob}) and (\ref{EQ:StabLoca}) to solve for the relaxed deformed state. The condition~(\ref{EQ:StabCont}) for the homogeneous contraction $\Contraction$ is unchanged, so we still obtain $\Contraction \approx 1-({\MeanSheaModu}/{d\BulkModuZero})$. For the stability condition for the random local deformations $\RelaRandL$ we follow a similar expansion to the one given in Eqs.~(\ref{EQ:DetPPExp}) and (\ref{EQ:BulkTermExp}), arriving at
\begin{eqnarray}
	&&	2(\Contraction \DefoGrad_{ai}z_i+(\RelaRandL)_{a}(z)) \int dz_2\, \NonLocaKern (z,z_2) \nonumber\\
	&&	\quad -2 \int dz_2\, \NonLocaKern (z,z_2)
			\,(\Contraction \DefoGrad_{ai}z_{2,i}+(\RelaRandL)_{a}(z_2)\,) \nonumber\\
	&&	\quad - \BulkModuZeroP \,
			 \DefoGrad^{-1}_{ia}\DefoGrad^{-1}_{jb}\,\partial_i \partial_j (\RelaRandL)_{b}(z) =0 .
\end{eqnarray}
As with the derivation given in Section \ref{SEC:relaxation}, we can solve this equation perturbatively, to leading order in $\NonLocaKernOne$ and $\RelaRandL$; see Appendix~\ref{APP:DefoRela} for details. The result is
\begin{eqnarray}
	(\vec{\RelaRandL})_p = \Bigg\lbrace \frac{\PPerpL}{2 D_p} + \frac{\PLongL}{\BulkModuZeroP \trOne \vert p \vert ^2+2 D_p}
	\Bigg\rbrace \cdot (\vec{\RandForcL})_p \, ,
\end{eqnarray}
where $\PPerpL$ and $\PLongL$, defined in Appendix~\ref{APP:DefoRela}, are the \emph{deformed} versions of the projection operators, and $\trOne$ is also defined in Appendix~\ref{APP:DefoRela}.

In the literature the nonaffine deformations are often characterized by the \lq\lq nonaffine deformation field\rq\rq~$\NADF$, which is defined in momentum space as
\begin{eqnarray}\label{EQ:DefNADF}
	(\NADF) _p &\equiv& \DefoGrad^{-1} \cdot (\tzL)_p - \tz_p \, , \nonumber\\
	&=& \DefoGrad^{-1} \cdot (\RelaRandL)_p - \RelaRand_p ,
\end{eqnarray}
where $\tz$ denotes the relaxed state of the undeformed system (as discussed in Section \ref{SEC:relaxation}), and $\tzL$ is the relaxed deformed state.

Inserting the solution for $\RelaRandL$ into the expression for nonaffine deformation field, Eq.~(\ref{EQ:DefNADF}), we have
\begin{eqnarray}\label{EQ:NADFp}
	(\NADF) _p &=& 2i\Contraction \Bigg\lbrace
			\frac{\BulkModuZeroP \vert p \vert ^2 }{(\BulkModuZeroP \vert p \vert ^2 \trOne + 2D_p)2D_p} \MetrTens^{-1}
			\nonumber\\
	&&	\quad - \frac{\BulkModuZeroP \vert p \vert ^2 }{(\BulkModuZeroP \vert p \vert ^2 + 2D_p)2D_p} \IdenT
		\Bigg\rbrace
		\cdot \PLongT \cdot S_p \, ,
\end{eqnarray}
where
\begin{eqnarray}
	S_p \equiv \frac{\partial}{\partial p_{1,a}} \NonLocaKernOne_{p_1,0}
		-\frac{\partial}{\partial p_{2,a}}\Big\vert_{p_2=0} \NonLocaKernOne_{p_1,p_2} ,
\end{eqnarray}
and
\begin{eqnarray}
	\MetrTens \equiv \DefoGrad\Tran\DefoGrad
\end{eqnarray}
In the incompressible limit, we have
\begin{eqnarray}\label{EQ:NADFpInco}
	(\NADF) _p &\approx& 2i\Big\lbrace
			\frac{1 }{2D_p \trOne} \MetrTens^{-1}
			- \frac{1 }{2D_p} \IdenT
		\Big\rbrace
		\cdot \PLongT \cdot S_p \,\, .
\end{eqnarray}
In Section~\ref{SEC:SNAD} we shall compute the mean value and disorder correlator of this nonaffine deformation field.

\section{Characterizing the physical elastic quenched disorder}\label{SEC:Heterogeneity}
In Section~\ref{SEC:EFERS} we addressed three elastic parameter fields---the residual stress $\StreT$ and the Lam\'e coefficients $\SheaModu_{\tp}$ and $\BulkModu_{\tp}$---which characterize the elastic energy relative to the \emph{relaxed state} and are therefore the physically relevant parameters for describing spatially heterogeneous elasticity.
There, we showed how these parameters are determined, for a given realization of the quenched disorder, by $\NonLocaKern$, i.e., the random nonlocal kernel of the phenomenological model, giving the connection in Eqs.~(\ref{EQ:RelaStre}).
In Section~\ref{SEC:Phen} we obtained a statistical characterization of $\NonLocaKern$ via a comparison with the semi-microscopic RLPM, giving the results in Eqs.~(\ref{EQ:CompDA}).
Thus, we have the ingredients for constructing a statistical characterization of the physical position-dependent elastic parameters, as we do in the present section.

\subsection{Disorder averages of the elastic parameters}
\label{SEC:DAEP}
Our first step is to determine the disorder average of the nonlocal kernel in the relaxed state $\tNonLocaKern$. To do this, we note that $\tNonLocaKern$ is related to $\NonLocaKern$ via Eq.~(\ref{EQ:tGtu}); the leading-order expansion of this relationship is given in Eq.~(\ref{EQ:tNonLocaKernText}). By taking the disorder average on both sides of Eq.~(\ref{EQ:tNonLocaKernText}), we find that only the first term on the RHS survives, because all other terms are linear in the fluctuation part of $\NonLocaKern$ and this vanishes upon disorder--averaging.
Thus, we find that the disorder average of $\tNonLocaKern$ is given by
\begin{align}
	\lda \tNonLocaKern(z_1,z_2) \rda = \lda \NonLocaKern(z_1,z_2) \rda = \KernOne(z_1,z_2) ,
\end{align}
which means that the disorder average of $\tNonLocaKern$ is the same as the disorder average of $\NonLocaKern$, where we have dropped the tilde on $\tz$ because now we discuss elastic parameters in the relaxed reference state only. It is worth noting that, as expected, because $\KernOne(z_1,z_2)$ is independent of the center of mass coordinate $(z_1+z_2)/2$, the disorder average of the nonlocal kernel $\lda \tNonLocaKern(z_1,z_2) \rda$ is translationally and rotationally invariant, depending only on $\vert z_1-z_2\vert$. This is a consequence of the macroscopic translational and rotational invariance of the random solid state discussed in Section~\ref{SEC:SSB}.

Second, we determine the disorder averages of the position-dependent elastic parameters in the local form of the elastic energy relative to the relaxed state, including the residual stress $\StreT$, the shear modulus $\SheaModu$, and the bulk modulus $\BulkModu$. For any given realization of the disorder, these elastic parameters are related to $\NonLocaKern$ and $\BulkModuZero$ via Eqs.~(\ref{EQ:RelaStre}). Thus, as with the nonlocal kernel, we obtain the disorder averages of these elastic parameters via the statistics of $\NonLocaKern$.

The disorder average of the residual stress $\StreT$ is straightforwardly seen to vanish:
\begin{eqnarray}
	\lda \Stre_{ij}(z) \rda =0 .
\end{eqnarray}
Thus, the residual stress is a quenched random field with zero mean.
As for the shear modulus $\SheaModu$, its disorder average is given by
\begin{eqnarray}
	\lda \SheaModu(z) \rda &=& \MeanSheaModu
	= -\frac{1}{\Dime} \int dz_2 \NonLocaKernZero(z-z_2) \vert z-z_2\vert^2 \nonumber\\
	&=& \PartDens \BoltCons T \, \UnivPara ,
\end{eqnarray}
with $\UnivPara$ given in Eq.~(\ref{EQ:theta}). This has been obtained in Ref.~\cite{Ulrich2006}. This mean shear modulus is linear in temperature $T$, reflecting its entropic nature, a result that confirms this aspect of the classical theory of rubber elasticity.
As for the disorder average of bulk modulus $\BulkModu$, it is obtained via Eq.~(\ref{EQ:BMZ}), which gives
\begin{eqnarray}
	\lda \BulkModu(z) \rda = \ExclVolu \PartDens^2 \, .
\end{eqnarray}
As one might expect, the mean bulk modulus depends on the particle number density $\PartDens$ and the strength of the excluded-volume interaction $\ExclVolu$.
The disorder average of the these three elastic parameters of the local form of the elastic energy relative to the relaxed state (viz., $\lda \Stre_{ij}(z) \rda$, $\lda \SheaModu(z) \rda$ and $\lda \BulkModu(z) \rda$) are all spatially homogeneous and isotropic; this is also a consequence of the macroscopic translational and rotational invariance of the random solid state discussed in Section~\ref{SEC:SSB}.

\subsection{Disorder correlators of the elastic parameters}
\label{SEC:Corr}
\subsubsection{Disorder correlator of the nonlocal kernel}
\label{SEC:CorrNK}
The nonlocal kernel $\tNonLocaKern$ characterizes the quenched random nonlocal interactions in the relaxed state. Its statistics can be described via its moments. In Section~\ref{SEC:DAEP}, we already determined the disorder average of $\tNonLocaKern$; in the present section we determine the disorder correlator of $\tNonLocaKern$.

To do this, we use Eq.~(\ref{EQ:tNonLocaKernText}), which relates $\tNonLocaKern$ to any given random configuration of $\NonLocaKern$. Using the disorder correlator of $\NonLocaKern$, Eq.~(\ref{EQ:KTwo}), we then arrive at the disorder correlator of $\tNonLocaKern$, viz., $\lda\tNonLocaKern(z_1,z_2)\tNonLocaKern(z_3,z_4)\rda$. This is a combination of Gaussian and delta-function factors in the separations of the six pairs formed by the four points $\{z_1,z_2,z_3,z_4\}$. The derivation and the momentum-space expression of the result for this disorder correlator is given in Appendix~\ref{APP:MM}.

To reveal the universal characteristics of the disorder correlator $\lda\tNonLocaKern(z_1,z_2)\tNonLocaKern(z_3,z_4)\rda$, we investigate its large-distance behavior.  The nonlocal kernel itself describes a short-distance attractive interaction, because the disorder average $\lda\tNonLocaKern(z_1,z_2)\rda$ is a short-ranged function in $\vert z_1-z_2\vert$ characterized by the typical localization length. Thus, to extract the long-distance behavior of $\lda\tNonLocaKern(z_1,z_2)\tNonLocaKern(z_3,z_4)\rda$, we take the limit that the two pairs $\{z_1,z_2\}$ and $\{z_3,z_4\}$ are far apart from one another, but $z_1$ is near $z_2$, and $z_3$ is near $z_4$.  This is precisely the construction of the local description of elasticity of the relaxed state introduced in Section~\ref{SEC:EFERS}, featuring the quenched random residual stress and the Lam\'e coefficients fields.  We shall now discuss the disorder correlators of these elastic parameters in the local description of elasticity.

\subsubsection{Disorder correlators of the elastic parameters in the local form of the elastic energy}
\label{SEC:Local}
The elastic parameters in the local form of the elastic energy, including the residual stress $\StreT$, the shear modulus $\SheaModu$, and the bulk modulus $\BulkModu$, are related to any given configuration $\NonLocaKern$ via Eq.~(\ref{EQ:RelaStre}). Using these relations and the disorder correlator of $\NonLocaKern$, Eq.~(\ref{EQ:KTwo}), we arrive at the disorder correlators of the elastic parameters. The details of this calculation are given in Appendix~\ref{APP:CFLL}; we summarize the results in Table~\ref{TABLE:CorrTable}.

\def\myshift{\,\,}
\begin{table}[h]
\begin{center}
\begin{tabular}{|r|c c c|}
\hline
    &$\myshift\Stre_{kl,p^{\prime}}\myshift$ \T
    &$\myshift\SheaModu_{p^{\prime}}\myshift$&
     $\myshift\BulkModu_{p^{\prime}}\myshift$
\\[4pt]
\hline
     $\myshift            \Stre_{ij,p}\myshift$ \T
    &$\myshift           \UnivPara A_{ijkl}\myshift$
    &$\myshift          -2\UnivPara\PPerp_{ij}\myshift$
    &$\myshift\phantom{-}4\UnivPara\PPerp_{ij}\myshift$
\\
     $\myshift            \SheaModu_{p}\myshift$ \T
    &$\myshift          -2\UnivPara\PPerp_{kl}\myshift$
    &$\myshift\phantom{-2} \Corr\myshift$
    &$\myshift          -2 \Corr\myshift$
\\
     $\myshift            \BulkModu_{p}\myshift$ \T
    &$\myshift\phantom{-}4\UnivPara\PPerp_{kl}\myshift$
    &$\myshift          -2 \Corr\myshift$
    &$\myshift\phantom{-}4 \Corr\myshift\myshift$
\\[4pt]
\hline
\end{tabular}
\caption[Long-wavelength variances and covariances of the elastic parameters of soft random solids in the relaxed state.]
{
Long-wavelength variances and covariances of the elastic properties of soft random solids in the relaxed state.
The entry in row R and column C, when multiplied by
$\PartDens(\BoltCons T)^2 \,\Volu_0\,\delta_{p+p^{\prime},0}$,
yields the connected disorder correlator
$[{\rm R}(p)\,{\rm C}(p^{\prime})]_{\rm c}\equiv
[{\rm R}(p)\,{\rm C}(p^{\prime})]-
[{\rm R}(p)]\,[{\rm C}(p^{\prime})]$.}
\label{TABLE:CorrTable}
\end{center}
\end{table}

The correlation function $\lda\StreT\StreT\rda$ features the tensor $A_{ijkl}$, which is defined as
\begin{eqnarray}
	A_{ijkl} \equiv 2 \PPerp_{ij} \PPerp_{kl} + \PPerp_{ik} \PPerp_{jl} + \PPerp_{il} \PPerp_{jk}\, .
\end{eqnarray}
where the projection operator $\PPerpT$ is defined in Section~\ref{SEC:relaxation}. The stability condition on the residual stress field $\StreT$ requires that its Fourier transform vanishes when contracted with the momentum $p$. It is straightforward to see that this feature is obeyed by the correlation function $\lda\StreT\StreT\rda$ given in Table~\ref{TABLE:CorrTable}, owing to the structure of $A$.

The parameters $\UnivPara$ and $\Corr$, on which the correlators in Table~\ref{TABLE:CorrTable} depend, are given by
\begin{subequations}
\begin{eqnarray}
	\UnivPara \!&\equiv&\! -\frac{\LinkDens \LocaPart^2}{2}+\LinkDens \LocaPart
	-1+e^{-\LinkDens \LocaPart} , \\
	\Corr \!&\equiv&\!
	 -\frac{3}{2}\LinkDens \LocaPart^2 \!+(\LinkDens \LocaPart)^2 +\LinkDens \LocaPart -1 + e^{\LinkDens \LocaPart} .
\end{eqnarray}
\end{subequations}

\begin{figure}[htbp]
	\centering
		\includegraphics[width=.45\textwidth]{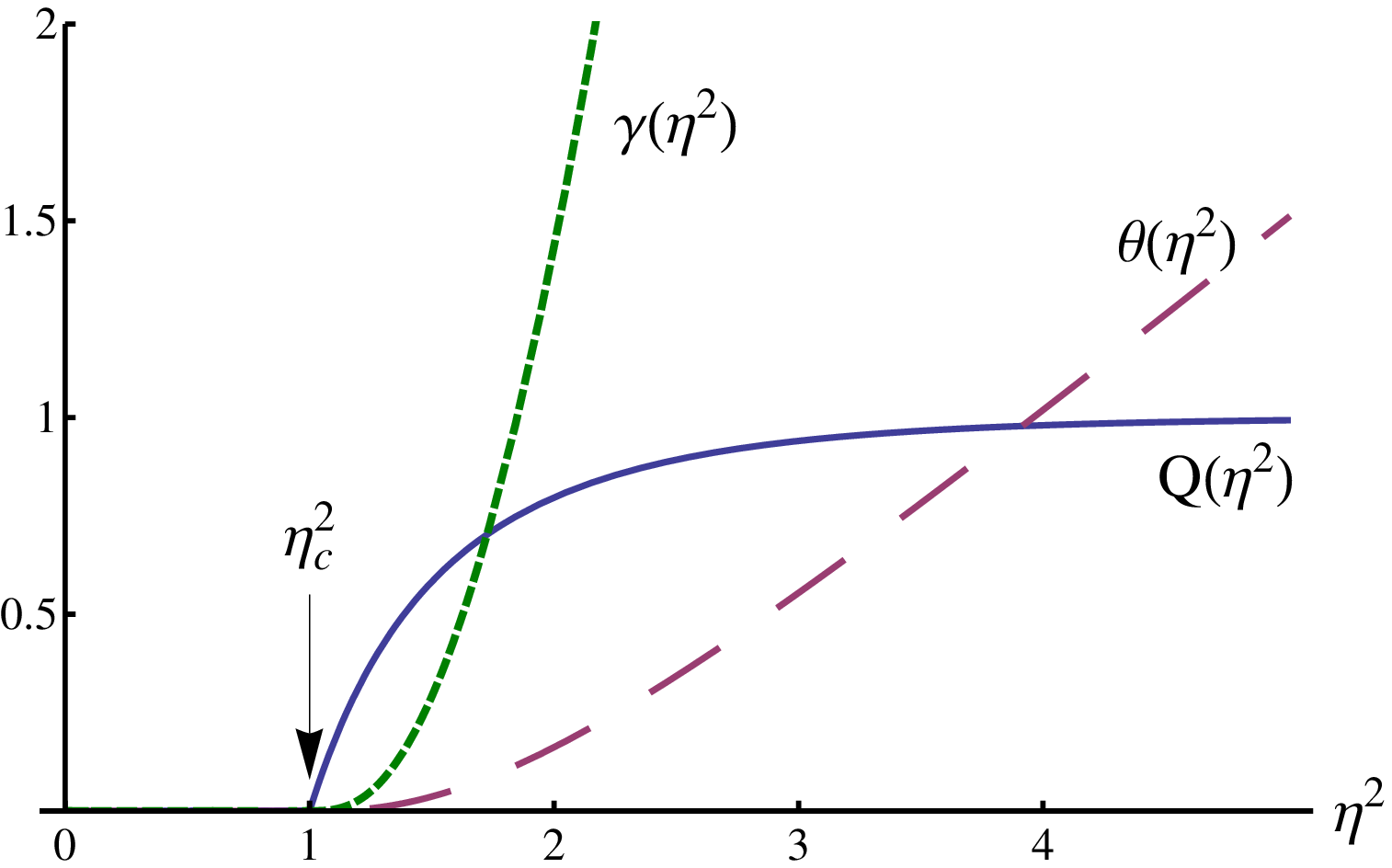}
	\caption{Plot of $\LocaPart$, $\UnivPara$ and $\Corr$ as functions of the links density parameter $\LinkDens$.}
	\label{FIG:ThetaGamma}
\end{figure}
\begin{figure}[htbp]
	\centering
		\includegraphics[width=.45\textwidth]{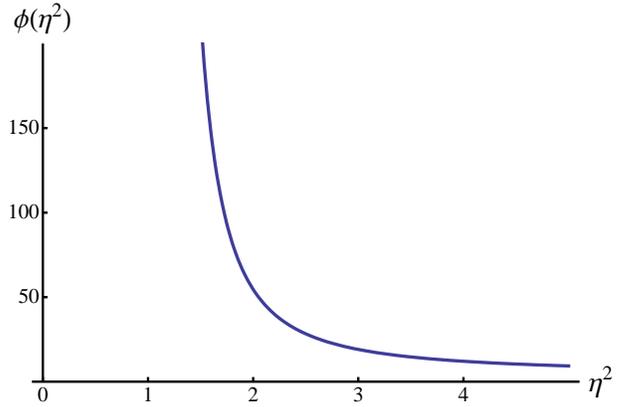}
	\caption{Plot of the function $\phi(\LinkDens)=\frac{\Corr}{\UnivPara^2}$, which characterizes the relative fluctuations of both the shear and bulk moduli, and also the connected correlator of the nonaffine deformations, as will be shown in Section~\ref{SEC:SNAD}.}
	\label{FIG:RelaFluc}
\end{figure}

The dependence of $\UnivPara$ and $\Corr$ on the density of link $\LinkDens$ is shown in Fig.~\ref{FIG:ThetaGamma}. The asymptotic behaviors of $\UnivPara$ and $\Corr$ are as follows:
\begin{eqnarray}
	\UnivPara =
	\left\{
		\begin{array}{ll}
			\frac{2}{3} (\LinkDens-1)^3 , & \textrm{for} \quad\LinkDens\gtrsim 1 ; \\
			\LinkDens/2 , & \textrm{for} \quad\LinkDens\gg 1 ;
		\end{array}
	\right.
\end{eqnarray}
\begin{eqnarray}
	\Corr =
	\left\{
		\begin{array}{ll}
			\frac{14}{3} (\LinkDens-1)^3 , & \textrm{for} \quad\LinkDens\gtrsim 1 ; \\
			\eta^4 , & \textrm{for} \quad\LinkDens\gg 1 ,
		\end{array}
	\right.
\end{eqnarray}
where $\LinkDens$ is equal to the mean coordination number of the particles.
Although the connected disorder correlators of the elastic parameters increase with the density of links $\LinkDens$, it is worth noting that the \emph{relative fluctuations} are decreasing functions of $\LinkDens$. For example, the relative fluctuation in the shear modulus, defined as $\lda \SheaModu \SheaModu\rda_c / \lda \SheaModu \rda^2$, scales as
\begin{eqnarray}
	\frac{\lda \SheaModu \SheaModu\rda_c}{ (\lda \SheaModu \rda)^2}
	\sim \frac{\Corr}{\UnivPara^2} ,
\end{eqnarray}
which is shown in Fig.~\ref{FIG:RelaFluc}. This is a decreasing function of $\LinkDens$, which means that the relative fluctuations of the shear modulus actually decrease as links are added.

It is interesting to look at the real-space behavior of disorder correlators of the elastic parameters in the local form of elastic energy. First, it is easy to see that the disorder correlators $\lda \SheaModu(0)\,\SheaModu(r) \rda$, $\lda \BulkModu(0)\,\BulkModu(r) \rda$ and $\lda \SheaModu(0)\,\BulkModu(r) \rda$ are short-ranged in real space: more precisely, they are proportional to $\delta_s(r)$, i.e., to a Dirac delta-function that has been smoothed on the scale of the short-distance cutoff. This cutoff should be taken to be the typical localization length, in order to validate the Goldstone-fluctuation framework for elastic deformations, because the Goldstone fluctuations in the RLPM are long-wavelength, low-energy excitations of the random solid state, and these do not touch lengthscales shorter than the typical localization length.

By contrast, entities involving the residual stress have a more interesting spatial correlations: in three dimensions and at large lengthscales we find that
\begin{subequations}
\begin{eqnarray}
	\!\!\!\!\!\!\lda \Stre_{ij}(0) \, \Stre_{kl}(\vec{r})\rda_c
		\!\!\!&=&\!\! \frac{(\BoltCons T)^2  \PartDens \UnivPara}{\pi \vert \vec{r}\vert^3} B_{ijkl}\, , \\
	\!\!\!\!\!\!\lda \Stre_{ij}(0) \, \SheaModu(\vec{r})\rda_c
		\!\!\!&=&\!\!\! - \frac{(\BoltCons T)^2  \PartDens \UnivPara}{\pi \vert \vec{r}\vert^3}
			\Big(\!\PLong_{ij}(\vec{r})\!-\!\PPerp_{ij}(\vec{r})\!\Big), \,\, \\
	\!\!\!\!\!\!\lda \Stre_{ij}(0) \, \BulkModu(\vec{r})\rda_c
		\!\!\!&=&\!\! \!\frac{2(\BoltCons T)^2  \PartDens \UnivPara}{\pi \vert \vec{r}\vert^3}
			\Big(\!\PLong_{ij}(\vec{r})\!-\!\PPerp_{ij}(\vec{r})\!\Big) , \,
\end{eqnarray}
\end{subequations}
where $\PLong_{ij}(\vec{r})$ and $\PPerp_{ij}(\vec{r})$ are, respectively, longitudinal and transverse projection operators in real space, which are given by
\begin{align}
	\PLong_{ij}(\vec{r})\equiv \frac{r_i r_j}{\vert \vec{r}\vert^2}, \quad\quad
	\PPerp_{ij}(\vec{r})\equiv \delta_{ij}-\PLong_{ij}(\vec{r}) .
\end{align}
The tensor $B_{ijkl}$ has a complicated structure comprising terms built from projection operators of $\vec{r}$, together with various index combinations, and also depends on the large-momentum cutoff, which can be identified with the inverse of the typical localization length
\footnote{The dependence on the large-momentum cutoff is a result of keeping only terms of leading order at small momentum $p$ in the calculation of the disorder correlators given in Table~\ref{TABLE:CorrTable}. This enables us to extract the small-momentum behavior in momentum-space, which corresponds to the large-distance behavior in real-space.
} .

{\colorred{
As the above results show, the mean shear modulus and the long-distance behavior of the disorder correlators depend only on the link density $\LinkDens$, the temperature $T$, and the particle density $\PartDens$, and do not depend on the details of the link potential.  This verifies the argument that the shear rigidity of the RLPM is a result of the network entropy.}}

\subsection{Statistics of nonaffine deformations}
\label{SEC:SNAD}
In this section we develop a statistical characterization of the nonaffine deformations of the soft random solid state. In Section \ref{SEC:NAD} we discussed why soft random solids undergo nonaffine deformations in presence of a given shear deformation $\DefoGrad$, and explained how to characterize these deformations in terms of the nonaffine deformation field $\NADF$.

The nonaffine deformation field is related to any given nonlocal random kernel $\NonLocaKern$ via Eq.~(\ref{EQ:NADFp}).
It is straightforward to see that the disorder average of the nonaffine deformation field vanishes, i.e., $\lda \NADF \rda =0$: it is proportional to the fluctuation part of the quenched random nonlocal kernel $\NonLocaKernOne$.

Next, we calculate the disorder correlator of the nonaffine deformations. For convenience, we take the incompressible limit, i.e., $\BulkModuZero\to\infty$, in which limit the nonaffine deformation field $\NADF$ is given by Eq.~(\ref{EQ:NADFpInco}). Using Eq.~(\ref{EQ:NADFpInco}), as well as the disorder correlator of the nonlocal kernel $\lda\NonLocaKern\NonLocaKern\rda_c$, the disorder correlator of the nonaffine deformation field $\NADF$ is found to be
\begin{eqnarray}\label{EQ:CorrNADF}
	\lda (\NADF)_p \cdot (\NADF)_{-p} \rda_c
	&=& \frac{1}{\vert \vec{p} \vert^2} \frac{1}{\PartDens} \frac{\Corr}{\UnivPara^2}
	\Big(\frac{\trTwo}{\trOne^2}-1\Big) ,
\end{eqnarray}
where
\begin{subequations}
\begin{eqnarray}
	\trOne &\equiv& \textrm{Tr}(\PLongT \MetrTens^{-1} ) , \\
	\trTwo &\equiv& \textrm{Tr}(\PLongT \MetrTens^{-1} \MetrTens^{-1}) , \\
	\MetrTens &\equiv& \DefoGrad\Tran\DefoGrad .
\end{eqnarray}
\end{subequations}
The dependence of the connected disorder-correlator of the nonaffine deformation field $\lda \NADF \NADF \rda_c$ on the density of links comes through the factor $\phi\equiv \Corr/(\UnivPara^2)$, which is shown in Fig.~\ref{FIG:RelaFluc}. It is evident that, as the density of links increases, the system has smaller relative fluctuations of its elasticity, i.e., the relative fluctuations in the elastic moduli decrease, and thus the nonaffine deformations also decrease, corresponding to the system becoming \emph{less heterogeneous}.

The disorder correlator of the nonaffine deformation field, Eq.~(\ref{EQ:CorrNADF}), is consistent with the disorder correlator of the nonaffine deformations given in Ref.~\cite{DiDonna2005}.
In Eq.~(3.22) of Ref.~\cite{DiDonna2005}, the disorder correlator of the nonaffine deformation $u^{\prime}$ (which corresponds to $\RelaRandL$ in our notation) was found to depend on the random local elastic modulus $K_{ijkl}$ in momentum-space as
\begin{align}
	\lda u^{\prime}(q) \, u^{\prime}(-q)\rda \propto \frac{\gamma^2}{q^2} \frac{\Delta^{K}(q)}{K^2} ,
\end{align}
where $\gamma$ represents the appropriate components of the tensorial externally applied deformation (i.e., $\DefoGrad$ in our notation), $\Delta^{K}$ represents components of the variance of the elastic-modulus tensor, and $K$ represents components of the average of the elastic-modulus tensor. Consistency with Ref.~\cite{DiDonna2005} is revealed by taking Eq.~(\ref{EQ:CorrNADF}), and recalling that $\lda\SheaModu\rda\sim\UnivPara$ and $\lda\SheaModu\SheaModu\rda_c\sim\Corr$, and therefore that our disorder correlator of the nonaffine deformation field can be written as
\begin{align}
	\lda (\NADF)_p (\NADF)_{-p} \rda_c \propto \frac{1}{\vert p \vert^2}
					\frac{\lda\SheaModu\SheaModu\rda_c}{\lda\SheaModu\rda^2} ,
\end{align}
which exhibits the same dependence on the mean and variance of the quenched random elastic modulus as Eq.~(3.22) of Ref.~\cite{DiDonna2005} does.

By transforming the disorder correlator of the nonaffine deformation field~(\ref{EQ:CorrNADF}) back to real space, we find that the large-distance behavior of the disorder correlator of the nonaffine deformation field $\lda \NADF(0) \cdot \NADF(r) \rda_c$ is proportional to $\vert r\vert^{-1}$ in three dimensions, which is also long-ranged.

\section{Concluding remarks}
\label{SEC:ConcDisc}
The heterogeneous elasticity of soft random solids has been investigated via a semi-microscopic approach. By starting with the Randomly Linked Particle Model (RLPM), which describes networks of particles randomly connected by soft links, and applying the concepts and techniques of vulcanization theory,
we have established a field-theoretic description of the liquid-to-random-solid transition, and have analyzed the corresponding pattern of spontaneous symmetry breaking and the structure of the associated Goldstone fluctuations. We have identified these Goldstone fluctuations as being related to shear deformations of the random solid state and, via this identification, we have obtained a statistical characterization of the quenched randomness exhibited by the heterogeneous elasticity of soft random solids, which features a random nonlocal kernel describing attractive interactions between mass-points.

The heterogeneous elasticity studied via the Goldstone fluctuations in the RLPM is a description of the elastic properties of the state right after linking (i.e., an elastic free energy that takes the state right after linking as its elastic reference state). We have shown that, after linking, the system relaxes to a stable state for any given realization of disorder (i.e., for any given heterogeneous configuration of the elastic parameters in the state right after linking), and this relaxed state, which is a state of mechanical equilibrium, is actually the state of experimental relevance. By solving for the relaxed state for any given realization of disorder, and expanding the elastic free energy for deformations relative to this relaxed state, we have obtained an elastic free energy relative to the relaxed state (i.e., taking the relaxed state as the new elastic reference state).
The statistics of the quenched randomness in this elastic free energy is subsequently determined.

The first statistical moments of the quenched random elastic parameters (i.e., the disorder averages of the elastic parameters), unveil the basic homogeneous macroscopic properties of the heterogeneous elastic medium. We have found that the disorder average of the nonlocal kernel of attractive interactions is characterized by the typical localization lengthscale of the RLPM, which is a scale smaller than the lengthscale of the elastic deformations that we are considering. Thus, it is reasonable to make a local expansion of the elastic energy, relative to the relaxed state. The resulting local form of the elastic energy is a version of Lagrangian elasticity, featuring heterogeneous (i.e., spatially randomly varying) residual stress and Lam\'e coefficients. The disorder average of the residual stress vanishes. The disorder average of the shear modulus is found to be proportional to temperature, reflecting the entropic nature of the shear rigidity of soft random solids. The disorder average of the bulk modulus depends on the particle number-density and the strength of the excluded-volume interaction. In particular, the disorder averages of these elastic parameters of the relaxed state are all translationally and rotationally invariant, reflecting the macroscopic translational and rotational invariance of the soft random solid state.

The second statistical moments of the quenched random elastic parameters (i.e., the spatial correlations of these elastic parameters) characterize the fluctuations of the quenched randomness in the elastic properties. The disorder correlators of the elastic parameters that appear in the local form of the elastic energy (relative to the relaxed state) exhibit interesting universal behaviors. In particular, the disorder-correlators involving the residual stress are found to be long-ranged and governed by a universal parameter that also determines the mean shear modulus, but the disorder-correlators of the shear and bulk moduli are found to be short ranged.

Because of the heterogeneity present in the elasticity of soft random solids, upon the application external stress, the system responds by adopting a strain field that is nonaffine (i.e., a strain field that is characterized by an inhomogeneous deformation gradient). We have also obtained a statistical description of these nonaffine deformations. The disorder average of the nonaffine deformations vanishes, and their disorder correlator is also found to be long ranged.

So far, we have studied the first two statistical moments of the quenched random elastic parameters of soft random solids. The entire probability distribution of the quenched random elastic parameters can also be explored using the formalism presented here, via the RLPM, and one can also progress beyond the local limit of the elasticity theory.


This approach to the heterogeneous elasticity of soft random solids can also be applied to the setting of liquid crystal elastomers, in which the constituent polymers of the random network possesses (or are capable of exhibiting) liquid-crystalline order~\cite{Warner2003,Lubensky2002,Xing2008,Clarke1998}. In liquid crystal elastomers, the strain field is coupled to the liquid-crystalline order, and this produces a rich collection of interesting phenomena, such as spontaneous sample-shape deformation upon changes of temperature, anomalously soft modes in the elasticity. The interplay of the heterogeneity of the random network and the liquid-crystalline order has interesting consequences: e.g., it can give rise to a polydomain structure in the liquid crystalline order. {\colorred These interesting topics shall be reserved for future studies. }

\begin{acknowledgments}
We thank Tom Lubensky for stimulating discussions.
This work was supported by
National Science Foundation Grant No.~DMR 06-05816 (X.M.\ and P.M.G.),
American Chemical Society Grant No.~PRF 44689-G7 (X.X.),
Deutsche Forschungsgemeinschaft through SFB 602 (standard) (A.Z.),
and the National Science Foundation Grant No.~DMR 0804900 (X.M.).
\end{acknowledgments}

\appendix
\section{Disorder average with the Deam-Edwards distribution}
\label{APP:DisoAver}
In this Appendix we calculate disorder averages weighted by the Deam-Edwards distribution, in particular, $Z_1$ and $Z_{1+n}$ of Section~\ref{SEC:RLPMReplica}.

First, we calculate the factor $Z_1$, which is defined as
\begin{eqnarray}\label{EQ:AZOne}
	Z_1 &\equiv& \sum_{\RealDiso}
		\frac{\left(\frac{\LinkDens \Volu_0}{2\PartNumb \LinkPote_0}\right)^{\LinkNumb}
		Z_{\RealDiso}(\Volu_0)}{ \LinkNumb !}.
\end{eqnarray}
The summation over the quenched disorder $\sum_{\RealDiso}$ includes two steps: a summation over the number of links $\LinkNumb$, and a summation over all possible ways of making these $\LinkNumb$ links, i.e., of assigning the $\LinkNumb$ links to different collections of pairs. Thus Eq.~(\ref{EQ:AZOne}) can be written as
\begin{widetext}
\begin{eqnarray}
	Z_1 &=& \sum_{\LinkNumb}
			\frac{\Big(\frac{\LinkDens \Volu_0}{2\PartNumb \LinkPote_0}\Big)^{\LinkNumb}
			Z_{\RealDiso}(\Volu_0)}{ \LinkNumb !} \nonumber\\
		&=& \sum_{\LinkNumb=0}^{\infty} \sum_{i_1\ne j_1}^{\PartNumb}
			\sum_{i_2\ne j_2}^{\PartNumb} \cdots
			\sum_{i_{\LinkNumb}\ne j_{\LinkNumb}}^{\PartNumb}
			\frac{\Big(\frac{\LinkDens \Volu_0}{2\PartNumb \LinkPote_0}\Big)^{\LinkNumb}}
			{ \LinkNumb !}  \PartitionLiquid(\Volu)
			\left\langle
			\prod_{e=1}^{\LinkNumb} \LinkPote \big(
			\vert \PartPosi_{i_e} - \PartPosi_{j_e} \vert
			\big) \right\rangle_{1}^{\HamiExcl} \nonumber\\
		&=& \PartitionLiquid(\Volu)\lthal
			\sum_{\LinkNumb=0}^{\infty} \frac{\big(\frac{\LinkDens \Volu_0}
			{2\PartNumb \LinkPote_0}\big)^{\LinkNumb}}{ \LinkNumb !}
			\Big(\sum_{i\ne j}^{\PartNumb}
			\LinkPote \big(
			\vert \PartPosi_{i} - \PartPosi_{j} \vert
			\big)\Big)^{\LinkNumb}\rthal_{1}^{\HamiExcl} \nonumber\\
        \noalign{\smallskip}
		&=& \PartitionLiquid(\Volu)\left\langle
			\exp\Big(
			\frac{\LinkDens \Volu_0}{2\PartNumb \LinkPote_0}
			\sum_{i\ne j}^{\PartNumb}
			\LinkPote \big(
			\vert \PartPosi_{i} - \PartPosi_{j} \vert
			\big)
			\Big)\right\rangle_{1}^{\HamiExcl}.
\end{eqnarray}
\end{widetext}
The mean-field approximation for $Z_1$ amounts to taking the number density of the unlinked liquid to be $\PartNumb/\Volu_0$, which is similar to the calculation that yields Eq.~(\ref{EQ:FLiquid}), and hence we arrive at
\begin{eqnarray}\label{EQ:Z1SP}
	Z_1=\exp\big(\PartNumb\ln\Volu_0
	-\frac{\ExclVolu\PartNumb^2}{2\Volu_0 \BoltCons T}+\frac{\PartNumb \LinkDens}{2} \big).
\end{eqnarray}

Second, we calculate $Z_{1+n}$, which is defined as
\begin{eqnarray}
	Z_{1+n}\equiv\sum_{\RealDiso}
			\frac{\big(\frac{\LinkDens \Volu_0}{2\PartNumb \LinkPote_0}\big)^{\LinkNumb}}
			{ \LinkNumb !} Z_{\RealDiso}(\Volu_0) Z_{\RealDiso}(\Volu)^n .
\end{eqnarray}
The factor $Z_{\RealDiso}(\Volu)^n$ can be written in terms of replicas as
\begin{eqnarray}
	Z_{\RealDiso}(\Volu)^n
	&=& \int_{\Volu} \ReplProd \prod_{j=1}^{\PartNumb} d\PartPosi_{j}\REPa
		\, e^{-\ReplSum \HamiExcl \REPa/\BoltCons T} \nonumber\\
	&&\quad\quad\times
		\ReplProd \prod_{e=1}^{\LinkNumb}
		  \LinkPote \big(
			\vert \PartPosi_{i_e}\REPa - \PartPosi_{j_e}\REPb \vert \big) ,
\end{eqnarray}
where $\HamiExcl \REPa$ is the part of the Hamiltonian $\HamiExcl$
(i.e., the excluded-volume interaction) for replica $\alpha$,
as defined in Eq.~(\ref{EQ:partition}).

We define the $\HamiExcl$ average for $1+n$ replicas as
\begin{eqnarray}
	\ltha \cdots \rtha_{1+n}^{\HamiExcl} &\equiv&
		\frac{1}{\PartitionLiquid(V_0) \PartitionLiquid(V)^n}\nonumber\\
	&&\quad\quad\times
		\int _{\Volu_0} \prod_{i=1}^{\PartNumb} d \PartPosi_i \REP0
		\int _{\Volu} \ReplProdOne \prod_{i=1}^{\PartNumb}
		d \PartPosi_i \REPa 	\nonumber\\
	&&\quad\quad\times
		e^{-\frac{\HamiExcl\REP0}{{\BoltCons}T}
			-\frac{\ReplSumOne \HamiExcl\REPa}{{\BoltCons}T}}\cdots .
\end{eqnarray}
Using this notation we arrive at
\begin{widetext}
\begin{eqnarray}
	Z_{1+n}&=&\sum_{\LinkNumb=0}^{\infty} \sum_{i_1\ne j_1}^{\PartNumb}
			\sum_{i_2\ne j_2}^{\PartNumb} \cdots
			\sum_{i_{\LinkNumb}\ne j_{\LinkNumb}}^{\PartNumb}
			\frac{\big(\frac{\LinkDens \Volu_0}{2\PartNumb \LinkPote_0}\big)^{\LinkNumb}}
			{ \LinkNumb !}  \PartitionLiquid(\Volu_0)\PartitionLiquid(\Volu)^n
			\ltha
			\ReplProd \prod_{e=1}^{\LinkNumb} \LinkPote \big(
			\vert \PartPosi_{i_e}\REPa - \PartPosi_{j_e}\REPa \vert
			\big) \rtha_{1+n}^{\HamiExcl} \nonumber\\
	&=& \PartitionLiquid(\Volu_0)\PartitionLiquid(\Volu)^n \lthal
			\sum_{\LinkNumb=0}^{\infty} \frac{\big(\frac{\LinkDens \Volu_0}
			{2\PartNumb \LinkPote_0}\big)^{\LinkNumb}}{ \LinkNumb !}
			\Big(\sum_{i\ne j}^{\PartNumb} \ReplProd
			\LinkPote  \big(
			\vert \PartPosi_{i}\REPa - \PartPosi_{j}\REPa \vert
			\big)\Big)^{\LinkNumb}\rthal_{1+n}^{\HamiExcl} \nonumber\\
	&=& \PartitionLiquid(\Volu_0)\PartitionLiquid(\Volu)^n
			\lthal \exp \Big(\frac{\LinkDens \Volu_0}
			{2\PartNumb \LinkPote_0}
			\sum_{i\ne j}^{\PartNumb} \ReplProd
			\LinkPote  \big(
			\vert \PartPosi_{i}\REPa - \PartPosi_{j}\REPa \vert
			\big)\Big)\rthal_{1+n}^{\HamiExcl}
\end{eqnarray}
\end{widetext}

\section{Hubbard-Stratonovich Transformation}
\label{APP:HSTransformation}
The effective Hamiltonian, Eq.~(\ref{EQ:HQHL}), can be analyzed via a Hubbard-Stratonovich (HS) transformation---a field-theoretic tool that is often applied to strongly coupled models to decouple interactions and develop a convenient representation in terms of functional integrals~\cite{Hubbard1959,Stratonovich1957}.

The version of the HS transformation that we use for the RLPM can be illustrated via the following simple example.  Consider a statistical-mechanical system having the following partition function:
\begin{eqnarray}
	Z(h) =\int dq\, e^{-H_0(q)}e^{Jq^2+hq} = Z_0\,\langle e^{Jq^2+hq} \rangle_{H_0(q)} ,
\end{eqnarray}
where $H(q)\equiv H_0(q)-Jq^2-hq$ is the total Hamiltonian for the variable $q$, with $H_0(q)$ being the leading-order term and $Jq^2$ being considered as a perturbation.  (Although it is just a simple quadratic term, we use it to illustrate the method.)\thinspace\  The factor $Z_0\equiv \int dq \, e^{-H_0(q)}$. The term $hq$ denotes the coupling to an external field, which generates the statistical moments of $q$ via
\begin{eqnarray}
	\langle q \rangle_{H(q)}=\frac{\partial }{\partial h}\Big\vert_{h=0} \ln Z(h) .
\end{eqnarray}
The $Jq^2$ term in the exponent can be decoupled using the following version of the HS transformation:
\begin{eqnarray}
	Z&=&\Big(\frac{J}{\pi}\Big)^{1/2} Z_0 e^{-\frac{h^2}{4J}}
		\int d \omega \, e^{-J\omega^2+h \omega}
		\langle e^{2J\omega q} \rangle_{H_0(q)} \nonumber\\
	&=&\Big(\frac{J}{\pi}\Big)^{1/2} \int d \omega\, e^{-\mathcal{H}(\omega)} ,
\end{eqnarray}
with
\begin{eqnarray}\label{EQ:Homega}
	\mathcal{H}(\omega)\equiv J\omega^2-\ln Z_0\,\langle e^{2J\omega q} \rangle_{H_0(q)}-h\omega +\frac{h^2}{4J} .
\end{eqnarray}
In this form, the partition function is expressed as an integral over the variable $\omega$, and the quadratic term in the original variable $q$ is now decoupled.  If fluctuations with large $q$ only appear with very small probabilities, as governed by $H_0(q)$, one can expand the $\ln\langle e^{2J\omega q} \rangle_{H_0(q)}$ term as a power series in $q$.  Thus, one can obtain an effective Hamiltonian $\mathcal{H}(\omega)$ via the  low-order terms in $\omega$, which has the form of a Landau free energy, and is convenient to analyze.

It is evident that the average of $\omega$, taken with the statistical weight defined by $\mathcal{H}(\omega)$, equals the average of $q$, taken with the statistical weight defined by $H(q)$:
\begin{eqnarray}\label{EQ:HSAver}
	\langle \omega \rangle_{\mathcal{H}(\omega)}
	=\frac{\partial }{\partial h}\Big\vert_{h=0} \ln Z(h)
	=\langle q \rangle_{H(q)}.
\end{eqnarray}
Thus, the statistical mechanics of $q$ can be examined by studying the statistical mechanics of $\omega$.  In the Hamiltonian $\mathcal{H}(\omega)$, $q$ appears linearly; therefore, in cases in which $q$ is a variable that involves a summation over many particles, this method will allow us to decouple the problem into a single-particle one, as will be seen in the following application of the HS transformation to vulcanization theory.

In the RLPM, the partition function we are going to decouple is Eq.~(\ref{EQ:ZQ}):
\begin{eqnarray}
	Z_{1+n} = \int_{\Volu_0} \prod_{i=1}^{\PartNumb} d\PartPosi_{i}^{0}
						\int_{\Volu} \ReplProdOne \prod_{j=1}^{\PartNumb} d\PartPosi_{j}\REPa
						\, e^{-\frac{H_{\DensFunc}\lbrack \DensFunc_{\REPP}\rbrack}{\BoltCons T}} ,
\end{eqnarray}
with
\begin{widetext}
\begin{eqnarray}
	H_{\DensFunc}\lbrack \DensFunc_{\REPP}\rbrack
	= -\frac{\PartNumb \LinkDens \BoltCons T}{2 \Volu^n \LinkPote_0}
			\sum_{\REPP\in HRS}\DensFunc_{\REPP}
			\DensFunc_{-\REPP}\LinkPoteRepl_{\REPP}
			+\frac{\ExclVoluRN_0 \PartNumb^2}{2\Volu_0} \sum_{p}\DensFunc_{p\BASE^{0}}
				\DensFunc_{-p\BASE^{0}}	
				+\frac{\ExclVoluRN \PartNumb^2}{2\Volu} \sum_{p} \ReplSumOne
				\DensFunc_{p\BASE\REPa} \DensFunc_{-p\BASE\REPa} .
\end{eqnarray}
The field $\DensFunc_{\REPP}=(1/N)\sum_{j=1}^{\PartNumb}e^{-i\REPP \cdot \hat{c}_j}$ is a complex field, so we apply the following equalities for the complex variables $q$ and $\omega$:
\begin{subequations}\label{EQ:HSComplex}
\begin{eqnarray}
	\!\!e^{-J\vert q \vert^2}\!\!
	&=&\!\! \frac{J}{\pi}\int\!\! d(\RealPart\, \omega)d(\ImagPart\, \omega) \label{EQ:HS1}
		e^{-J\vert \omega \vert^2+2iJ \RealPart\, q\omega^{*}} , \\
	\!\!e^{+J\vert q \vert^2}\!\!
	&=&\!\! \frac{J}{\pi}\int \!\!d(\RealPart\, \omega)d(\ImagPart\, \omega)
		e^{-J\vert \omega \vert^2+2J \RealPart\, q\omega^{*}} , \label{EQ:HS2}
\end{eqnarray}
\end{subequations}
noticing that the product $\RealPart \, (q\omega^{*}) = (\RealPart\, q)(\RealPart\,\omega)+ (\ImagPart\, q)(\ImagPart\,\omega)$. We use Eq.~(\ref{EQ:HS1}) for the HS transformation for the LRS fields, and Eq.~(\ref{EQ:HS2}) for the HS transformation for the HRS field, and hence arrive at the form (For more details, see Section V in~\cite{Goldbart1996}.)
\begin{eqnarray}\label{EQ:HSPartition}
	Z_{1+n} = \int \mathcal{D}\VOP_{\hat{p}}\ReplProd\mathcal{D}\VOP_{p}\REPa
	e^{-\frac{H_{\VOP}\lbrack\VOP_{p}\REPa,\VOP_{\hat{p}}\rbrack}{\BoltCons T}} ,
\end{eqnarray}
with
\begin{eqnarray}
	H_{\VOP}\lbrack\VOP_{p}\REPa,\VOP_{\hat{p}}\rbrack
	= \frac{\PartNumb \LinkDens \BoltCons T}{2 \Volu^n \LinkPote_0}
			\!\sum_{\REPP\in HRS}\!\!\VOP_{\REPP}
			\VOP_{-\REPP}\LinkPoteRepl_{\REPP}
			+\!\frac{\ExclVoluRN_0 \PartNumb^2}{2\Volu_0}\!\sum_{p}\VOP_{p\BASE^{0}}
				\VOP_{-p\BASE^{0}}	
				+\!\frac{\ExclVoluRN \PartNumb^2}{2\Volu} \sum_{p}\!\ReplSumOne
				\VOP_{p\BASE\REPa} \VOP_{-p\BASE\REPa} \!
				- \! N\BoltCons T \ln \MyZzero,
\end{eqnarray}
where the $N\BoltCons T \ln \MyZzero$ term is analog of the $\ln \langle e^{2J\omega q} \rangle_{H_0(q)}$ term in Eq.~(\ref{EQ:Homega}) and, using $\DensFunc_{-\REPP}=(1/\PartNumb)\sum_{j=1}^{\PartNumb} e^{i\REPP\cdot\hat{c}_j}$
and
$\DensFunc_{-p}\REPa=(1/\PartNumb)\sum_{j=1}^{\PartNumb} e^{ip\cdot c_j\REPa}$, we have
\begin{eqnarray}
	\PartNumb \ln \MyZzero
	&\equiv& 	
	\ln \Big\lbrace \int_{\Volu_0} \prod_{i=1}^{\PartNumb} d\PartPosi_{i}^{0}
						\int_{\Volu} \ReplProdOne \prod_{j=1}^{\PartNumb} d\PartPosi_{j}\REPa
			\exp\big\lbrack
			\frac{\PartNumb \LinkDens }{ \Volu^n \LinkPote_0}
			\sum_{\REPP\in HRS}\VOP_{\REPP}
			\frac{1}{\PartNumb}\sum_{j=1}^{\PartNumb}
			e^{i\REPP\cdot\hat{c}_j}\LinkPoteRepl_{\REPP} \nonumber\\
	&&\quad\quad + \frac{i\ExclVoluRN_0 \PartNumb^2}{\Volu_0\BoltCons T}
	 \sum_{p}\VOP_{p\BASE^{0}}
			\frac{1}{\PartNumb}\sum_{j=1}^{\PartNumb} e^{ip\cdot c_j^{0}}
			+\frac{i\ExclVoluRN \PartNumb^2}{\Volu\BoltCons T} \sum_{p} \ReplSumOne
				\VOP_{p\BASE\REPa} \frac{1}{\PartNumb}\sum_{j=1}^{\PartNumb} e^{ip\cdot c_j^{0}}
				\big\rbrack\Big\rbrace \nonumber\\
	&=&\PartNumb \ln \Bigg\lbrace \int_{\Volu_0} \! d\PartPosi^{0} \!\!
						\int_{\Volu} \!\ReplProd  d\PartPosi\REPa
			\exp\Big\lbrack
			\frac{ \LinkDens}{ \Volu^n \LinkPote_0}
			\sum_{\REPP\in HRS}\VOP_{\REPP}\LinkPoteRepl_{\REPP}e^{i\REPP \cdot \hat{c}}
			\nonumber\\
	&& \quad		 + \frac{i\ExclVoluRN_0 \PartNumb}{\Volu_0\BoltCons T}
			\sum_{p}\VOP_{p\BASE^{0}}e^{ip^{0}c^{0}}
		\!\!+\frac{i\ExclVoluRN \PartNumb}{\Volu\BoltCons T} \sum_{p} \ReplSumOne
				\VOP_{p\BASE\REPa} e^{ip\REPa c\REPa} \Big\rbrack\Bigg\rbrace .	
\end{eqnarray}
\end{widetext}
In this form it is evident that the $\PartNumb$ particles are actually \emph{decoupled}.  Notice that in Eq.~(\ref{EQ:HSPartition}) the functional integrals
$\int \mathcal{D}\,\VOP_{\hat{p}}\ReplProd\mathcal{D}\VOP_{p}\REPa $
have carefully chosen prefactors [as in Eq.~(\ref{EQ:HSComplex})], to ensure that the integration is properly normalized.

\section{Hamiltonian of the stationary point}
\label{APP:HSP}
In this Appendix we calculate the value of the Hamiltonian at the stationary point by inserting the stationary point order parameter~(\ref{EQ:VOPAnsatzM}) into the Hamiltonian~(\ref{EQ:HVOPHRS}).

The first term in the Hamiltonian is
\begin{widetext}
\begin{eqnarray}
	&& \frac{\PartNumb \LinkDens \BoltCons T}{2 \Volu^n \LinkPote_0}
			\sum_{\REPP\in HRS}\VOP_{\REPP} \VOP_{-\REPP}\LinkPoteRepl_{\REPP} \nonumber\\
	&=&\frac{\PartNumb \LinkDens \BoltCons T}{2 \Volu^n \LinkPote_0}
			\sum_{\REPP} (\LinkPote)^{1+n}
			e^{-\frac{\LinkScal^2\vert\REPP\vert^2}{2\BoltCons T}}
			\big\lbrace \LocaPart \int \frac{dz_1}{\Volu_0}\int _{\ISLL_1}
			e^{-\frac{\vert\REPP\vert^2}{2\ISLL_1}
			-i \REPP \cdot \hat{z}_{\Contraction,1} }
			-\LocaPart\delta^{((1+n)d)}_{\REPP}\big\rbrace \nonumber\\
	&&\quad\times		\big\lbrace \LocaPart \int \frac{dz_2}{\Volu_0}\int _{\ISLL_2}
			e^{-\frac{\vert\REPP\vert^2}{2\ISLL_2}
			+i \REPP \cdot \hat{z}_{\Contraction,2} }
			-\LocaPart\delta^{((1+n)d)}_{\REPP}\big\rbrace \nonumber\\
	&=&\frac{\PartNumb \LinkDens \BoltCons T\LocaPart^2}{2 \Volu^n \LinkPote_0}
			(\LinkPote)^{1+n}\nonumber\\
	&&		+\frac{\PartNumb \LinkDens \BoltCons T\LocaPart^2}{2 \Volu^n \LinkPote_0}
			(\LinkPote)^{1+n}
			\sum_{\REPP}
			\int\frac{dz_1 dz_2}{\Volu_0^2}\int _{\ISLL_1,\ISLL_2}
			e^{\big(\frac{1}{2\ISLL_1}+\frac{1}{2\ISLL_2}+
				\frac{\LinkScal^2}{2\BoltCons T}\big)\vert\REPP\vert^2
				-i\REPP\cdot (\hat{z}_{\Contraction,1}-\hat{z}_{\Contraction,2})} \nonumber\\
	&=&\frac{\PartNumb \LinkDens \BoltCons T\LocaPart^2(\LinkPote)^{n}}{2 \Volu^n}
			+\frac{\PartNumb \LinkDens \BoltCons T\LocaPart^2(\LinkPote)^{n}}{2 \Volu^n }
			(1+n\Contraction^2)^{-d/2}\int _{\ISLL_1,\ISLL_2}
			\Big\lbrace 2\pi \Big( \frac{1}{2\ISLL_1}+\frac{1}{2\ISLL_2}+
				\frac{\LinkScal^2}{2\BoltCons T}\Big)
			\Big\rbrace^{-nd/2} ,
\end{eqnarray}
where $\hat{z}_{\Contraction}\equiv \{z,\Contraction z,\ldots,\Contraction z\}$. The sum $\sum_{\REPP\in HRS}$ can be changed into the $\sum_{\REPP}$ because the order parameter we have inserted vanishes for $\REPP\in\,$LRS.
We have also changed momentum summation into an integral by using $\frac{1}{\Volu_0\Volu^n}\sum_{\REPP}=\int\frac{d^{(1+n)d}\REPP}{(2\pi)^{(1+n)d}}$.

The free energy of the system is related to the $O(n)$ term of this Hamiltonian, as given by Eqs.~(\ref{EQ:HFEDisoAver}, \ref{EQ:HFEDisoAverDeri}). Thus, we make the small-$n$ expansion.  It is straightforward to see that the $O(1)$ terms cancel, and that the leading-order term is given by $O(n)$
\begin{eqnarray}
	&& \frac{\PartNumb \LinkDens \BoltCons T}{2 \Volu^n \LinkPote_0}
			\sum_{\REPP\in HRS}\VOP_{\REPP} \VOP_{-\REPP}\LinkPoteRepl_{\REPP} \nonumber\\
	&=& n \frac{\PartNumb \LinkDens \BoltCons T\LocaPart^2}{2 \Volu^n}
			\Big\lbrace
				\ln \Volu -\frac{d}{2}\big(\ln(2\pi)+\Contraction^2\big)
				-\frac{d}{2} \int _{\ISLL_1,\ISLL_2}
				\Big( \frac{1}{2\ISLL_1}+\frac{1}{2\ISLL_2}+
				\frac{\LinkScal^2}{2\BoltCons T}\Big)
			\Big\rbrace .
\end{eqnarray}

Similarly, we can calculate the $\ln \MyZzero$ term.  By inserting the saddle-point value of the order parameter into $\MyZzero$,
and summing (or, more precisely, integrating) over momentum $\REPP$, we have
\begin{eqnarray}
	\MyZzero\!&=&\!\int_{\Volu_0} \! d\PartPosi^{0} \!\!
						\int_{\Volu} \!\ReplProd  d\PartPosi\REPa
			\exp\Big\lbrace
			\frac{ \LinkDens}{ \Volu^n \LinkPote_0}
			\sum_{\REPP\in HRS}\VOP_{\REPP}\LinkPoteRepl_{\REPP}e^{i\REPP \cdot \hat{c}}
		 \Big\rbrace \nonumber\\
	&=& e^{-\LinkDens \LocaPart (\LinkPote_0/\Volu)^n}
			\int d\hat{c} \exp \big\lbrace
				\LinkDens \LocaPart (\LinkPote_0)^n
				\int dz \int_{\ISLL} \big(\frac{\tISLL}{2\pi}\big)^{\frac{(1+n)d}{2}}
				e^{-\frac{\tISLL}{2}(\hat{z}_{\Contraction}-\hat{c})^2}
			\big\rbrace ,
\end{eqnarray}
where $\tISLL\equiv\big(\frac{1}{\ISLL}+\LinkScal^2\big)^{-1}$.  We then Taylor-expand the exponential (keeping all orders) and integrate out $\hat{c}$ to get
\begin{eqnarray}\label{EQ:Z0intec}
	\MyZzero &=& e^{-\LinkDens \LocaPart (\LinkPote_0/\Volu)^n}\Volu_0\Volu^n
					\Big\lbrace
					1+\LinkDens \LocaPart (\LinkPote_0/\Volu)^n
					+\frac{1}{\Volu_0\Volu^n} \sum_{m=2}^{\infty}
					\frac{\big(\LinkDens \LocaPart (\LinkPote_0)^n\big)^m}{m!}
					\int dz_1 \cdots dz_m
					\nonumber\\
			&&\quad \times \int_{\ISLL_1,\ldots,\ISLL_m}
				\prod _{j=1}^{m} \Big(\frac{\tISLL_j}{2\pi}\Big)^{\frac{(1+n)d}{2}}
					\Big(\frac{2\pi}{\tISLL_1+\tISLL_m}\Big)^{\frac{(1+n)d}{2}}
					e^{-\frac{\tISLL_1\tISLL_2
					(\hat{z}_{\Contraction,1}-\hat{z}_{\Contraction,2})^2+\cdots}
					{2(\tISLL_1+\cdots+\tISLL_m)}}
				\Big\rbrace ,
\end{eqnarray}
where in the exponential the terms following $(\hat{z}_{\Contraction,1}-\hat{z}_{\Contraction,2})^2$ include all pairs of the $m$ variables [there are $m(m-1)/2$ such terms].
By using $(\hat{z}_{\Contraction,1}-\hat{z}_{\Contraction,2})^2
=(1+n\Contraction^2)(z_1-z_2)^2$
[recall that $\hat{z}_{\Contraction,1}$ is a $(1+n)d$-dimensional vector,
and that $z_1$ is a $d$-dimensional vector],
the integration $\int dz_1 \cdots dz_m $ can be readily performed,
and we thus obtain
\begin{eqnarray}\label{EQ:Z0intez}
	\MyZzero&=& e^{-\LinkDens \LocaPart (\LinkPote_0/\Volu)^n}\Volu_0\Volu^n
					\Big\lbrace
					1+\LinkDens \LocaPart (\LinkPote_0/\Volu)^n \nonumber\\
					&& \quad +\frac{1}{\Volu^n} \sum_{m=2}^{\infty}
					\frac{\big(\LinkDens \LocaPart (\LinkPote_0)^n\big)^m}{m!}
					\int_{\ISLL_1,\ldots,\ISLL_m}
					\prod _{j=1}^{m} \Big(\frac{\tISLL_j}{2\pi}\Big)^{\frac{nd}{2}}
					\Big(\frac{2\pi}{\tISLL_1+\tISLL_m}\Big)^{\frac{nd}{2}}
					(1+n\Contraction^2)^{\frac{(1-m)d}{2}}
				\Big\rbrace .
\end{eqnarray}
From this, $-\ln \MyZzero$ can obtained by making the small-$n$ expansion,
using the following equality:
\begin{eqnarray}
	\ln(x+ny+O(n^2))=\ln(x(1+n(y/x)+O(n^2)))=\ln x + n (y/x) +O(n^2),
\end{eqnarray}
so that we have
\begin{eqnarray}
	-\ln \MyZzero
	&=& -\ln\Volu_0 +n \Big\lbrace
		-\ln \Volu +(\LinkDens \LocaPart + e^{-\LinkDens \LocaPart}-1)
		\Big(\frac{d}{2}\big(\ln(2\pi)+\Contraction^2\big)-\ln \Volu \Big) \nonumber\\
		&&\quad -e^{-\LinkDens \LocaPart} \frac{d}{2} \sum_{m=1}^{\infty}
		\frac{\big(\LinkDens \LocaPart \big)^m}{m!}
		\ln \Big(\frac{\tISLL_1\cdots\tISLL_m}{\tISLL_1+\cdots+\tISLL_m}\Big)
	\Big\rbrace .
\end{eqnarray}

Therefore, the small-$n$ expansion of the stationary-point Hamiltonian $\HVOPSP$ to $O(n)$ is given by
\begin{eqnarray}
	\HVOPSP&=&\frac{\ExclVoluRN_0(0)\PartNumb^2}{2\Volu_0}
		+\frac{n\ExclVoluRN(0)\PartNumb^2}{2\Volu}
	-\PartNumb \BoltCons T \ln \Volu_0 -n\PartNumb \BoltCons T \ln \Volu
	+n\PartNumb \BoltCons T \Bigg\lbrace
		\UnivPara \big\lbrack
			\frac{d}{2}\big(\ln(2\pi)+\Contraction^2\big)-\ln\Volu
		\big\rbrack \nonumber\\
	&&-	\frac{\LinkDens \LocaPart^2}{2}\cdot\frac{d}{2}\int_{\ISLL_1,\ISLL_2}
		\ln\big( \frac{1}{\ISLL_1}+\frac{1}{\ISLL_2}+\frac{\LinkScal^2}{\BoltCons T} \big)
		-e^{-\LinkDens \LocaPart}\frac{d}{2}
		\sum_{m=1}^{\infty}\frac{(\LinkDens \LocaPart)^m}{m!}
		\int_{\ISLL_1,\ldots,\ISLL_2} \ln
		\Big( \frac{\tISLL_1\cdots\tISLL_m}{\tISLL_1 +\cdots +\tISLL_m} \Big)
	\Bigg\rbrace ,
\end{eqnarray}
\end{widetext}
where the parameter $\UnivPara$ is defined via
\begin{eqnarray}\label{EQ:UnivParaDef}
	\UnivPara\equiv -\frac{\LinkDens \LocaPart^2}{2}+\LinkDens \LocaPart
	+e^{-\LinkDens \LocaPart}-1.
\end{eqnarray}

\section{Hamiltonian of the Goldstone deformed order parameter}
\label{APP:HGS}
In this Appendix we calculate the value of the Hamiltonian for the Goldstone-deformed order parameter by inserting the Goldstone-deformed order parameter~(\ref{EQ:VOPGSR}) into the Hamiltonian~(\ref{EQ:HVOPHRS}), following a calculation similar to that in Appendix~\ref{APP:HSP}.
To obtain a description of the elasticity, we shall expand the Hamiltonian for small deformations, specifically in a series in the small, scalar variable that characterizes the replicated deformation field, viz., $\DefoScalPsi(z_1,z_2)\equiv
(\hat{\DefoPosi}(z_1)-\hat{\DefoPosi}(z_2))^2-(1+n)(z_1-z_2)^2$.

The quadratic term in the Hamiltonian is given by
\begin{widetext}
\begin{eqnarray}
	&& \frac{\PartNumb \LinkDens \BoltCons T}{2 \Volu^n \LinkPote_0}
			\sum_{\REPP\in HRS}\VOP_{\REPP} \VOP_{-\REPP}\LinkPoteRepl_{\REPP} \nonumber\\
	&=&\frac{\PartNumb \LinkDens \BoltCons T}{2 \Volu^n \LinkPote_0}
			\sum_{\REPP} (\LinkPote)^{1+n}
			e^{-\frac{\LinkScal^2\vert\REPP\vert^2}{2\BoltCons T}}
			\big\lbrace \LocaPart \int \frac{dz_1}{\Volu_0}\int _{\ISLL_1}
			e^{-\frac{\vert\REPP\vert^2}{2\ISLL_1}
			-i \REPP \cdot \hat{\DefoPosi}(z_1) }
			-\LocaPart\delta^{((1+n)d)}_{\REPP}\big\rbrace  \nonumber\\
	&&	\quad\times		\big\lbrace \LocaPart \int \frac{dz_2}{\Volu_0}\int _{\ISLL_2}
			e^{-\frac{\vert\REPP\vert^2}{2\ISLL_2}
			+i \REPP \cdot \hat{\DefoPosi}(z_2) }
			-\LocaPart\delta^{((1+n)d)}_{\REPP}\big\rbrace \nonumber\\
	&=&\frac{\PartNumb \LinkDens \BoltCons T\LocaPart^2(\LinkPote)^{n}}{2 \Volu^n} \nonumber\\
	&&		+\frac{\PartNumb \LinkDens \BoltCons T\LocaPart^2(\LinkPote)^{n}}{2 \Volu^n }
			\Volu_0 \Volu^n \int\frac{dz_1 dz_2}{\Volu_0^2}\int _{\ISLL_1,\ISLL_2}
			\Big\lbrace 2\pi \Big( \frac{1}{2\ISLL_1}+\frac{1}{2\ISLL_2}+
				\frac{\LinkScal^2}{2\BoltCons T}\Big)
			\Big\rbrace^{-\frac{(1+n)d}{2}}
			e^{-\frac{(\hat{\DefoPosi}(z_1)-\hat{\DefoPosi}(z_2))^2}
				{2\big( \frac{1}{2\ISLL_1}+\frac{1}{2\ISLL_2}+
					\frac{\LinkScal^2}{2\BoltCons T}\big)}} .
\end{eqnarray}
Next, we expand for small $\DefoScalPsi$, adopting the notation $\DefoScalPsi(z_1,z_2)\equiv
(\hat{\DefoPosi}(z_1)-\hat{\DefoPosi}(z_2))^2-(1+n)(z_1-z_2)^2$.
Note that $\DefoScalPsi$ is \emph{not} related to the deformation relative to the stationary point, this stationary point being characterized by the mean positions of the replicas of the particle
$\hat{z}_{\Contraction}=\{z,\Contraction z,\ldots,\Contraction z\}$.
Instead, $\DefoScalPsi$ describes deformations relative to the \lq\lq state right after linking\rq\rq\ (i.e., prior to relaxation), this state being characterized by the mean positions of the replicas of the particle
$\hat{z}=\{z,z,\ldots,z\}$,
as discussed in Sections~\ref{SEC:EnerGoldSton} and \ref{SEC:relaxation}.
The expansion of the quadratic term for small $\DefoScalPsi$ is given by
\begin{eqnarray}
	&& \frac{\PartNumb \LinkDens \BoltCons T}{2 \Volu^n \LinkPote_0}
			\sum_{\REPP\in HRS}\VOP_{\REPP} \VOP_{-\REPP}\LinkPoteRepl_{\REPP} \nonumber\\
	&=&\frac{\PartNumb \LinkDens \BoltCons T\LocaPart^2(\LinkPote)^{n}}{2 \Volu^n}
			\nonumber\\
	&&		+\frac{\PartNumb \LinkDens \BoltCons T\LocaPart^2(\LinkPote)^{n}}{2  }
			\int\frac{dz_1 dz_2}{\Volu_0}\int _{\ISLL_1,\ISLL_2}
			\Big\lbrace 2\pi \Big( \frac{1}{2\ISLL_1}+\frac{1}{2\ISLL_2}+
				\frac{\LinkScal^2}{2\BoltCons T}\Big)
			\Big\rbrace^{-(1+n)d/2}
			e^{-\frac{(1+n)(z_1-z_2)^2}
				{2\big( \frac{1}{2\ISLL_1}+\frac{1}{2\ISLL_2}+
					\frac{\LinkScal^2}{2\BoltCons T}\big)}} \nonumber\\
	&&\times \Bigg\lbrace
			1-\frac{\DefoScalPsi(z_1,z_2)}{2\big( \frac{1}{2\ISLL_1}+\frac{1}{2\ISLL_2}+
				\frac{\LinkScal^2}{2\BoltCons T}\big)}
			+\frac{1}{2}\Big(\frac{\DefoScalPsi(z_1,z_2)}
				{2\big( \frac{1}{2\ISLL_1}+\frac{1}{2\ISLL_2}+
				\frac{\LinkScal^2}{2\BoltCons T}\big)}\Big)^2
			+O(\DefoScalPsi(z_1,z_2)^3)
		\Bigg\rbrace .
\end{eqnarray}
The small-$n$ expansion on this quadratic term is then given by
\begin{eqnarray}
	&& \frac{\PartNumb \LinkDens \BoltCons T}{2 \Volu^n \LinkPote_0}
			\sum_{\REPP\in HRS}\VOP_{\REPP} \VOP_{-\REPP}\LinkPoteRepl_{\REPP} \nonumber\\
	&=&\frac{\PartNumb \LinkDens \BoltCons T\LocaPart^2}{2} \Bigg\lbrace
				n \Big\lbrace \ln \Volu -\frac{d}{2}\big(\ln(2\pi)+1\big)
				-\frac{d}{2}\int _{\ISLL_1,\ISLL_2}
				\ln \Big( \frac{1}{2\ISLL_1}+\frac{1}{2\ISLL_2}+
				\frac{\LinkScal^2}{2\BoltCons T}\Big) \Big\rbrace \nonumber\\
	&&\quad		+\int\frac{dz_1 dz_2}{\Volu_0}\int _{\ISLL_1,\ISLL_2}
				\Big\lbrace 2\pi \Big( \frac{1}{2\ISLL_1}+\frac{1}{2\ISLL_2}+
					\frac{\LinkScal^2}{2\BoltCons T}\Big)
				\Big\rbrace^{-d/2}
				e^{-\frac{(1+n)(z_1-z_2)^2}
				{2\big( \frac{1}{2\ISLL_1}+\frac{1}{2\ISLL_2}+
					\frac{\LinkScal^2}{2\BoltCons T}\big)}} \nonumber\\
	&&\quad\quad\times \Big\lbrace
				-\frac{\DefoScalPsi(z_1,z_2)}{2\big( \frac{1}{2\ISLL_1}+\frac{1}{2\ISLL_2}+
					\frac{\LinkScal^2}{2\BoltCons T}\big)}
				+\frac{1}{2}\Big(\frac{\DefoScalPsi(z_1,z_2)}
					{2\big( \frac{1}{2\ISLL_1}+\frac{1}{2\ISLL_2}+
					\frac{\LinkScal^2}{2\BoltCons T}\big)}\Big)^2
				+O(\DefoScalPsi(z_1,z_2)^3)
				\Big\rbrace
			\Bigg\rbrace .
\end{eqnarray}
The calculation for the $\ln \MyZzero$ term is similar to the above calculation of the quadratic term.  The expansion in small quantity $\DefoScalPsi$ reads
\begin{eqnarray}
	\MyZzero
			&=& e^{-\LinkDens \LocaPart (\LinkPote_0/\Volu)^n}\Volu_0\Volu^n
					\Big\lbrace
					1+\LinkDens \LocaPart (\LinkPote_0/\Volu)^n \nonumber\\
			&&\quad
					+\frac{1}{\Volu_0\Volu^n} \sum_{m=2}^{\infty}
					\frac{\big(\LinkDens \LocaPart (\LinkPote_0)^n\big)^m}{m!}
					\int dz_1 \cdots dz_m \int_{\ISLL_1,\ldots,\ISLL_m}
					\prod _{j=1}^{m} \Big(\frac{\tISLL_j}{2\pi}\Big)^{\frac{(1+n)d}{2}}
					\Big(\frac{2\pi}{\tISLL_1+\cdots+\tISLL_m}\Big)^{\frac{(1+n)d}{2}}
					e^{-\frac{\tISLL_1\tISLL_2(z_1-z_2)^2+\cdots}
					{2(\tISLL_1+\cdots+\tISLL_m)}} \nonumber\\
			&&\quad\times\big\lbrace
					1-\frac{\tISLL_1\tISLL_2 \DefoScalPsi(z_1,z_2)+\cdots}
					{2(\tISLL_1+\cdots+\tISLL_m)}
					+\frac{1}{2}\big(\frac{\tISLL_1\tISLL_2 \DefoScalPsi(z_1,z_2)+\cdots}
						{2(\tISLL_1+\cdots+\tISLL_m)}
					\big)^2
				\big\rbrace
				\Big\rbrace ,
\end{eqnarray}
where the summations that we have abbreviated with $\cdots$ include all pairs formed by $\{z_1,\ldots,z_m\}$.  Next, we expand for small $n$, keep terms to $O(n)$ in the $\ln \MyZzero$ term,
{\colorgreen assuming that both $\DefoScalPsi$ and $\DefoScalPsi^2$ contain $O(n)$ terms. }
After a tedious calculation we have
\begin{eqnarray}\label{EQ:AppLnz}
	-\ln \MyZzero
	&=& -\LinkDens \LocaPart (\frac{\LinkPote}{\Volu})^n
			-\ln \Volu_0 -n\ln \Volu - \LinkDens \LocaPart \nonumber\\
	&& -n e^{-\LinkDens \LocaPart} \Big\lbrace
				(1-e^{\LinkDens \LocaPart})\ln \Volu
				+\LinkDens\LocaPart e^{\LinkDens \LocaPart} \ln \Volu_0
				+\big(e^{\LinkDens \LocaPart}-1
					-\LinkDens\LocaPart e^{\LinkDens \LocaPart}\big)\frac{d}{2}
					\big( \ln(2\pi)+1 \big) \nonumber\\
	&&\quad\quad +\frac{d}{2}\sum_{m=1}^{\infty}\int_{\ISLL _1,\ldots\ISLL _m}
				\ln\Big(\frac{\tISLL_1\cdots\tISLL_m}{\tISLL_1+\cdots+\tISLL_m}\Big)
			\Big\rbrace			\nonumber\\
	&& -e^{-\LinkDens\LocaPart} \frac{1}{\Volu_0}
			\sum_{m=2}^{\infty}\frac{(\LinkDens \LocaPart)^m}{m!}\int dz_1\cdots dz_m
			\int_{\ISLL_1,\ldots,\ISLL_m} \prod_{j=1}^{m}
			\Big(\frac{\tISLL_j}{2\pi}\Big)^{\frac{d}{2}}
					\Big(\frac{2\pi}{\tISLL_1+\cdots+\tISLL_m}\Big)^{\frac{d}{2}}
					e^{-\frac{\tISLL_1\tISLL_2(z_1-z_2)^2+\cdots}
					{2(\tISLL_1+\cdots+\tISLL_m)}} \nonumber\\
	&&\quad\quad\times \Bigg\lbrace
					-\frac{\tISLL_1\tISLL_2 \DefoScalPsi(z_1,z_2)+\cdots}
					{2(\tISLL_1+\cdots+\tISLL_m)}
					+\frac{1}{2}\Big(\frac{\tISLL_1\tISLL_2 \DefoScalPsi(z_1,z_2)+\cdots}
						{2(\tISLL_1+\cdots+\tISLL_m)}
					\Big)^2
				\Bigg\rbrace .
\end{eqnarray}
To further simplify the expression, first consider the $O(\DefoScalPsi)$ terms in the expansion,
$-\frac{\tISLL_1\tISLL_2 \DefoScalPsi(z_1,z_2)+\cdots}{2(\tISLL_1+\cdots+\tISLL_m)}$.
The first term has a factor of $\DefoScalPsi(z_1,z_2)\equiv
(\hat{\DefoPosi}(z_1)-\hat{\DefoPosi}(z_2))^2-(1+n)(z_1-z_2)^2$,
which only involves two variables $z_1$ and $z_2$,
so we can integrate out the other $(m-2)$ variables, i.e., $z_3,\ldots,z_m$.
(Of course, for $m=2$, no integrals are needed.)\thinspace\
In total, there are $\frac{m(m-1)}{2}$ such terms
(i.e., the number of pairs among $m$ variables).
Thus, the $O(\DefoScalPsi)$ term in $-\ln \MyZzero$ is given by
\begin{eqnarray}\label{EQ:FirsOrdeInte}
	&& -e^{-\LinkDens\LocaPart} \frac{1}{\Volu_0}
			\sum_{m=2}^{\infty}\frac{(\LinkDens \LocaPart)^m}{m!}\int dz_1\cdots dz_m
			\int_{\ISLL_1,\ldots,\ISLL_m} \prod_{j=1}^{m}
			\Big(\frac{\tISLL_j}{2\pi}\Big)^{\frac{d}{2}}
					\Big(\frac{2\pi}{\tISLL_1+\cdots+\tISLL_m}\Big)^{\frac{d}{2}}
	\nonumber\\
	&&	\quad\times				e^{-\frac{\tISLL_1\tISLL_2(z_1-z_2)^2+\cdots}
					{2(\tISLL_1+\cdots+\tISLL_m)}}
			\Bigg\lbrace
					-\frac{\tISLL_1\tISLL_2 \DefoScalPsi(z_1,z_2)+\cdots}
					{2(\tISLL_1+\cdots+\tISLL_m)}
			\Bigg\rbrace \nonumber\\
	&=& -e^{-\LinkDens\LocaPart} \frac{1}{\Volu_0}
			\sum_{m=2}^{\infty}\frac{(\LinkDens \LocaPart)^m}{m!}\int dz_1\cdots dz_m
			\int_{\ISLL_1,\ldots,\ISLL_m} \prod_{j=1}^{m}
			\Big(\frac{\tISLL_j}{2\pi}\Big)^{\frac{d}{2}}
			\int dc
			\nonumber\\
	&&	\quad\times e^{-\frac{\tISLL_1}{2}(z_1-c)^2-\frac{\tISLL_2}{2}(z_2-c)^2-\cdots}
			\Bigg\lbrace
					-\frac{\tISLL_1\tISLL_2 \DefoScalPsi(z_1,z_2)+\cdots}
					{2(\tISLL_1+\cdots+\tISLL_m)}
			\Bigg\rbrace \nonumber\\
	&=& -e^{-\LinkDens\LocaPart} \frac{1}{\Volu_0}
			\sum_{m=2}^{\infty}\frac{(\LinkDens \LocaPart)^m}{m!} \frac{m(m-1)}{2}
			\int dz_1 dz_2 \int_{\ISLL_1,\ldots,\ISLL_m}
			\Big(\frac{\tISLL_1 \tISLL_2}{2\pi(\tISLL_1+\tISLL_2)}\Big)^{\frac{d}{2}}
			\nonumber\\
	&&\quad\times		e^{-\frac{\tISLL_1 \tISLL_2 (z_1-z_2)^2}{2(\tISLL_1+\tISLL_2)}}
			\Big\lbrace
				-\frac{\tISLL_1\tISLL_2 }
					{2(\tISLL_1+\cdots+\tISLL_m)}
			\Big\rbrace \DefoScalPsi(z_1,z_2) ,
\end{eqnarray}
where in the last line here we have used the fact that $\{z_1,z_2,\ldots,z_m\}$ appear symmetrically, so that the $\frac{m(m-1)}{2}$ terms are identical.

Similarly, for the $O(\DefoScalPsi^2)$ terms in the expansion in Eq.~(\ref{EQ:AppLnz})  there are
terms involving two points, such as $\DefoScalPsi(z_1,z_2)^2$,
three points, such as $\DefoScalPsi(z_1,z_2)\DefoScalPsi(z_1,z_3)$, and
four points, such as $\DefoScalPsi(z_1,z_2)\DefoScalPsi(z_3,z_4)$.
(Of course, for $m=3$ there are no four-point terms,
and for $m=2$ there are no three or four points terms.)\thinspace\
Thus, the $O(\DefoScalPsi^2)$ terms can be written as
\begin{eqnarray}\label{EQ:QuadExpa}
	&& 
		\Big\lbrace\frac{\tISLL_1\tISLL_2 \DefoScalPsi(z_1,z_2)+\cdots}
						{2(\tISLL_1+\cdots+\tISLL_m)}\Big\rbrace ^2 \nonumber\\
	&\to& 
		\frac{1} {4(\tISLL_1+\cdots+\tISLL_m)^2}
			\Big\lbrace \frac{m(m-1)}{2}\tISLL_1^2\tISLL_2^2 \DefoScalPsi(z_1,z_2)^2 \nonumber\\
	&&+  m(m-1)(m-2)\tISLL_1\tISLL_2^2\tISLL_3 \DefoScalPsi(z_1,z_2)\DefoScalPsi(z_2,z_3) \nonumber\\
	&&+ \frac{m(m-1)(m-2)(m-3)}{4}\tISLL_1\tISLL_2\tISLL_3\tISLL_4
			\DefoScalPsi(z_1,z_2)\DefoScalPsi(z_3,z_4)
			\Big\rbrace .
\end{eqnarray}
Following a calculation similar to that in Eq.~(\ref{EQ:FirsOrdeInte}), we can integrate out the integration variables that are not present in $\DefoScalPsi$, and thus obtain the $O(\DefoScalPsi^2)$ term in $\ln \MyZzero$.

Summing up the contributions from the quadratic term and the $\ln \MyZzero$ term, we arrive at the Hamiltonian of the Goldstone deformed state:
\begin{eqnarray}
	H_{\VOP}^{(G)}= \HVOPSP+H_{\VOP}^{\DefoScalPsi},
\end{eqnarray}
with $\HVOPSP$ being the Hamiltonian of the stationary point, and the increase of the Hamiltonian due to Goldstone deformation is given by
\begin{eqnarray}\label{EQ:HPsi}
	H_{\VOP}^{\DefoScalPsi}
	&=& -n\PartNumb \BoltCons T \frac{\UnivPara d}{2}(\Contraction^2-1)
		+ \frac{1}{2}\int dz_1 dz_2 	
			\KernOne(z_1,z_2)\DefoScalPsi(z_1,z_2) \nonumber\\
	&&	-\frac{1}{8\BoltCons T}\int dz_1 dz_2 dz_3 dz_4 \KernTwo(z_1,z_2,z_3,z_4)
		\DefoScalPsi(z_1,z_2)\DefoScalPsi(z_3,z_4)  .
\end{eqnarray}
The first term here, i.e.,
$-\PartNumb \BoltCons T \frac{\UnivPara d}{2}(\Contraction^2-1)$,
is present due to the fact that the expansion variable $\DefoScalPsi(z_1,z_2)$
measures departures from
the \emph{state right after linking},
not the stationary point, as we have previously discussed.
Note that $H_{\VOP}^{\DefoScalPsi}$ involves only the energy of shear deformation, because the Goldstone modes contain only pure shear deformation.
The energy of volume variations is in the stationary-point Hamiltonian part, which contains the variable contraction parameter $\Contraction$.

The kernels in Eq.~(\ref{EQ:HPsi}) are given by
\begin{eqnarray}\label{EQ:KoneApp}
	\frac{1}{2}\KernOne(z_1,z_2)
	&=& 	\frac{\PartNumb \LinkDens \BoltCons T \LocaPart^2}{4 \Volu_0} \int_{\tISLL_1,\tISLL_2}
				 \Big(2\pi\Big(\frac{1}{\tISLL_1}+\frac{1}{\tISLL_2}+\frac{\LinkScal^2}{\BoltCons T}\Big)\Big)^{-d/2}
				 \Big(\frac{1}{\tISLL_1}+\frac{1}{\tISLL_2}+\frac{\LinkScal^2}{\BoltCons T}\Big)^{-1}
				 e^{-\frac{(z_1-z_2)^2}{2\big(\frac{1}{\tISLL_1}+\frac{1}{\tISLL_2}+\frac{\LinkScal^2}{\BoltCons T}\big)}}
				\nonumber\\
	&& 		+\frac{\PartNumb \BoltCons T}{2 \Volu_0} e^{-\LinkDens \LocaPart}
				\sum_{m=2}^{\infty}\frac{(\LinkDens \LocaPart)^m}{m!}\frac{m(m-1)}{2}\int_{\tISLL_1,\ldots,\tISLL_m}
				\Big(\frac{\tISLL_1\tISLL_2}{2\pi(\tISLL_1+\tISLL_2)}\Big)^{d/2} \nonumber\\
	&&		\quad\quad\times
				e^{-\frac{\tISLL_1\tISLL_2(z_1-z_2)^2}{2(\tISLL_1+\tISLL_2)}}
				\frac{\tISLL_1\tISLL_2}{\tISLL_1+\cdots+\tISLL_m} ,
\end{eqnarray}
and
\begin{eqnarray}\label{EQ:KtwoApp}
	&&	-\frac{1}{8\BoltCons T}\KernTwo(z_1,z_2,z_3,z_4) \nonumber\\
	&=&	\frac{\PartNumb \LinkDens \BoltCons T \LocaPart^2}{16 \Volu_0} \int_{\tISLL_1,\tISLL_2}
			 \Big(2\pi\Big(\frac{1}{\tISLL_1}+\frac{1}{\tISLL_2}+\frac{\LinkScal^2}{\BoltCons T}\Big)\Big)^{-d/2}
			\Big(\frac{1}{\tISLL_1}+\frac{1}{\tISLL_2}+\frac{\LinkScal^2}{\BoltCons T}\Big)^{-1} \nonumber\\
	&&	\quad\quad\times
			 e^{-\frac{(z_1-z_2)^2}{2\big(\frac{1}{\tISLL_1}+\frac{1}{\tISLL_2}+\frac{\LinkScal^2}{\BoltCons T}\big)}}
			\delta^{(d)}(z_1-z_3)	\delta^{(d)}(z_2-z_4) \nonumber\\
	&&	-\frac{\PartNumb \BoltCons T}{8 \Volu_0} e^{-\LinkDens \LocaPart}
			\sum_{m=2}^{\infty}\frac{(\LinkDens \LocaPart)^m}{m!}\frac{m(m-1)}{2}\int_{\tISLL_1,\ldots,\tISLL_m}
			\Big(\frac{\tISLL_1\tISLL_2}{2\pi(\tISLL_1+\tISLL_2)}\Big)^{d/2} \nonumber\\
	&&	\quad\quad\times
			e^{-\frac{\tISLL_1\tISLL_2(z_1-z_2)^2}{2(\tISLL_1+\tISLL_2)}}
			 \frac{\tISLL_1^2\tISLL_2^2}{(\tISLL_1+\cdots+\tISLL_m)^2}\delta^{(d)}(z_1-z_3)	 \delta^{(d)}(z_2-z_4)
			\nonumber\\
	&&	-\frac{\PartNumb \BoltCons T}{8 \Volu_0} e^{-\LinkDens \LocaPart}
			\sum_{m=3}^{\infty}\frac{(\LinkDens \LocaPart)^m}{m!}m(m-1)(m-2)\int_{\tISLL_1,\ldots,\tISLL_m}
			 \Big(\frac{\tISLL_1\tISLL_2\tISLL_3}{4\pi^2(\tISLL_1+\tISLL_2+\tISLL_3)}\Big)^{d/2} \nonumber\\
	&&	\quad\quad\times
			 e^{-\frac{\tISLL_1\tISLL_2(z_1-z_2)^2+\tISLL_2\tISLL_3(z_2-z_3)^2+\tISLL_3\tISLL_1(z_3-z_1)^2}
				{2(\tISLL_1+\tISLL_2+\tISLL_3)}}
			 \frac{\tISLL_1\tISLL_2^2\tISLL_3}{(\tISLL_1+\cdots+\tISLL_m)^2}\delta^{(d)}(z_2-z_4)
			\nonumber\\
	&&	-\frac{\PartNumb \BoltCons T}{8 \Volu_0} e^{-\LinkDens \LocaPart}
			\sum_{m=3}^{\infty}\frac{(\LinkDens \LocaPart)^m}{m!}\frac{m(m-1)(m-2)(m-3)}{4}\int_{\tISLL_1,\ldots,\tISLL_m}
			 \Big(\frac{\tISLL_1\tISLL_2\tISLL_3\tISLL_4}{8\pi^3(\tISLL_1+\tISLL_2+\tISLL_3+\tISLL_4)}\Big)^{d/2} \nonumber\\
	&&	\quad\quad\times
			 e^{-\frac{\tISLL_1\tISLL_2(z_1-z_2)^2+\tISLL_1\tISLL_3(z_1-z_3)^2+\tISLL_1\tISLL_4(z_1-z_4)^2
					 +\tISLL_2\tISLL_3(z_2-z_3)^2+\tISLL_2\tISLL_4(z_2-z_4)^2+\tISLL_3\tISLL_4(z_3-z_4)^2}
				{2(\tISLL_1+\tISLL_2+\tISLL_3+\tISLL_4)}}
			\frac{\tISLL_1\tISLL_2\tISLL_3\tISLL_4}{(\tISLL_1+\cdots+\tISLL_m)^2} .
\end{eqnarray}
Strictly speaking, the kernel $\KernTwo$ should be symmetric under the exchanges of variables $z_1 \leftrightarrow z_2$ or $z_3 \leftrightarrow z_4$. Here, to save space, we have written the above non-symmetric form.  The true (i.e., symmetric) form can be recovered by averaging:
\begin{eqnarray}\label{EQ:KSym}
	\KernTwo(z_1,z_2,z_3,z_4) \to
    \frac{1}{4} \big(
	\KernTwo(z_1,z_2,z_3,z_4)+
    \KernTwo(z_1,z_2,z_4,z_3)+
    \KernTwo(z_2,z_1,z_3,z_4)+
    \KernTwo(z_2,z_1,z_4,z_3)
	\big) .
\end{eqnarray}

\end{widetext}

\section{Relaxation of the phenomenological elastic free
energy for a given realization of disorder}
\label{APP:Relaxation}
In this Appendix we solve the stationarity condition for the random local deformations $\RelaRand$.  First, we need to calculate the variation of the \lq\lq bulk term\rlap,\rq\rq\ which can be expanded, to leading order in small $\RelaRand$, as \footnote{A similar expansion, but to higher order in $\RelaRand$, is performed in Eq.~(\ref{EQ:EXPDet}), in terms of the strain tensor $\StraTensT$.}
\begin{eqnarray}\label{EQ:DetPPExp}
	\textrm{det}\Big(\frac{\partial\DefoPosi_i(z)}{\partial z_j}\Big)
	&=& \textrm{det} \Big( \Contraction \delta_{ij} + \partial_j \RelaRand_i(z) \Big) \nonumber\\
	&=& \Contraction^{d} \textrm{det} \Big( \delta_{ij} + \Contraction^{-1}\partial_j \RelaRand_i(z) \Big) \nonumber\\
	&\simeq& \Contraction^{d} \big(1+\Contraction^{-1} \partial_i \RelaRand_i(z)\big).
\end{eqnarray}
Using this expansion, we have
\begin{eqnarray}\label{EQ:BulkTermExp}
	&& \Big\lbrace \textrm{det}\Big(\frac{\partial\DefoPosi_i(z)}{\partial z_j}\Big)-1 \Big\rbrace^2 \nonumber\\
	&=& (\Contraction^{d}-1)^2 + 2(\Contraction^d-1)\Contraction^{d-1}\partial_i \RelaRand_i(z) \nonumber\\
	&&	+ \Contraction^{2d-2}\partial_i \RelaRand_i(z)\partial_j \RelaRand_j(z) .
\end{eqnarray}
Thus, the stationarity equation reads
\begin{widetext}
\begin{eqnarray}
	0 &=& 2(\Contraction z_a + \RelaRand_a(z))\int dz_2 \NonLocaKern(z,z_2)
		-2\!\int dz_2\NonLocaKern(z,z_2)(\Contraction z_{2,a} + \RelaRand_a(z_2))
			-\BulkModuZeroP \partial_{a} (\partial_i \RelaRand_i(z)) ,
\end{eqnarray}
where
\begin{align}
	\BulkModuZeroP \equiv \BulkModuZero \Contraction^{2d-2} .
\end{align}
We take the disorder average of the nonlocal kernel $\NonLocaKernZero$ to be a zeroth-order quantity, and the fluctuation part $\NonLocaKernOne$ to be a first-order quantity and, thus, $\RelaRand(z)$ is also of first order.  The he zeroth-order equation is then
\begin{eqnarray}
	0=2z_a\int dz_2 \NonLocaKernZero(z-z_2)-2\int dz_2 \NonLocaKernZero(z-z_2)z_{2,a}\, ,
\end{eqnarray}
which is automatically satisfied, given that $\NonLocaKernZero(z-z_2)$ is even in $(z-z_2)$.

The first order equation reads
\begin{eqnarray}
	\! 0\! = \Contraction z_a \! \int\! dz_2 \NonLocaKernZero(z,z_2)
			\!+ \RelaRand_a(z)\!\! \int\! dz_2 \NonLocaKernZero(z-z_2)
			\!- \!\!\int \!\! dz_2 \NonLocaKernOne(z,z_2)\Contraction z_{2,a}
			\!- \!\int \!\! dz_2 \NonLocaKernZero(z-z_2) \RelaRand_a(z)
			\!- \frac{\BulkModuZeroP}{2} \partial_{a} (\partial_i \RelaRand_i(z)) . \,
\end{eqnarray}
We address this equation in momentum space.  We define the following Fourier transforms (on a specific finite volume---the volume of the state right after linking, viz., $\Volu_0$):
\begin{eqnarray}
	\NonLocaKernZero_{p} &=& \int dx e^{-ipx} \NonLocaKernZero (x) , \nonumber\\
	\NonLocaKernOne_{p_1,p_2} &=& \int dx e^{-ip_1 x_1-ip_2 x_2}
			\NonLocaKernZero (x_1,x_2) ,
\end{eqnarray}
so that the momentum-space stationarity equation becomes
\begin{eqnarray}
\label{mom-space-stat}
	0 &=& i \Contraction \frac{\partial}{\partial p_{1,a}} \NonLocaKernOne_{p_1,0}
				-i \Contraction \frac{\partial}{\partial p_{2,a}} \Big\vert_{p_2=0}
						 \NonLocaKernOne_{p_1,p_2}
				+ (\NonLocaKernZero_{0}-\NonLocaKernZero_{p_1})\RelaRand_{a,p_1}
				+ \frac{\BulkModuZeroP}{2}p_{1,a} p_{1,b}\RelaRand_{b,p_1} .
\end{eqnarray}
\end{widetext}
Strictly speaking, the derivatives here should instead be understood as difference quotients because we are using a finite-volume version of the Fourier transform; but for convenience we write it a derivatives.

Equation~(\ref{mom-space-stat}) can be written in the tensorial form
\begin{eqnarray}\label{EQ:OrdeOneMomeTens}
	\Big\lbrace
		2\Big( \frac{\DiffG_p}{p^2}\Big) \IdenT
		+\BulkModuZeroP \PLongT
	\Big\rbrace \cdot \vert p \vert ^2 \vec{\RelaRand}(p)
	= \vec{\RandForc}(p) ,
\end{eqnarray}
where
\begin{eqnarray}\label{EQ:VectRelaRand}
	\RandForc_{a,p_1}\equiv -2\Contraction \Big(
		i\frac{\partial}{\partial p_{1,a}} \NonLocaKernOne_{p_1,0}
		-i\frac{\partial}{\partial p_{2,a}}\Big\vert_{p_2=0}  \NonLocaKernOne_{p_1,p_2}
	\Big).
\end{eqnarray}
This quantity $\RandForc_{a,p_1}$ is actually the random force in the state that is contracted but has not yet been equilibrated for the randomness, and
\begin{eqnarray}
	\DiffG_p &\equiv& \NonLocaKernZero_{0}-\NonLocaKernZero_{p}.
\end{eqnarray}
Furthermore, $\IdenT$ is the $\Dime$-dimensional identity matrix, and the projection operators in momentum space, $\PLong_{ij}$ and $\PPerp_{ij}$, are defined as
\begin{eqnarray}\label{EQ:DefiProj}
	\PLong_{ij} &\equiv& (p_i\,p_j)/p^2 , \nonumber\\
	\PPerp_{ij} &\equiv& \delta_{i,j}-(p_i\,p_j)/p^2;
\end{eqnarray}
They satisfy the following relations:
\begin{eqnarray}
	(\PLongT)^2 = \PLongT, \quad
	(\PPerpT)^2 = \PPerpT, \quad
	\PLongT \cdot \PPerpT =0 .
\end{eqnarray}
In the following we shall use bold-face letters to denote rank-2 tensors, and letters with an overhead arrow (such as $\vec{\RelaRand}(p)$) to denote a vector, when needed.

By this decomposition we arrive at the solution to Eq.~(\ref{EQ:OrdeOneMomeTens}):
\begin{eqnarray}\label{EQ:SoluRelaRand}
	\vec{\RelaRand}_{p}
	&=& \frac{\PPerpT \cdot \vec{\RandForc}_{p}}
				{2\DiffG_p}
			+\frac{\PLongT \cdot \vec{\RandForc}_{p}}
				{\BulkModuZeroP + 2\DiffG_p} .
\end{eqnarray}
Notice that the second term is much smaller than the first term, due to the large bulk modulus $\BulkModuZeroP$. In the incompressible limit (i.e., $\BulkModuZero \to \infty$), we have
\begin{eqnarray}
	\vec{\RelaRand}_{p}=\frac{\PPerpT \cdot \vec{\RandForc}_{p}}
				{2\DiffG_p} ,
\end{eqnarray}
which is a purely transverse field, meaning that it satisfies $p_i\,\RelaRand_{i,p}=0$ or, equivalently, $\partial_i\,\RelaRand_{i}(x)=0$, which is the only deformation allowed in an incompressible medium.

\section{Re-expanding the elastic energy around the
equilibrium reference state}
\label{APP:ReExpandFreeEner}
In this Appendix we re-expand the elastic energy for deformations relative to the relaxed state, $\tz=\Contraction z + \RelaRand(z)$, as discussed in Section~\ref{SEC:EFERS}. The small variable in this expansion is the deformation field $\tu(\tz)$ relative to the relaxed state.
Furthermore, to obtain a continuum description of the elasticity, we adopt a notation involving the strain tensor $\StraTens_{ij}(x)$:
\begin{eqnarray}
	\!\StraTens_{ij}(x)\!&\equiv&\! \frac{1}{2}(\DefoGrad_{ij}(x)\DefoGrad_{ij}(x)-\delta_{ij})
	\nonumber\\
	\!&=&\! \frac{1}{2}(\partial_i \deformation_j(x)\!+\!\partial_j \deformation_i(x)
		\!+\!\partial_i \deformation_l(x)\partial_j \deformation_l(x) ).
\end{eqnarray}
where $\DefoGrad_{ij}(x)\equiv \partial \DefoPosi_i(x)/\partial x_j$ is the deformation gradient tensor. This strain tensor transforms as a tensor in the reference space labeled by $x$, and as a scalar in the target space labeled by $\DefoPosi$.

\subsection{Expanding the nonlocal kernel $\tNonLocaKern$
in the relaxed state}
\label{SEC:NLKRR}
The definition of $\tNonLocaKern$, given in Section~\ref{SEC:EFERS}, is
\begin{eqnarray}
	\tNonLocaKern(\tz_1,\tz_2)&\equiv&\NonLocaKern(z(\tz_1),z(\tz_2)) .
\end{eqnarray}
It can be expanded for small $\RelaRand$ to yield a direct expression for $\tNonLocaKern$.  In momentum-space this reads
\begin{widetext}
\begin{align}
	\tNonLocaKern_{\tp_1,\tp_2}
	=& \int d\tz_1 d\tz_2 e^{-i\tp_1 \tz_1-i\tp_2 \tz_2}
			\tNonLocaKern(\tz_1,\tz_2)
			= \!\int d\tz_1 d\tz_2 e^{-i\tp_1 \tz_1-i\tp_2 \tz_2}
				\NonLocaKern(z(\tz_1),z(\tz_2)) \nonumber\\
	=& \int dz_1 dz_2 \Jaco(z_1) \Jaco(z_2)
			e^{-i\tp_1 (\Contraction z_1+\RelaRand(z_1))
					-i\tp_2 (\Contraction z_2+\RelaRand(z_2))}
			\NonLocaKern(z_1,z_2) ,
\end{align}
where, in the first line, $z(\tz_1)$ is the mapping of a mass point $\tz_1$ in the relaxed state back to the position $z(\tz_1)$, at which it was located in the state right after linking.
Inserting in the expressions for $\Contraction$ and $\RelaRand(z)$,
given in Eqs.~(\ref{EQ:SoluCont}, \ref{EQ:SoluV}),
and keeping terms to $O((1/\BulkModuZero)^0)$
[which gives $\Contraction \simeq 1$ and $\Jaco(z)\simeq 1$],
we can expand $\RelaRand$ down from the exponent, and keep terms to first order in $\NonLocaKernOne$
(noting that $\RelaRand$ is the same order as $\NonLocaKernOne$),
and thus arrive at
\begin{align}\label{EQ:tNonLocaKern}
	\tNonLocaKern_{\tp_1,\tp_2}
	\simeq& \int dz_1 dz_2
			\big(1-i\tp_1 \RelaRand(z_1)
					-i\tp_2 \RelaRand(z_2)\big) e^{-i\tp_1 z_1-i\tp_2 z_2}
			(\NonLocaKernZero(z_1,z_2)+\NonLocaKernOne(z_1,z_2)) \nonumber\\
	\simeq& \,\,\NonLocaKernZero_{\tp_1,\tp_2}+\NonLocaKernOne_{\tp_1,\tp_2}
			-i \int dz_1 dz_2 \big(\tp_1 \RelaRand(z_1)
					+\tp_2 \RelaRand(z_2)\big) e^{-i\tp_1 z_1-i\tp_2 z_2}
						\NonLocaKernZero(z_1,z_2) \nonumber\\
	=& \,\, \NonLocaKernZero_{\tp_1,\tp_2}+\NonLocaKernOne_{\tp_1,\tp_2}
			-i \big(
				\tp_1\cdot\RelaRand_{(\tp_1+\tp_2)} \NonLocaKernZero_{\tp_2}
				+\tp_2\cdot\RelaRand_{(\tp_1+\tp_2)} \NonLocaKernZero_{\tp_1}
			\big) .
\end{align}

\subsection{Local expansion of the harmonic attraction}
\label{APP:REHT}
In this section we make a local expansion of the nonlocal term in the elastic free energy of the equilibrium reference state, Eq.~(\ref{EQ:REExpaEner}), i.e., the term
\begin{align}\label{EQ:FNL}
	\FreeEnerPhen_{\textrm{nonlocal}} = & \,
		\frac{1}{2} \!\int\! d\tz_1 d\tz_2 \Jaco(z_1)^{-1}\Jaco(z_2)^{-1}
			\tNonLocaKern (\tz_1,\tz_2)  	
		\Big(
				\big\vert \tDefoPosi(\tz_1)-\tDefoPosi(\tz_2)\big\vert^2
				- \big\vert \tz_1-\tz_2\big\vert^2
			\Big) .
\end{align}
For convenience of notation, we define the following change of variables:
\begin{eqnarray}
	z&=&z_1 ,\nonumber\\
	y&=&z_2-z_1 ,\nonumber\\
	\NLM(z,y)&\equiv&\tNonLocaKern(z_1,z_2),
\end{eqnarray}
so that the nonlocal kernel in the relaxed state, Eq.~(\ref{EQ:tNonLocaKernText}), can be written (in momentum space) as
\begin{eqnarray}\label{EQ:tMinM}
	\tNLM_{\tp,\tq}&\simeq&\NLMZero_{\tp,\tq}+\NLMOne_{\tp,\tq}
	 + \tg_{\tp,\tq}\big( \RandForc_{\tp} \cdot \PPerpT \cdot \tq \big) ,
\end{eqnarray}
with the definition of $\tg_{\tp,\tq}$,
and then its leading-order expansion in momentum,
being given by
\begin{eqnarray}
	\tg_{\tp,\tq}&\equiv&\frac{i(\NonLocaKernZero_{\tq}-\NonLocaKernZero_{\tp-\tq})}
			{2(\NonLocaKernZero_{0}-\NonLocaKernZero_{\tp})} \simeq \frac{i(\tp^2-2\tp\cdot\tq)}{2\tp^2} .
\end{eqnarray}
The local expansion of Eq.~(\ref{EQ:FNL}) then becomes
\begin{eqnarray}
		\FreeEnerPhen_{\textrm{nonlocal}}
	&=&
		\frac{1}{2} \int d\tz d\ty \, \tNLM (\tz,\ty)  \Big(
				\big\vert \tDefoPosi(\tz)-\tDefoPosi(\tz+\ty)\big\vert^2
				- \big\vert y \big\vert^2
		\Big) \nonumber\\
	&\simeq& \frac{1}{2} \int d\tz
		\big(	\partial _i \tDefoPosi_l(\tz)\partial_j \tDefoPosi_l(\tz) -\delta_{ij}
		\big)  \int d\ty \,\, \ty_{i}\,\ty_{j} \, \tNLM(\tz,\ty),
\end{eqnarray}
where the factor $\Jaco(z_1)^{-1}\Jaco(z_2)^{-1}$ is ignored because its difference from unity is of $O(1/\BulkModuZero)$.
Now it is straightforward to express the elastic energy $\FreeEnerPhen_{\textrm{nonlocal}}$ in the standard form of Lagrangian elasticity using the strain tensor $\tStraTens_{ij}(\tz)=\frac{1}{2} \big( \partial _i \tDefoPosi_l(\tz)\partial_j \tDefoPosi_l(\tz) -\delta_{ij} \big)$.
		
The complete expression of the local form of the elastic energy for deformations relative to the relaxed state, including the contribution from the bulk term, will be calculated in Appendix~\ref{APP:Lagr}.

\subsection{Expansion of the bulk term}
The \lq\lq bulk term\rq\rq~in the elastic free energy Eq.~(\ref{EQ:REExpaEner}) is given by
\begin{eqnarray}
	\FreeEnerPhen_{\textrm{bulk}} \equiv \frac{\BulkModuZero}{2} \int d\tz \Jaco(z)^{-1}
			\Big\lbrace
				\Jaco(z)\, \textrm{det} \Big( \frac{\partial \tDefoPosi_i(\tz)}{\partial \tz_j}
					\Big)-1
			\Big\rbrace^2 .
\end{eqnarray}
The determinant in this equation can be expanded
using the strain tensor $\tStraTensT$:
\begin{eqnarray}\label{EQ:EXPDet}
	\textrm{det} \Big( \frac{\partial \tDefoPosi_i(\tz)}{\partial \tz_j}\Big)
	&=& \textrm{det} \big(\tDefoGradT(\tz)\big)
		= \big\lbrace \textrm{det} \big(\IdenT+2\tStraTensT(\tz)\big)\big\rbrace^{1/2}
		= e^{\frac{1}{2}\textrm{Tr}\, \ln \big(\IdenT+2\tStraTensT(\tz)\big)} \nonumber\\
	&=& 1+ \textrm{Tr} \tStraTensT(\tz) -\textrm{Tr} \tStraTensT(\tz)^2
			+\frac{1}{2} \big(\textrm{Tr} \tStraTensT(\tz)\big)^2 + O(\tStraTensT(\tz)^3).
\end{eqnarray}
Thus, we have
\begin{align}
	\FreeEnerPhen_{\textrm{bulk}}
	=&\, \frac{\BulkModuZero}{2} \int d\tz \Jaco(z)^{-1}
			\Big\lbrace
				\Jaco(z) \textrm{det} \Big( \frac{\partial \tDefoPosi_i(\tz)}{\partial \tz_j}
					\Big)-1
			\Big\rbrace^2 \nonumber\\
	\simeq&\, \frac{\BulkModuZero}{2} \int d\tz \Jaco(z)^{-1}
			\Big\lbrace
				(\Jaco(z)-1)^2 +2(\Jaco(z)-1)\Jaco(z) \textrm{Tr} \tStraTensT (\tz)
				 \nonumber\\
	& \quad\quad\quad		- 2(\Jaco(z)-1)\Jaco(z) \textrm{Tr} \tStraTensT (\tz)^2
		+ (2\Jaco(z)-1)\Jaco(z) (\textrm{Tr} \tStraTensT (\tz))^2
			\Big\rbrace  .
\end{align}
Inserting the solutions for $\Contraction$ and $\RelaRand$, given in Eqs.~(\ref{EQ:SoluCont}, \ref{EQ:SoluV}), into the Jacobian $\Jaco (z) \equiv \big\vert \frac{\partial \tz_i}{\partial z_j}\big\vert$, we arrive at
\begin{eqnarray}
	\FreeEnerPhen_{\textrm{bulk}}
	= \int d\tz \big\lbrace
				\textrm{Tr}(\StreBulkT(\tz)\cdot\tStraTensT(\tz))
						+ \SheaModu(\tz) \textrm{Tr} \tStraTensT (\tz)^2
				+ \frac{\BulkModu(\tz)}{2} (\textrm{Tr} \tStraTensT (\tz))^2
			\big\rbrace ,
\end{eqnarray}
with the elastic parameters (in momentum space) being given by
\begin{subequations}
\begin{eqnarray}
	\StreBulk_{ij,p}
	&=&\delta_{ij}\Big\lbrace
										\frac{i\tp\cdot\vec{\RandForc}_{\tp}}{\tp^2}
										-\MeanSheaModu\Volu_0\delta_{\tp}
								\Big\rbrace , \\
	\SheaModu_{\tp}
	&=& \MeanSheaModu\Volu_0\delta_{\tp} -
			\frac{i\tp\cdot\vec{\RandForc}_{\tp}}{\tp^2} , \\
	\BulkModu_{\tp}
	&=& \BulkModuZero\Volu_0\delta_{\tp} + 2\Big\lbrace
										\frac{i\tp\cdot\vec{\RandForc}_{\tp}}{\tp^2}
										-\MeanSheaModu\Volu_0\delta_{\tp}
								\Big\rbrace .
\end{eqnarray}
\end{subequations}

\subsection{Local form of the elastic energy relative
to the relaxed state}
\label{APP:Lagr}
Summing up the contributions from the nonlocal term $\FreeEnerPhen_{\textrm{nonlocal}}$ and the bulk term $\FreeEnerPhen_{\textrm{bulk}}$ to the elastic free energy~(\ref{EQ:REExpaEner}), we arrive at the local form of the elastic energy for deformations relative to the relaxed state:
\begin{eqnarray}
	\FreeEnerPhen
	= \int d\tz \big\lbrace
				\textrm{Tr}(\StreT(\tz)\cdot\tStraTensT(\tz))
					+ \SheaModu(\tz) \textrm{Tr} \tStraTensT (\tz)^2
				+ \frac{\BulkModu(\tz)}{2} (\textrm{Tr} \tStraTensT (\tz))^2
			\big\rbrace ,
\end{eqnarray}
with the elastic parameters being given by
\begin{subequations}
\begin{align}
	\Stre_{ij,\tp}
	=&\, -\frac{\partial^2}{\partial \tq_i \partial \tq_j}
		\Big\vert_{q=0} \NonLocaKernOne_{\tp-\tq,\tp}
			+ i \delta_{ij} \frac{i\tp\cdot\vec{\RandForc}_{\tp}}{\vert \tp \vert ^2}
			-\frac{\RandForc_{a,\tp}}{\vert \tp \vert ^2}\big(
				\tp_{i} \PPerp_{ja,\tp} + \tp_{j} \PPerp_{ia,\tp}
			\big) , \\
	\SheaModu_{\tp}
	=&\, \MeanSheaModu \Volu_0 \delta_{\tp} -
			\frac{i\tp\cdot\vec{\RandForc}_{\tp}}{\vert \tp \vert ^2} , \\
	\BulkModu_{\tp}
	=&\, \BulkModuZero \Volu_0 \delta_{\tp} + 2\Big\lbrace
										\frac{i\tp\cdot\vec{\RandForc}_{\tp}}{\vert \tp \vert ^2}
										-\MeanSheaModu \Volu_0 \delta_{\tp}
								\Big\rbrace .
\end{align}
\end{subequations}
\end{widetext}

\section{Relaxation of the deformed state}
\label{APP:DefoRela}
In this section we solve the stationarity equation with a given macroscopic deformation $\DefoGrad$, as discussed in Section~\ref{SEC:NAD}, and thus obtain information about nonaffine deformations.
The stationarity condition is given by
\begin{widetext}
\begin{eqnarray}
	2(\Contraction \DefoGrad_{ai}z_i+(\RelaRandL)_{a}(z)) \int dz_2 \NonLocaKern (z,z_2)
	-2 \int dz_2 \NonLocaKern (z,z_2) (\Contraction \DefoGrad_{ai}z_{2,i}+(\RelaRandL)_{a}(z_2))
	- \BulkModuZeroP \DefoGrad^{-1}_{ia}\DefoGrad^{-1}_{jb}\partial_i \partial_j (\RelaRandL)_{b}(z) =0 .
\end{eqnarray}
We take $\NonLocaKernZero$ to be of zeroth order, and $\NonLocaKernOne$ and $\RelaRandL(z)$ to be first-order quantities.  Thus, the zeroth order equation reads
\begin{eqnarray}
	0=2\DefoGrad_{ai}z_i\int dz_2 \NonLocaKernZero(z-z_2)-2\DefoGrad_{ai}\int dz_2 \NonLocaKernZero(z-z_2)z_{2,i} ,
\end{eqnarray}
which is already satisfied given $\NonLocaKernZero(z-z_2)$ is even in $(z-z_2)$.

The first-order equation reads
\begin{eqnarray}
	&&	\Contraction \DefoGrad_{ai}z_i \int dz_2 \NonLocaKernOne (z,z_2)
			+(\RelaRandL)_{a}(z) \int dz_2 \NonLocaKernZero (z,z_2)
			- \int dz_2 \NonLocaKernOne (z,z_2) \Contraction \DefoGrad_{ai}z_{2,i}
			- \int dz_2 \NonLocaKernZero (z,z_2) (\RelaRandL)_{a}(z_2) \nonumber\\
	&&	- \frac{\BulkModuZeroP}{2} \DefoGrad^{-1}_{ia}\DefoGrad^{-1}_{jb}\partial_i \partial_j (\RelaRandL)_{b}(z) =0 .
\end{eqnarray}
We can solve this equation in momentum space, where it is expressed as
\begin{eqnarray}
	0 &=& i \Contraction \DefoGrad_{ai} \frac{\partial}{\partial p_{1,i}} \NonLocaKernOne_{p_1,0}
				-i \Contraction \DefoGrad_{ai} \frac{\partial}{\partial p_{2,i}} \Big\vert_{p_2=0}
						 \NonLocaKernOne_{p_1,p_2}
				+ (\NonLocaKernZero_{0}-\NonLocaKernZero_{p_1})(\RelaRandL)_{a,p_1}
				+ \frac{\BulkModuZeroP}{2} \DefoGrad^{-1}_{ia}\DefoGrad^{-1}_{jb} p_{1,i} p_{1,j}\RelaRand_{b,p_1} .
\end{eqnarray}
Writing this equation in tensorial form, we have
\begin{eqnarray}\label{EQ:DefoRelaOOne}
	\Bigg\lbrace
		2\Big( \frac{D_p}{\vert p \vert ^2}\Big) \IdenT
		+\BulkModuZeroP \DefoGradT^{-T} \PLongT \DefoGradT^{-1}
	\Bigg\rbrace \cdot \vert p \vert ^2 (\Vect{\RelaRandL})_p
	= (\Vect{\RandForcL})_p ,
\end{eqnarray}
where $\MetrTens=\DefoGrad^{T} \DefoGrad$ is the metric tensor, and
\begin{subequations}
\begin{eqnarray}
	D_p &\equiv& \NonLocaKernZero_{0}-\NonLocaKernZero_{p}, \\
	(\RandForcL)_{a,p_1} &\equiv& -2\Contraction \Big(
		i \DefoGrad_{ai} \frac{\partial}{\partial p_{1,i}} \NonLocaKernOne_{p_1,0}
		-i \DefoGrad_{ai} \frac{\partial}{\partial p_{2,i}} \NonLocaKernOne_{p_1,p_2}
	\Big) .
\end{eqnarray}
\end{subequations}
\end{widetext}
To solve Eq.~(\ref{EQ:DefoRelaOOne}),
it is useful to define the $\DefoGradT$-deformed versions
of the projection operators, i.e.,
\begin{subequations}
\begin{eqnarray}
	\PLongL &\equiv& \frac{1}{\textrm{Tr}(\PLongT \MetrTens^{-1})}
    \DefoGradT^{-\textrm{T}} \PLongT \DefoGradT^{-1} , \\
	\PPerpL &\equiv& \IdenT - \PLongL .
\end{eqnarray}
\end{subequations}
It is straightforward to verify that they obey
\begin{eqnarray}
	(\PLongL)^2 = \PLongL, \quad
	(\PPerpL)^2 = \PPerpL, \quad
	\PLongL \cdot \PPerpL =0 .
\end{eqnarray}
By using these projection operators we can write Eq.~(\ref{EQ:DefoRelaOOne}) as
\begin{eqnarray}
	\Big\lbrace
		 \frac{2 D_p}{\vert p \vert ^2} \PPerpL
		+\Big(\frac{2 D_p}{\vert p \vert ^2}+\BulkModuZeroP \trOne \Big)  \PLongL
	\Big\rbrace\! \cdot \vert p \vert ^2 (\Vect{\RelaRandL})_p
	= (\Vect{\RandForcL})_p ,
\end{eqnarray}
where we have defined
\begin{eqnarray}
	\trOne  \equiv  \textrm{Tr}(\PLongT\MetrTens^{-1}) .
\end{eqnarray}
Thus, it is straightforward to arrive at the solution:
\begin{eqnarray}
	(\Vect{\RelaRandL})_p = \Bigg\lbrace \frac{\PPerpL}{2 D_p}
		+ \frac{\PLongL}{\BulkModuZeroP \trOne \vert p \vert ^2+2 D_p}
	\Bigg\rbrace \cdot (\Vect{\RandForcL})_p .
\end{eqnarray}

\section{Correlation functions of the elastic parameters in the
equilibrium reference state}
\subsection{Correlation function of the non-local kernel
in the equilibrium reference state}
\label{APP:MM}
The nonlocal kernel in the equilibrium reference state $\tNonLocaKern$
is related to
the nonlocal kernel in the state right after linking $\NonLocaKern$
via Eq.~(\ref{EQ:tGtu});
to leading-order in the small quantity $\RelaRand$ we have
\begin{widetext}
\begin{eqnarray}
	\tNonLocaKern_{\tp_1,\tp_2}
	\simeq  \NonLocaKernZero_{\tp_1,\tp_2}+\NonLocaKernOne_{\tp_1,\tp_2}
		-i \big(
				\tp_1\cdot\vec{\RelaRand}_{(\tp_1+\tp_2)} \NonLocaKernZero_{\tp_2}
				+\tp_2\cdot\vec{\RelaRand}_{(\tp_1+\tp_2)} \NonLocaKernZero_{\tp_1}
			\big) .
\end{eqnarray}
By using this relation, we can derive the correlation function of $\tNonLocaKern$ from the correlation function of $\NonLocaKern$ [which is given in Eq.~(\ref{EQ:KTwo})], and thus we arrive at the correlation function
\begin{eqnarray}
	&&	\lda \tNLM_{p_1,q_1}\tNLM_{p_2,q_2} \rda_c \nonumber\\
	&=& \delta_{p_1+p_2} \Big\lbrace
				-\frac{\PartNumb \LinkDens \LocaPart^2}{2\BoltCons T} \int_{\ISLL_1,\ISLL_2}
				\Big(\frac{1}{\ISLL_1}+\frac{1}{\ISLL_2}+\LinkScal^2\Big)^{-2}
				\Big(
					e^{-\frac{1}{2}\big(\frac{1}{\ISLL_1}
						+\frac{1}{\ISLL_2}+\LinkScal^2 \big)\vert q_1+q_2\vert^2}
					+ e^{-\frac{1}{2}\big(\frac{1}{\ISLL_1}
						+\frac{1}{\ISLL_2}+\LinkScal^2 \big)\vert p_1-q_1+q_2\vert^2}
				\Big)/2 \nonumber\\
	&&	\quad	+ \frac{\PartNumb}{(\BoltCons T)^2}
			e^{-\LinkDens \LocaPart} \sum_{m=2}^{\infty} \frac{(\LinkDens \LocaPart)^m}{m!}
				\frac{m(m-1)}{2}\int_{\ISLL_1,\ldots,\ISLL_m}\Big(
					\frac{\tISLL_1 \tISLL_2}{\tISLL_1+\cdots+\tISLL_m}\Big)^2
				\Big(
					e^{-\frac{(\tISLL_1+\tISLL_2)\vert q_1+q_2\vert^2}{2\tISLL_1\tISLL_2}}
					+ e^{-\frac{(\tISLL_1+\tISLL_2)\vert p_1-q_1+q_2\vert^2}{2\tISLL_1\tISLL_2}}
				\Big)/2 \nonumber\\
	&& 	\quad + \frac{\PartNumb}{(\BoltCons T)^2}
			e^{-\LinkDens \LocaPart} \sum_{m=3}^{\infty} \frac{(\LinkDens \LocaPart)^m}{m!}
				m(m-1)(m-2)\int_{\ISLL_1,\ldots,\ISLL_m}
					\frac{\tISLL_1 \tISLL_2^2 \tISLL_3}{(\tISLL_1+\cdots+\tISLL_m)^2}
			\Big(
					e^{-\frac{1}{2}
							\big(\frac{\vert p_1-q_1\vert^2}{\tISLL_1}
								+\frac{\vert p_1-q_1+q_2\vert^2}{\tISLL_2}
								+\frac{\vert q_2 \vert^2}{\tISLL_3}\big)} \nonumber\\
	&&	\quad\quad	+ e^{-\frac{1}{2}
							\big(\frac{\vert q_1\vert^2}{\tISLL_1}
									+\frac{\vert q_1+q_2\vert^2}{\tISLL_2}
									+\frac{\vert q_2 \vert^2}{\tISLL_3}\big)}
					+ e^{-\frac{1}{2}
							\big(\frac{\vert p_1-q_1\vert^2}{\tISLL_1}
									+\frac{\vert q_1+q_2\vert^2}{\tISLL_2}
									+\frac{\vert p_1+q_2 \vert^2}{\tISLL_3}\big)}
					+e^{-\frac{1}{2}
							\big(\frac{\vert q_1\vert^2}{\tISLL_1}
								+\frac{\vert p_1-q_1+q_2\vert^2}{\tISLL_2}
								+\frac{\vert p_1+q_2 \vert^2}{\tISLL_3}\big)}
			\Big)/4 \nonumber\\
	&& 	\quad + \frac{\PartNumb}{(\BoltCons T)^2}
				e^{-\LinkDens \LocaPart} \sum_{m=4}^{\infty} \frac{(\LinkDens \LocaPart)^m}{m!}
				\frac{m(m-1)(m-2)(m-3)}{4}\nonumber\\
	&&\quad\times			\int_{\ISLL_1,\ldots,\ISLL_m}
					\frac{\tISLL_1 \tISLL_2 \tISLL_3 \tISLL_4}{(\tISLL_1+\cdots+\tISLL_m)^2}
				e^{-\frac{1}{2}
							\big(\frac{\vert p_1-q_1\vert^2}{\tISLL_1}
								+\frac{\vert q_1\vert^2}{\tISLL_2}
								+\frac{\vert p_1+q_2 \vert^2}{\tISLL_3}
								+\frac{\vert q_2 \vert^2}{\tISLL_4}\big)}\Big\rbrace \nonumber\\
	&&	+2i \delta_{p_1+p_2} q_1 \cdot \PPerpT_1 \cdot q_2 \Big\lbrace
				-\frac{\PartNumb \LinkDens \LocaPart^2}{2\BoltCons T} \int_{\ISLL_1,\ISLL_2}
				\Big(\frac{1}{\ISLL_1}+\frac{1}{\ISLL_2}+\LinkScal^2\Big)^{-1} \nonumber\\
	&& \quad\quad \times
				\Big\lbrack t_{p_1,q_1}
				\big(
					 e^{-\frac{1}{2}\big(\frac{1}{\ISLL_1}
					 		+\frac{1}{\ISLL_2}+\LinkScal^2 \big)\vert q_2\vert^2}
					+ e^{-\frac{1}{2}\big(\frac{1}{\ISLL_1}
							+\frac{1}{\ISLL_2}+\LinkScal^2 \big)\vert p_1+q_2\vert^2}
				\big)/2 \nonumber\\
	&&\quad			-t_{-p_1,q_2}
				\big(
					 e^{-\frac{1}{2}\big(\frac{1}{\ISLL_1}
					 	+\frac{1}{\ISLL_2}+\LinkScal^2 \big)\vert q_1\vert^2}
					+ e^{-\frac{1}{2}\big(\frac{1}{\ISLL_1}
						+\frac{1}{\ISLL_2}+\LinkScal^2 \big)\vert -p_1+q_1\vert^2}
				\big)/2
				\Big\rbrack \nonumber\\
	&&	\quad	+ \frac{\PartNumb}{(\BoltCons T)^2}
			e^{-\LinkDens \LocaPart} \sum_{m=2}^{\infty} \frac{(\LinkDens \LocaPart)^m}{m!}
				\frac{m(m-1)}{2}\int_{\ISLL_1,\ldots,\ISLL_m}
					\frac{\tISLL_1 \tISLL_2(\tISLL_1 + \tISLL_2)}{(\tISLL_1+\cdots+\tISLL_m)^2}
					 \nonumber\\
	&&	\quad\quad\times		\Big\lbrack	t_{p_1,q_1}
					\big(e^{-\frac{(\tISLL_1+\tISLL_2)\vert q_2\vert^2}{2\tISLL_1\tISLL_2}}
						+ e^{-\frac{(\tISLL_1+\tISLL_2)\vert p_1+q_2\vert^2}{2\tISLL_1\tISLL_2}}
					\big)/2
				- t_{-p_1,q_2}
					\big(e^{-\frac{(\tISLL_1+\tISLL_2)\vert q_1\vert^2}{2\tISLL_1\tISLL_2}}
						+ e^{-\frac{(\tISLL_1+\tISLL_2)\vert -p_1+q_1\vert^2}{2\tISLL_1\tISLL_2}}
					\big)/2
				\Big\rbrack \nonumber\\
	&&	\quad + \frac{\PartNumb}{(\BoltCons T)^2}
			e^{-\LinkDens \LocaPart} \sum_{m=3}^{\infty} \frac{(\LinkDens \LocaPart)^m}{m!}
				m(m-1)(m-2)\int_{\ISLL_1,\ldots,\ISLL_m}
					\frac{\tISLL_1 \tISLL_2^2 \tISLL_3}{(\tISLL_1+\cdots+\tISLL_m)^2} \nonumber\\
	&&	\quad\quad\times		\Big\lbrack t_{p_1,q_1} \big(
						- e^{-\frac{1}{2} \big(\frac{1}{\tISLL_2}
							+\frac{1}{\tISLL_3}\vert q_2\vert^2 \big)}
						+ e^{-\frac{1}{2} \big(\frac{1}{\tISLL_2}
							+\frac{1}{\tISLL_3}\vert p_1+q_2\vert^2 \big)}\big)/4 \nonumber\\
	&&\quad				- t_{-p_1,q_1} \big(
						- e^{-\frac{1}{2} \big(\frac{1}{\tISLL_2}
							+\frac{1}{\tISLL_3}\vert q_2\vert^2 \big)}
						+ e^{-\frac{1}{2} \big(\frac{1}{\tISLL_2}
							+\frac{1}{\tISLL_3}\vert - p_1+q_2\vert^2 \big)}\big)/4 \Big\rbrack
			\Big\rbrace ,
\end{eqnarray}
\end{widetext}
where we have used the notation $\NLM(z,y)\equiv\tNonLocaKern(z_1,z_2)$ defined in Appendix~\ref{APP:REHT}.

\subsection{Disorder correlators of the elastic parameters
in the local form}
\label{APP:CFLL}
In this appendix we calculate the disorder correlators of the quenched random elastic parameters in the local form of the elastic energy for deformations relative to the relaxed state.

First, we calculate the disorder correlator of the residual stress $\lda \Stre\Stre \rda_c$.  The residual stress $\StreT$ in the relaxed state is related to the nonlocal kernel $\NonLocaKern$ via Eq.~(\ref{EQ:Stress}). Thus, the correlator of the residual stress can be expressed as
\begin{widetext}
\begin{eqnarray}\label{EQ:SSNN}
	&&	\lda \Stre_{ij,p_1}\Stre_{kl,p_2} \rda_c  \nonumber\\
	\!\! &=& \!\!\frac{\partial}{\partial q_{1,i}}\Big\vert_{q_1=0}
			\frac{\partial}{\partial q_{2,j}}\Big\vert_{q_2=0} \lda N_{j,p_1,q_1} N_{l,p_2,q_2} \rda_c \nonumber\\
	&&\!\!
			- \frac{2}{\vert p_1 \vert^2} \big(p_{1,k}\PPerpT_{bl}(p_1)+p_{1,l}\PPerpT_{bk}(p_1)+p_{1,b}\PPerpT_{kl}(p_1)\big)
			\frac{\partial}{\partial q_{1,i}}\Big\vert_{q_1=0}  \lda N_{j,p_1,q_1} N_{l,p_2,0} \rda_c \nonumber\\
	&&\!\!
			+ \frac{2}{\vert p_1 \vert^2} \big(p_{1,i}\PPerpT_{aj}(p_1)+p_{1,j}\PPerpT_{ai}(p_1)+p_{1,a}\PPerpT_{ij}(p_1)\big)
			\frac{\partial}{\partial q_{2,k}}\Big\vert_{q_2=0}  \lda N_{j,p_1,0} N_{l,p_2,q_2} \rda_c \nonumber\\
	&&\!\!	- \frac{2}{(\vert p_1 \vert^2)^2}
			 \big(p_{1,i}\PPerpT_{aj}(p_1)\!+p_{1,j}\PPerpT_{ai}(p_1)\!+p_{1,a}\PPerpT_{ij}(p_1)\big) \!
			 \big(p_{1,k}\PPerpT_{bl}(p_1)\!+p_{1,l}\PPerpT_{bk}(p_1)\!+p_{1,b}\PPerpT_{kl}(p_1)\big)
			\lda N_{j,p_1,0} N_{l,p_2,0} \rda_c  , \,
\end{eqnarray}
\end{widetext}
where we have defined
  $N_{j,p,q} \equiv {\partial \NLM_{p,q}}/{\partial q_j}$,
and the notation $\NLM(z,y)\equiv\tNonLocaKern(z_1,z_2)$ is defined in Appendix~\ref{APP:REHT}.

We then insert in the disorder correlator
$\lda \NLM_{p_1,q_1}\NLM_{p_2,q_2} \rda _c$,
given in Eq.~(\ref{EQ:KTwo})
(in the form of $\lda\NonLocaKern\NonLocaKern\rda$) into Eq.~(\ref{EQ:SSNN}).
After a lengthy calculation, and making use of the identity
\begin{eqnarray}
	&& m\int_{\ISLL_1,\ldots,\tISLL_m}\frac{\tISLL_1^2}{(\tISLL_1+\cdots+\tISLL_m)^2} \nonumber\\
	&& + m(m-1)\int_{\ISLL_1,\ldots,\tISLL_m}
    \frac{\tISLL_1\,\tISLL_2}{(\tISLL_1+\cdots+\tISLL_m)^2}=1,
\end{eqnarray}
we arrive at the correlator
\begin{eqnarray}
	&& \lda \Stre_{ij,p_1}\Stre_{kl,p_2}\rda_c
    \nonumber\\
	&=& \delta_{p_1+p_2} \frac{\PartNumb \UnivPara}{(\BoltCons T)^2}
	(2\PPerpT_{ij}\PPerpT_{kl}+\PPerpT_{il}\PPerpT_{jk}+\PPerpT_{ik}\PPerpT_{jl}),
\end{eqnarray}
where $\UnivPara\equiv -\frac{1}{2}{\LinkDens \LocaPart^2}+\LinkDens \LocaPart
	+e^{-\LinkDens \LocaPart}-1$ is given in Eq.~(\ref{EQ:UnivParaDef}).

By following a similar scheme, we have also calculated the
other disorder correlators and cross-correlators of the
quenched random elastic parameters
in the local form of elasticity of the relaxed state.
Hence, we arrive at the correlators of the shear modulus and bulk modulus, which read
\begin{subequations}
\begin{align}
	\lda \SheaModu_{p_1}\SheaModu_{p_2} \rda_c =& \Corr \,\delta_{p_1+p_2} \PartNumb(\BoltCons T)^2 , \\
	\lda \BulkModu_{p_1}\BulkModu_{p_2} \rda_c =& 4 \Corr\, \delta_{p_1+p_2} \PartNumb(\BoltCons T)^2 ,
\end{align}
\end{subequations}
in which the dimensionless scale factor $\Corr$ is given by
\begin{eqnarray}\label{EQ:AppDefiCorr}
	\Corr \equiv -\frac{3}{2}\LinkDens \LocaPart^2 + e^{\LinkDens \LocaPart}-1+\LinkDens \LocaPart+(\LinkDens \LocaPart)^2 .
\end{eqnarray}
We also arrive at the cross-correlators, which are given by
\begin{subequations}
\begin{eqnarray}
	\lda \Stre_{ij,p_1} \SheaModu_{p_2}\rda_c &=&
		 -2 \PartNumb (\BoltCons T)^2  \UnivPara\, \delta_{p_1+p_2}\PPerpT_{ij}(p_1), \\
	\lda \Stre_{ij,p_1} \BulkModu_{p_2} \rda_c &=&
		4 \PartNumb (\BoltCons T)^2 \UnivPara\, \delta_{p_1+p_2} \PPerpT_{ij}(p_1), \\
	\lda \SheaModu_{p_1} \BulkModu_{p_2} \rda_c &=&
		-2 \PartNumb(\BoltCons T)^2\Corr\,\delta_{p_1+p_2};
\end{eqnarray}
\end{subequations}
the scale factor  $\UnivPara$ is defined in Eq.~(\ref{EQ:UnivParaDef}).


\end{document}